\documentclass[prd,onecolumn,amsmath,amssymb,floatfix,nofootinbib]{revtex4}

\usepackage{amssymb}
\usepackage{amsmath}

\usepackage{graphicx}
\usepackage{graphics}
\usepackage{dcolumn}
\usepackage{color}
\usepackage{rotate}
\usepackage{fancyhdr} 
\usepackage{hyperref}
\usepackage{indentfirst}

\usepackage{wasysym}

\usepackage{umoline}
\usepackage{ulem}

\def\be{\begin{eqnarray}}
\def\ee{\end{eqnarray}}
\def\no{\nonumber}

\definecolor{darkred}{rgb}{.743,0,0}

\newcommand\eqspla[2]{\begin{equation}\label{#1}
\begin{split}
#2
\end{split}
\end{equation}}

\begin{document}
\title{Neutrino flavour as a test of the explosion mechanism of core-collapse supernovae}

\author{Nitsan Bar}
\email{nitsan.bar@weizmann.ac.il}
\affiliation{Weizmann Institute of Science, Rehovot, Israel 7610001}
\author{Kfir Blum}
\email{kfir.blum@cern.ch}
\affiliation{Weizmann Institute of Science, Rehovot, Israel 7610001}
\affiliation{Theory department, CERN, CH-1211 Geneve 23, Switzerland}
\author{Guido D'Amico}
\email{guido.damico@cern.ch}
\affiliation{Theory department, CERN, CH-1211 Geneve 23, Switzerland}
\affiliation{Stanford Institute for Theoretical Physics, Stanford University, Stanford, CA 94306, USA}


\begin{abstract}
We study the ratio of neutrino-proton elastic scattering to inverse beta decay event counts, measurable in a scintillation detector like JUNO, as a key observable for identifying the explosion mechanism of a galactic core-collapse supernova. 
If the supernova is not powered by the core but rather, e.g., by collapse-induced thermonuclear explosion, then a prolonged period of accretion-dominated neutrino luminosity is predicted. 
Using 1D numerical simulations, we show that the distinct resulting flavour composition of the neutrino burst can be tested in JUNO with high significance, overcoming theoretical uncertainties in the progenitor star profile and equation of state. 
%
%
\end{abstract}

\maketitle


\section{Introduction}\label{sec:intro}
How do core-collapse supernovae (CCSNe) explode? 
A popular belief is that the explosion follows the delayed neutrino mechanism (D$\nu$M)~\cite{Bethe:1984ux,Kotake:2005zn,Burrows:2012ew,Janka:2016fox}. In the D$\nu$M, explosion is hypothesized to start a fraction of a second after core-collapse with the neutrino-assisted  revival of a stalled core bounce shock, about $\sim100$~km above the surface of a proto-neutron star (PNS). 
This paradigm, however, has not yet been fully corroborated by numerical simulations. Numerically converged spherically symmetric (1D) D$\nu$M simulations of iron-core stars consistently fail to explode~\cite{OConnor:2018sti}. Simulations in 2D and 3D pose an extreme numerical challenge and, as a result, present somewhat controversial conclusions. Numerical convergence has not yet been demonstrated for simulations that produce explosions. 
Typically, numerical noise is introduced in order to seed non-spherical instabilities\footnote{\label{fn:noise}See, e.g., the discussion in App.~1 and  Sec.2.1 of~\cite{Nakamura:2014caa}; in Sec.5 of~\cite{Seadrow:2018ftp}; in Sec.2 of~\cite{Tamborra:2014hga}; or in Sec.~4 of~\cite{Vartanyan:2018iah} and references there.}, instead of deriving these instabilities from the actual physical conditions during collapse~\cite{Couch:2015gua,2016ApJ...833..124M}. 
To date, a self-consistent exploding simulation that reproduces the energetics of, e.g., SN1987A, has not been reported. 
Success could be just around the corner~\cite{Vartanyan:2018xcd}; nevertheless, it may be useful to keep an open mind to the possibility that the D$\nu$M fails.

While not yet at the same level of sophistication as D$\nu$M simulations, there are competing alternative models to explain CCSNe~\cite{Janka:2012wk}. Some of these proposals, including the MHD mechanism of~\cite{LeBlanc:1970kg,1979SvA....23..705A,Ardeljan:2004fq,Burrows:2007yx,Takiwaki:2010cf}, the acoustic mechanism of~\cite{Burrows:2005dv}, or the jet activity suggested in~\cite{Gilkis:2015adr}, share with the D$\nu$M the attempt to couple a small fraction (of order a percent) of the gravitational binding energy release of the collapsing core into an outward explosion of the envelope. Another, qualitatively different model is collapse-induced thermonuclear explosion (CITE)~\cite{Burbidge:1957vc,Hoyle:1960zz,Fowler:1964zz,Kushnir:2014oca,Kushnir:2015mca}. In CITE, the explosion occurs $\mathcal{O}(10)$~sec after core collapse due to thermonuclear detonation, triggered some $\sim10^4$~km out of the core when a rotation-induced accretion shock traverses an explosive layer in the envelope. 
Unlike the D$\nu$M, the source of energy here is not the gravitational energy of the core. Instead, the observed explosion kinetic energies of CCSNe, $E_{K}\sim10^{51}-10^{52}$~erg, are reproduced by the $\sim$MeV nuclear binding energy per nucleon, released in burning a few M$_\odot$ of the progenitor star~\cite{Kushnir:2015vka}. 
On the one hand, the initial conditions required for CITE have not yet been demonstrated in self-consistent stellar evolution codes~\cite{Gofman:2018ldp}. On the other hand, converged 2D numerical simulations of CITE with a wide variety of initial conditions explode stars with correct energetics~\cite{Kushnir:2015vka,Kushnir:2015mca}. 

The Rosetta Stone of CCSNe is the neutrino burst. The neutrino burst of SN1987A~\cite{Hirata:1987hu,Bionta:1987qt} was consistent with the broad characteristics of core-collapse~\cite{Spergel:1987ch,Lattimer:1989zz,Loredo2002,Pagliaroli2009,Vissani:2014doa}. However, the sparse data were not enough to identify the explosion mechanism: CITE and the D$\nu$M are equally compatible with the neutrino burst of SN1987A~\cite{Blum:2016afe}. 
The situation would change with the occurrence of a neutrino burst from a galactic CCSN~\cite{Adams:2013ana}, that would trigger thousands of events in terrestrial detectors (see~\cite{Mirizzi:2015eza,Horiuchi:2017sku} for recent reviews). 
Still, core-collapse is complicated and the neutrino burst  exhibits temporal behaviour reflecting the density, composition, and internal rotation profile, that vary between different progenitor stars~\cite{OConnor2011a,Woosley:2002zz,Sukhbold:2013yca}. It is therefore important to identify robust signatures of the main physical phenomena in the explosion.

In this work we study neutrino flavour -- specifically the $\bar\nu_e$ fraction of the total neutrino flux -- as a key diagnostic of a CCSN neutrino burst. 
We apply this diagnostic to the two classes of models discussed above. The first class of models posits that the explosion is powered by the collapsing core; we use the D$\nu$M to represent this idea, noting that it may be more generally applicable and include e.g.~\cite{LeBlanc:1970kg,1979SvA....23..705A,Ardeljan:2004fq,Burrows:2007yx,Takiwaki:2010cf} and~\cite{Burrows:2005dv}. The second class of models posits that the explosion may be decoupled from the dynamics of the core; this will be represented by CITE. 
The physics point we wish to study hinges on a difference in the neutrino source of the two classes of models on times larger than a few hundred milliseconds post-bounce. 

In both classes of models, core bounce is immediately followed by a quasi-stationary shock forming at $r_{shock}\lesssim100$~km. Matter falls through this shock with an accretion rate $\dot M\sim$(0.1-1)~M$_\odot$/s. During this accretion phase\footnote{The accretion phase is preceded by a short ($\sim$~ms duration) burst of $\nu_e$ from the de-leptonisation of the core. This initial $\nu_e$ burst is common to all models and we do not discuss it further, focusing instead on the post-bounce dynamics.}, the neutrino source contains a  component of $\nu_e$ and $\bar\nu_e$ coming from $e^+e^-$ annihilation\footnote{The cross section for $e^+e^-\to\nu_e\bar\nu_e$ is $\sim4.5$ times larger than that for $e^+e^-\to\nu_x\bar\nu_x$ in the energy range of interest ($E_{e^\pm}<100$~MeV).} and nucleon conversion in an optically thin region above the PNS neutrinosphere and below the shock. This accretion luminosity is augmented by PNS cooling luminosity of neutrinos of all flavours. Overall, numerical simulations find a source flavour ratio of 
\be\label{eq:fe} \frac{L_{\bar\nu_e}}{L_{\nu_x}}=f_{\bar e},\;\;\;\;\frac{L_{\nu_e}}{L_{\nu_x}}= f_e,\;\;\;\;\;f_{\bar e}\approx f_e\gtrsim2,\ee 
during the accretion phase, where $ L_{\nu_x}=L_{\nu_\mu}=L_{\nu_\tau}=L_{\bar{\nu}_\mu}=L_{\bar{\nu}_\tau}$. 

The initial accretion phase is the same in CITE and in the D$\nu$M. What comes next, however, is  different. 
In the D$\nu$M, within a fraction of a second from core bounce, the stalled shock above the PNS must transition into an outward explosion, if the star is to explode at all. The explosion strips off the accreting material above the PNS neutrinosphere. From this point on, the neutrino source in the D$\nu$M is predicted to be a bare PNS, emitting a comparable flux of neutrinos of all flavours. 
Importantly, the D$\nu$M predicts that $f_{\bar e}$ and $f_e$ decrease in time, approaching $f_{\bar e}\sim f_e\sim1$, once the explosion gets on its way.

\begin{figure}[htb!]
	\centering
        \includegraphics[clip, trim=0.9cm 7.7cm 0.2cm 1.5cm, width=0.7\textwidth]{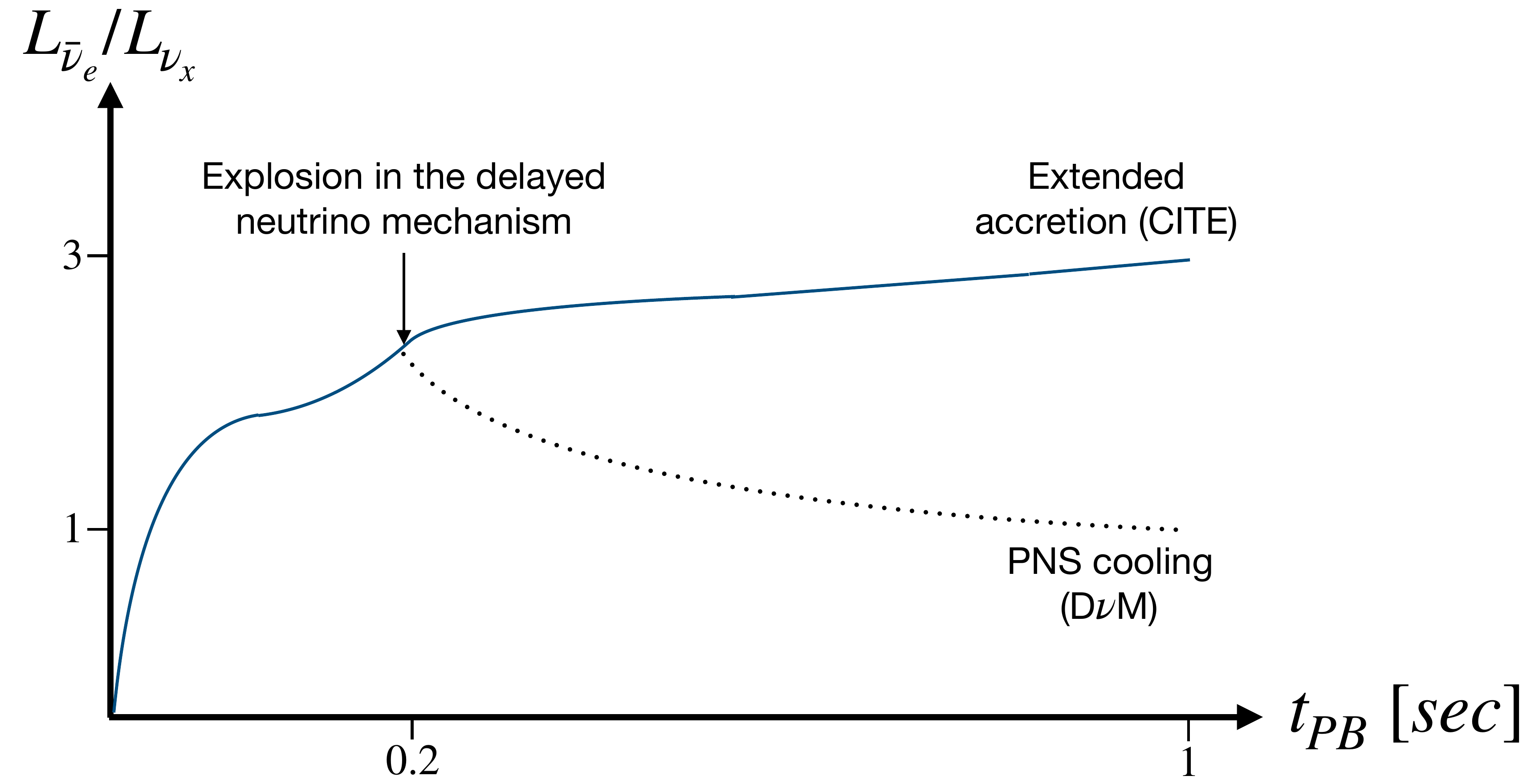}
	\caption{Cartoon of luminosity ratio vs. post-bounce time. For post-bounce times $t_{PB}\lesssim0.2$~sec different explosion mechanisms agree. Around $t_{PB}\sim0.2-0.4$~sec, the D$\nu$M predicts that the explosion should start, after which neutrino emission is thought to be dominated by PNS cooling (dashed line). On the other hand, CITE predicts an extended period of accretion that may last for several seconds. In this work we focus on the observational identification of $L_{\bar\nu_e}/L_{\nu_x}>1$ during $t_{PB}\gtrsim 0.2$~sec.}\label{fig:cartoon}
\end{figure}

In CITE, the same initial period of accretion luminosity at $t<1$~sec continues on to $t>1$~sec and can last up to a few seconds, much longer than is expected in the D$\nu$M. In some cases, especially for strong explosions with $E_K\gtrsim10^{51}$~erg, the extended accretion phase can lead to the direct formation of a black hole (BH)~\cite{Kushnir:2015mca,Blum:2016afe}\footnote{Ref.~\cite{Gofman:2018ldp} points out that the $E_K\sim1.9\times10^{51}$~erg inferred for SNR Kes 73 (G27.4+0.0)~\cite{Borkowski:2017nkc}, which hosts a NS remnant, may pose a problem for CITE.}. BH formation abruptly cuts-off neutrino emission due to the PNS cooling {\it and} the spherical accretion above it. However, if the star is rotating (as is needed for CITE), an accretion disk forms. Further neutrino emission comes from accretion disk luminosity, that is again dominated by $\nu_e$ and $\bar\nu_e$ at the source (see, e.g.~\cite{Popham:1998ab,MacFadyen:1998vz,Liu:2015prx,Blum:2016afe}). 
This scenario leads to $f_{\bar e}\approx f_e\gtrsim2$, most likely rising with time, throughout the neutrino burst.   A cartoon of $ L_{\bar{\nu}_e}/L_{\nu_x} $ during the first 1~sec post-bounce is shown in Fig.~\ref{fig:cartoon}.

Can the neutrino flavour at the source be measured by terrestrial detectors? the answer is yes, for specific flavour observables, given the appropriate detector, and assuming that some theoretical features of neutrino oscillations are understood. 
Consider the ratio: 
\be\label{eq:R}
R &=&\frac{\rm number\;of\;pES\; events}{\rm number\;of\;IBD\; events},
\ee
measurable in a scintillation detector like JUNO~\cite{An:2015jdp}, currently under construction (see also the LENA~\cite{Wurm:2010ny} proposal). 
Here, pES denotes neutrino-proton elastic scattering, $\nu+p\to\nu+p$, sensitive to all neutrino species\footnote{{\it Almost} equally; see Sec.~\ref{ssec:pes} and App.~\ref{app:xs}.}, and IBD stands for inverse beta decay, $ \bar{\nu}_e+p\to e^++n$. 
Obviously, $R$ measures neutrino flavour at Earth. Connecting this information to flavour at the source is a nontrivial task, requiring knowledge of neutrino propagation and detection cross sections. Even once this is achieved, theoretical uncertainties in the source itself could affect the interpretation of the results. Our goal in this work is to study both sources of uncertainty, those related to propagation and detection and those related to source modelling. 
%
%
The end result of our analysis is that with a realistic treatment of all of these effects, in the event of a galactic CCSN JUNO could identify $f_{\bar e}\gtrsim2$ (predicted in CITE throughout the neutrino burst for $t\gtrsim0.3$~sec post-bounce) from $f_{\bar e}\sim1$ (predicted in the D$\nu$M once explosion starts), with high significance. 


There are several caveats. 
The most important caveat is due to neutrino self-induced oscillations at the source, which is an open theoretical problem. We will assume that self-induced oscillations do not play an important role in the neutrino propagation. This assumption appears realistic for the prolonged accretion  scenario (CITE)~\cite{Chakraborty:2011nf,Sarikas2012b,Chakraborty:2014nma}, but it may be less realistic for the D$\nu$M. This makes the predictions for CITE more robust than for the D$\nu$M. Even with this caveat, our results would be useful also if self-induced oscillations do play a role: in that case, the measurement of $R$ can help to identify the self-induced oscillations effect. 
A second caveat is the theoretical uncertainty on the pES cross section. There is a proposal to eliminate this problem~\cite{Pagliaroli2013}, possibly in JUNO itself, and we assume that this program succeeds. We find that all of the other issues, including the theoretical uncertainty in progenitor star properties and EoS, are under control, in the sense that accretion-dominated neutrino emission leads to robust predictions and can be distinguished from PNS cooling.

CCSN neutrino flavour and the relevant capabilities of scintillation detectors were studied in a number of previous works~\cite{Beacom:2002hs,Dasgupta:2011wg,Laha:2013hva,Laha:2014yua,Nikrant2017,Minakata:2001cd,Barger:2001yx,Minakata:2008nc,GalloRosso:2017hbp,GalloRosso:2017mdz,Lu:2016ipr,Horiuchi2017,Seadrow:2018ftp}. Ref.~\cite{Beacom:2002hs,Dasgupta:2011wg,Nikrant2017} highlighted the importance of pES as a probe of the total neutrino flux at the source. Ref.~\cite{Laha:2014yua} analysed the prospects for measuring the spectrum of CCSN $\nu_e$ via various sub-leading reactions in scintillation detectors (following~\cite{Laha:2013hva} which considered the measurement of $\nu_e$ in water Cherenkov detectors doped with Gadolinium). 
Still focusing on $\nu_e$, Ref.~\cite{Nikrant2017} took a similar approach to ours, in that the neutrino spectra were based on numerical simulations~\cite{Nakamura:2014caa}
and the need to fit for an a-priori unknown spectral shape was accounted for. The neutronization, accretion, and PNS cooling phases of the neutrino burst were analysed w.r.t. the capabilities of the water Cherenkov detectors Super-K~\cite{Fukuda:2002uc} and Hyper-K~\cite{Abe:2018uyc} and the liquid Ar detector DUNE~\cite{Acciarri:2016ooe}. 
%
Ref.~\cite{Minakata:2001cd,Barger:2001yx,Minakata:2008nc,GalloRosso:2017hbp,GalloRosso:2017mdz} presented likelihood analyses of spectral reconstruction in water Cherenkov detectors, 
assuming analytic equipartition (PNS cooling) source models. 

Ref.~\cite{Lu:2016ipr} considered the accuracy by which JUNO could reconstruct the mean energy per neutrino $\langle E_{\nu_\alpha}\rangle$ and the total energy carried by neutrinos of each flavour $E^{\rm tot}_{\nu_\alpha}$, assuming an analytic equipartition source model. 
Accretion-dominated neutrino emission, in which energy equipartition is violated, was not studied. In comparison, we use numerical simulations aiming to go beyond an illustration of the detector capabilities, and study the actual predicted neutrino burst.

Ref.~\cite{Horiuchi2017} used 2D CCSN simulations to analyse the expected signals in Super-K,  Hyper-K, and DUNE, focusing on the relation between the progenitor core structure and the neutrino signal intensity during the accretion phase. 
Ref.~\cite{Seadrow:2018ftp} used 1D and 2D CCSN simulations~\cite{Burrows:2016ohd,Radice:2017ykv,Vartanyan:2018xcd} to analyse the expected signals in Super-K, DUNE, JUNO, and IceCube. The dependence on progenitor mass in the range 9-21~M$_\odot$ was studied. The characteristics of the accretion phase were discussed, emphasizing the $\bar\nu_e$ and $\nu_e$ dominance, and arguing that the cessation of the accretion component due to the explosion would provide key information to test the D$\nu$M. To demonstrate this point, Ref.~\cite{Seadrow:2018ftp} compared the detected neutrino light-curves obtained for 1D ``non-exploding" and 2D ``exploding" simulations 
of the same progenitor.
The analysis in~\cite{Seadrow:2018ftp} therefore closely overlaps with our current work. There are, however, important differences. First, Ref.~\cite{Seadrow:2018ftp} altogether ignored pES in JUNO. As we discuss (see also~\cite{Beacom:2002hs,Dasgupta:2011wg,An:2015jdp,Lu:2016ipr}), pES is actually the second most important class of events (after IBD) in JUNO, assuming a realistic lower energy threshold, and it plays a key role in the current work. Second, while Ref.~\cite{Seadrow:2018ftp} presented qualitative comparisons between the expected neutrino signals for ``exploding" vs. ``non-exploding" simulations, it focused solely on integrated total event rates and did not analyse in detail the prospects to actually discriminate between the two cases in a detection. Using total event rates, this discrimination would be challenging\footnote{Compare, for example, the JUNO event rates presented in the left (``non-exploding") and right (``exploding") panels in Fig.~11 in~\cite{Seadrow:2018ftp}, and recall that in a detection, the actual structure of the progenitor would be unknown.}. One of our main results in this work is that flavour information, specifically the pES/IBD event rate ratio, provides a robust test of a continued accretion phase.

This paper is structured as follows. 
In Sec.~\ref{sec:accphase} we demonstrate the basic physical features of the accretion-dominated neutrino luminosity, expected to characterise the neutrino light-curve in CITE and also in the early stages of the CCSN in the D$\nu$M. We use numerical simulations to illustrate the discussion. For completeness, in App.~\ref{app:comp} we review analogous results from numerical simulations conducted by different groups and using different codes. 

In Sec.~\ref{sec:detect} we discuss channels for CCSN neutrino detection in a scintillation detector, focusing on JUNO. We highlight the IBD and pES channels, which are dominant in terms of event statistics, and explain our analysis strategy including cuts in the deposited energy spectrum. The importance of achieving a low quenched proton recoil energy threshold is explained. Some cross section formulae are collected in App.~\ref{app:xs}. 

In Sec.~\ref{sec:osc} we describe our treatment of neutrino flavour conversion in the CCSN, relating flavour-specific neutrino source luminosities to the luminosities at Earth. In the case of CITE, neutrino propagation proceeds in the deep adiabatic regime throughout the first few seconds after core-collapse. We explain this result in App.~\ref{app:osc}. 

In Sec.~\ref{sec:juno} we analyse results from numerical simulations. Our main goal in this section is to explore (i) the theoretical uncertainty arising from different progenitor stars and EoS, and (ii) realistic statistical uncertainties for a galactic CCSN. We also study the impact of uncertainties in neutrino oscillation parameters. 

In Sec.~\ref{sec:sum} we summarise our results.

%
%

\section{$\bar\nu_e/\nu_x$ ratio during the accretion phase}\label{sec:accphase}

\begin{figure}[htb!]
\centering
\includegraphics[width=0.495\textwidth]{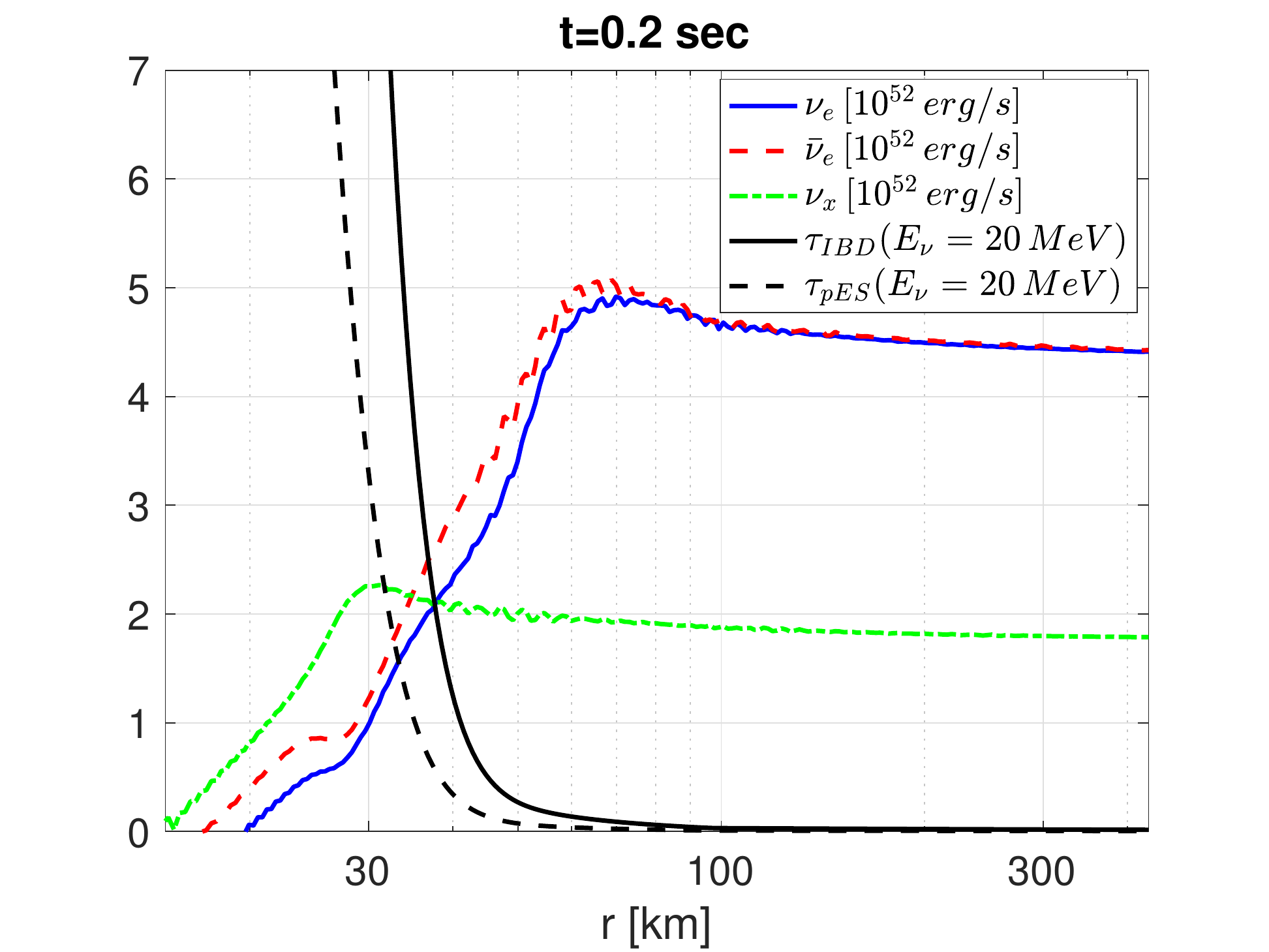}
\includegraphics[width=0.495\textwidth]{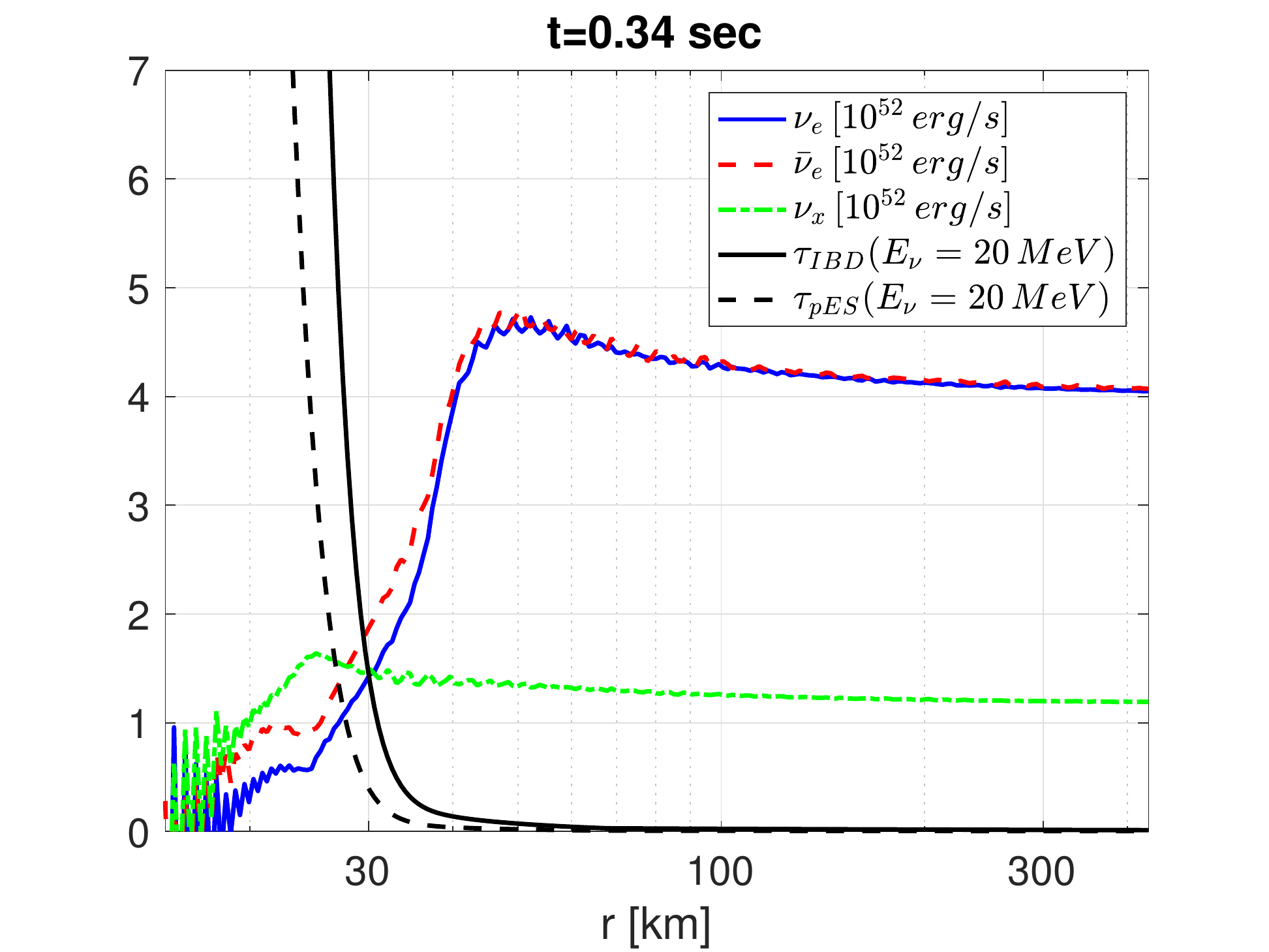}
\caption{Snapshots of the radial profiles of the neutrino luminosity per flavour (solid blue, dashed red, dot-dashed green referring to $\nu_e,\bar\nu_e,\nu_x$, respectively), during the accretion phase, calculated with GR1D. Solid (dashed) black line shows the optical depth to scattering for IBD (pES) for a neutrino with energy $E_\nu=20$~MeV. {\bf Left:} $t=0.2$~sec post-bounce. {\bf Right:} $t=0.34$~sec post-bounce.
}\label{fig:Lr}
\end{figure}

Our goal in this section is to clarify the origin of the accretion luminosity, that leads to excess $\nu_e,\bar\nu_e$ compared to the x-flavours. 
In Fig.~\ref{fig:Lr} we show radial profiles of the neutrino luminosity per flavour at fixed time, for two post-bounce times $t=0.2,0.34$~sec (left and right panels, respectively) during the accretion phase, calculated with the (non-exploding) open-source general-relativistic hydrodynamics 1D code GR1D~\cite{2010CQGra..27k4103O,OConnor:2014sgn} for a 15~M$_\odot$ progenitor star\footnote{Our progenitor profiles are taken from the non-rotating, solar metallicity sample of~\cite{Woosley:2002zz} (see https://2sn.org/stellarevolution/data.shtml). Masses refer to zero-age main-sequence.}, with the SLy4 EoS~\cite{Chabanat:1997un,daSilvaSchneider:2017jpg}. Black solid (dashed) lines show the optical depth to IBD (pES), the first being the key quantity for $\bar\nu_e$ and the second for $\nu_x$. Both are computed for neutrino energy $E_\nu=20$~MeV. 

The $\nu_e,\bar\nu_e$ luminosities in Fig.~\ref{fig:Lr} are dominated by nucleon conversion, $p+e^-\rightarrow n+\nu_e$ and $n+e^+\rightarrow p+\bar\nu_e$, and $ e^-e^+ $ annihilation taking place in the region below the accretion shock and above the neutrinosphere. $L_{\bar\nu_e}$ and $L_{\nu_e}$ continue to build up in the optically thin region ($\tau_{pES},\,\tau_{IBD}<1$), all the way to the location of the accretion shock, at which point the luminosity saturates. The contribution to the luminosity coming from the optically-thin region, is what we refer to as accretion luminosity. In contrast, x-flavour emission saturates on smaller radii near the x-flavour neutrinosphere ($\tau_{pES}\approx1$). 

\begin{figure}[htb!]
\centering
\includegraphics[width=0.6\textwidth]{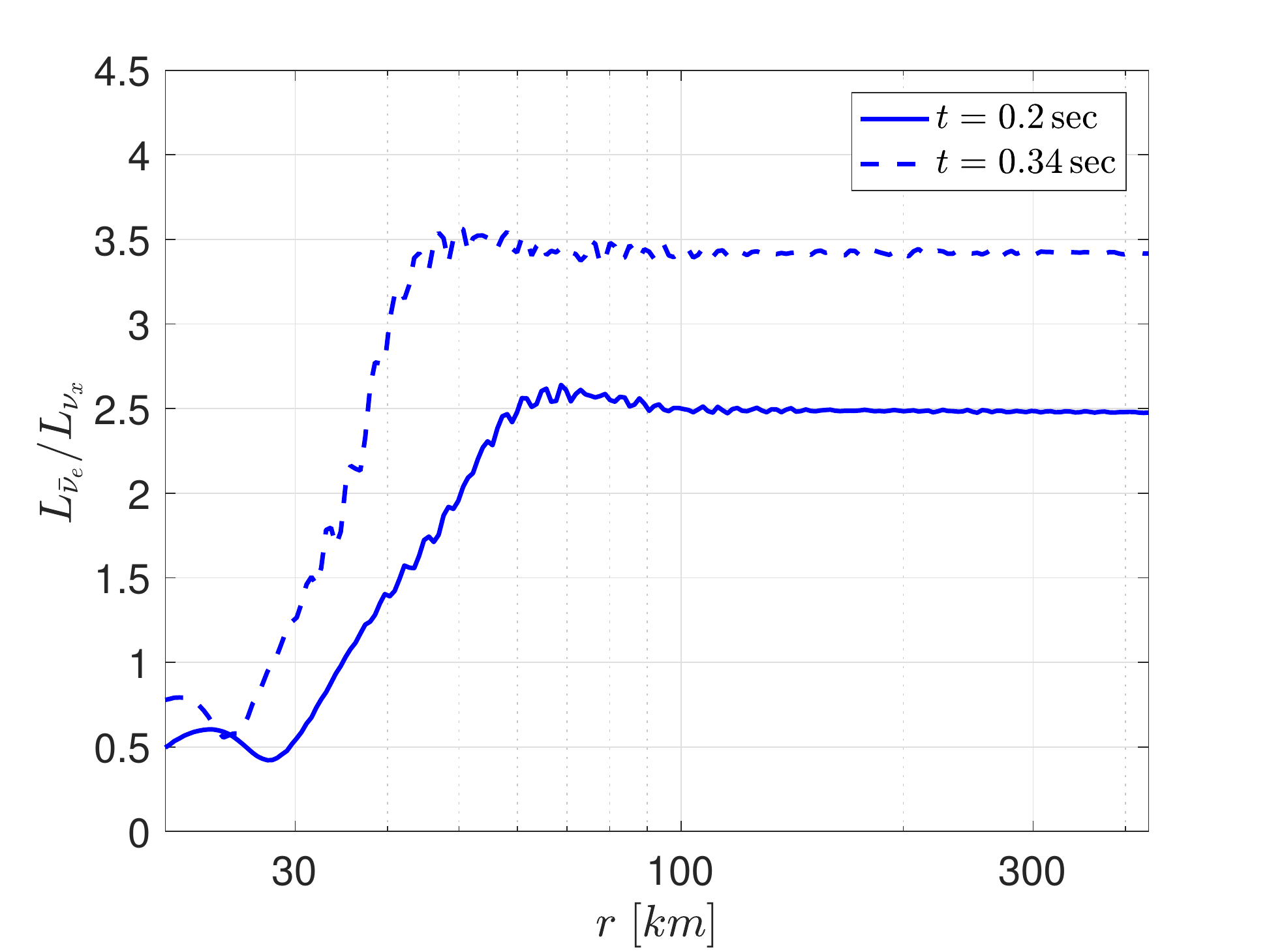}
\caption{Snapshots of the radial profile of $f_e=L_{\bar\nu_e}/L_{\nu_x}$ during the accretion phase, calculated with GR1D. Smooth (dashed) lines refer to snapshots at $t=0.2$~sec ($t=0.34$~sec) post-bounce.
}\label{fig:Lanue2Lnuxr}
\end{figure}

\begin{figure}[htb!]
\centering
\includegraphics[width=0.6\textwidth]{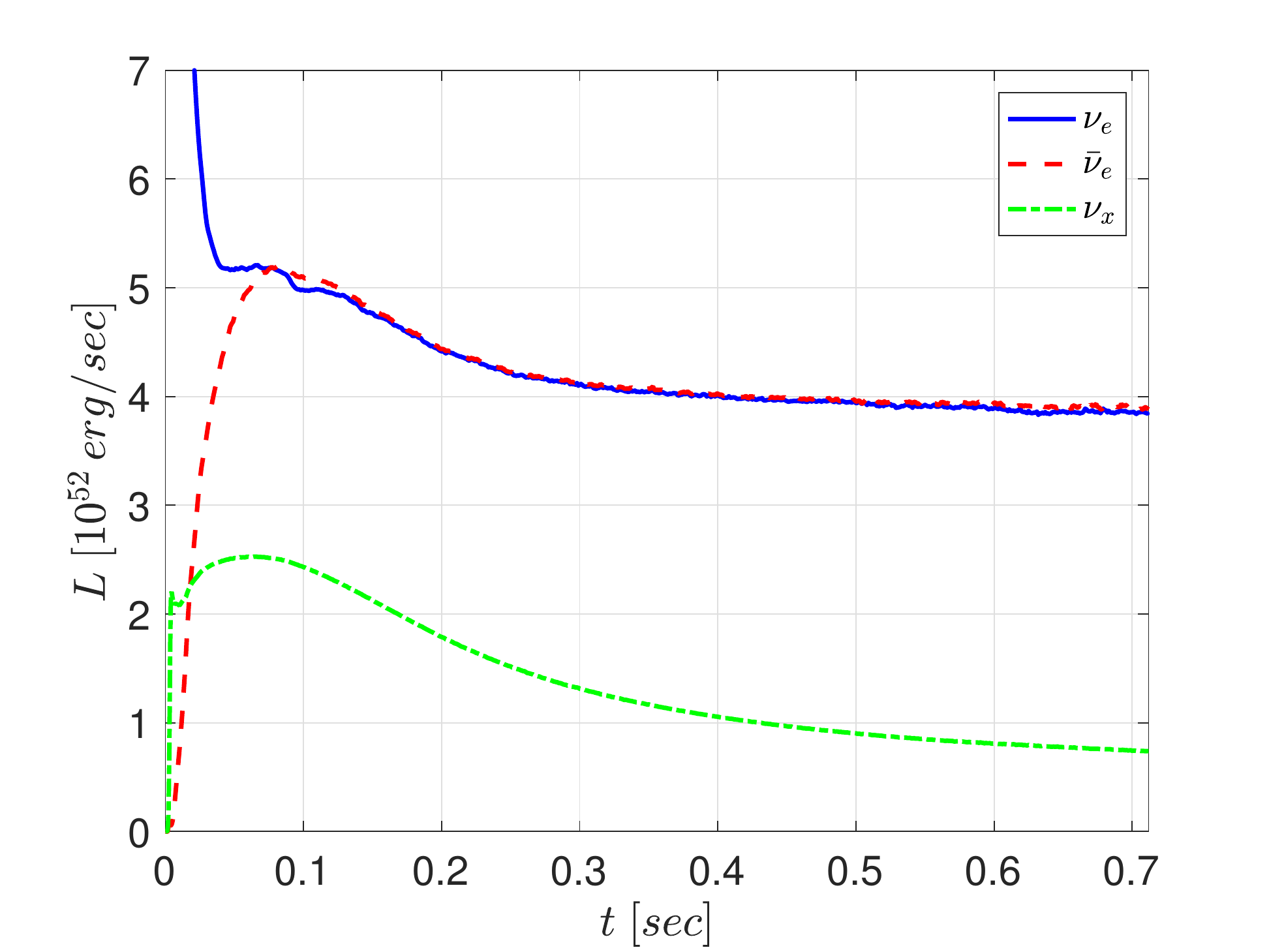}
\caption{
Time-dependent neutrino luminosity crossing through $r=400$~km during the accretion phase, calculated with GR1D. In this plot, we do not include neutrino oscillations, the effect of which will be accounted for later. $t$ is time post-bounce.
}\label{fig:L1}
\end{figure}

The ratio $f_{\bar e}=L_{\bar\nu_e}/L_{\nu_x}$, computed for both panels of Fig.~\ref{fig:Lr}, is shown in Fig.~\ref{fig:Lanue2Lnuxr} as a function of distance from the centre of the star. In Fig.~\ref{fig:L1} we show the per-flavour neutrino luminosity crossing through $r=400$~km as a function of time, for the same simulation. Note that  Fig.~\ref{fig:L1} does not include the effect of neutrino oscillations. The effect of oscillations will be accounted for below.

\begin{figure}[htb!]
\centering
\includegraphics[width=0.495\textwidth]{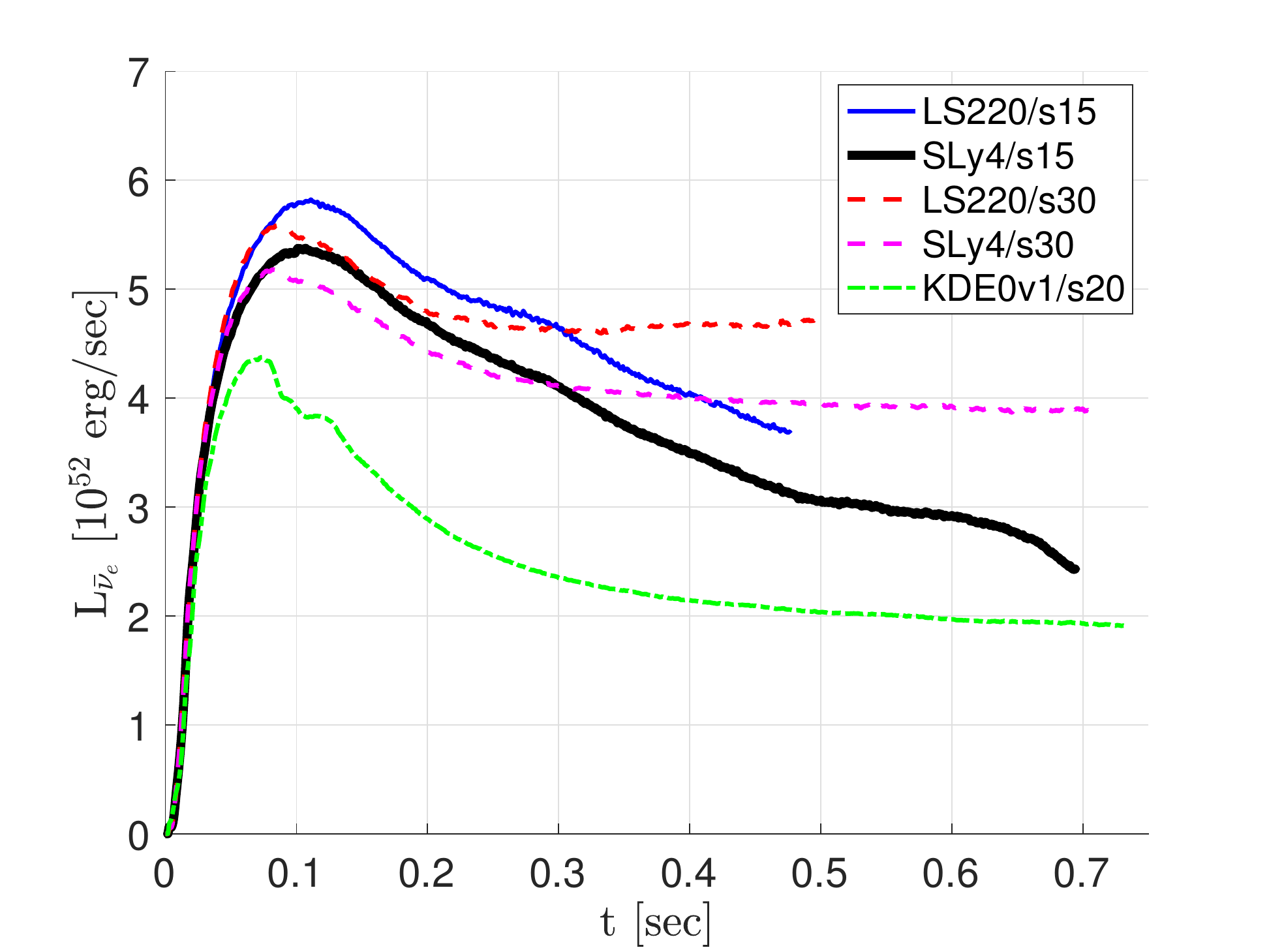}
\includegraphics[width=0.495\textwidth]{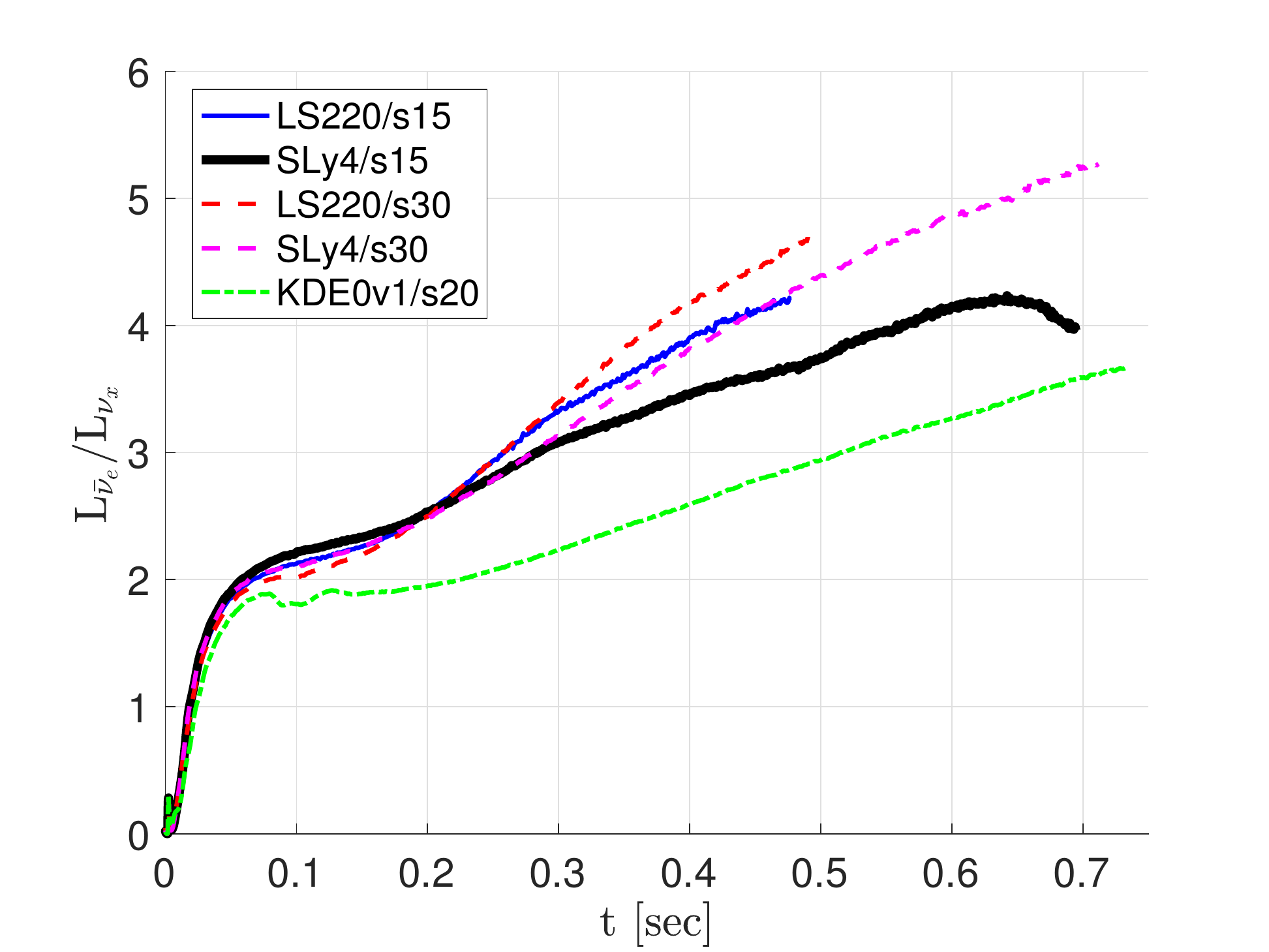}
\caption{
Time-dependent neutrino luminosity at the source ($r=400$~km). $t$ is time post-bounce. {\bf Left:} $L_{\bar\nu_e}$. {\bf Right:} The ratio $L_{\bar\nu_e}/L_{\nu_x}$. We do not include neutrino oscillations, the effect of which will be accounted for later. Results are shown for the five simulations described in the text.
}\label{fig:L2}
\end{figure}

The details of the neutrino burst vary between progenitor stars and depend also on the assumed EoS. However, e-flavour dominance during the accretion phase is a generic phenomenon. We illustrate this point in Fig.~\ref{fig:L2}, where we show results calculated for a set of five simulations: (i) SLy4 EoS, 15~M$_{\odot}$ star (thick black, the same simulation used in Figs.~\ref{fig:Lr}-\ref{fig:L1}), (ii) SLy4 EoS, 30~$M_{\odot}$ (dashed magenta), (iii) LS220~\cite{Lattimer:1991nc,daSilvaSchneider:2017jpg}, 15~M$_{\odot}$ (blue) (iv) LS220, 30~M$_\odot$ (dashed red), (v) KDE0v1 EoS~\cite{Agrawal:2005ix}, 20~M$_{\odot}$ (dotted green).
On the left panel we show the $\bar\nu_e$ luminosity, and on the right we show the ratio $L_{\bar\nu_e}/L_{\nu_x}$, both calculated at the source ($r=400$~km). 

In App.~\ref{app:comp} we review two additional examples from the literature~\cite{Perego:2015zca,Seadrow:2018ftp}, using different simulation codes. 
%
Ref.~\cite{Perego:2015zca} compared the neutrino light-curves of failed explosions, to the light-curves of simulations in which an explosion was set-off by hand. Similarly, Ref.~\cite{Seadrow:2018ftp} compared 1D ``failed" and 2D ``successful" explosions in simulations of the same progenitor stars\footnote{Seed perturbations and numerical noise were used to initiate turbulence in the 2D models analysed in~\cite{Seadrow:2018ftp}; see Sec.~5 there.}. 
The failed explosions in~\cite{Perego:2015zca,Seadrow:2018ftp} give similar results to the GR1D simulations that we explore here: the different calculations agree on $f_{\bar e}>2$ during the accretion phase. 

The successful explosions in~\cite{Perego:2015zca,Seadrow:2018ftp} serve to illustrate the D$\nu$M PNS cooling scenario. We cannot test this scenario directly with GR1D (without setting-off artificial explosions). Instead, we will use an indirect prescription to estimate the D$\nu$M scenario, which we explain below. As a result of this indirect prescription, our D$\nu$M analysis will be less robust than that for CITE.

\section{Neutrino detection channels in a scintillation detector}\label{sec:detect}

The JUNO detector~\cite{An:2015jdp}\footnote{Our analysis applies equally well to the proposed $50$~kton LENA~\cite{Wurm:2010ny} detector.}, currently under construction, will contain 20~kton of Linear Alkylbenzene (LAB, composed of C$_6$H$_5$C$_n$H$_{2n+1}$). In calculations, we round off the available number of target protons to $N_p=1.5\times 10^{33}$. For $n$ in the range $10-13$, the available number of target electrons is $N_e=(4.63\pm0.065)\times N_p$. In what follows, we describe the neutrino detection channels that are relevant for the CCSN analysis~\cite{Lu:2016ipr,Nikrant:2017nya}. 

The detection channels available in JUNO include IBD, pES, eES, and neutrino-nucleus ($\nu$N) charged- and neutral-current scattering. We mainly focus on the IBD and pES channels, which are expected to dominate the event rate and which contain the flavour information we are after in this study. The different channels are characterised by distinct deposited energy spectra. This, combined with additional event-by-event information (such as the gamma-rays due to neutron capture and $e^+$ annihilation for IBD), allows an analysis in JUNO to effectively tag the different channels. 

Considering background contamination, this would not affect the result for IBD~\cite{An:2015jdp} but could be significant for low-threshold pES~\cite{Beacom:2002hs,Dasgupta:2011wg}. Given that a detailed spectral background measurement will become available with the completion of JUNO,  we account for backgrounds effectively by considering different values for the lower threshold deposited energy in proton recoil, discussed in Sec.~\ref{ssec:pes}. As we will show, the lower the threshold, the better are the capabilities of the flavour analysis.

\subsection{IBD ($ \bar{\nu}_e+p\to e^++n $)}\label{ssec:ibd}
	We adopt the approximation used in \cite{Vissani:2014doa}, exploiting the weak dependence of the cross section on the scattering angle. The incoming neutrino energy $E_\nu$ is well-reconstructed by the measured  positron energy $E_e$, via\footnote{The exact result is $ E_\nu=(E_e+\delta)/[1-(E_e-p_e\cos\theta)/m_p] $, where $ \delta\equiv (m_n^2-m_p^2-m_e^2)/2m_p$, $ p_e $ is the positron momentum and $ \theta $ is the scattering angle of the positron.}  
	\be\label{eq:EvEeIBD} E_\nu &\approx& \frac{E_e+Q}{1-E_e/m_p}, \;\;\;\; Q = m_n-m_p=1.293\, {\rm MeV}.\ee	
	The rate of detected events is then 
	\be\label{ibdevents}
	\dot N_{IBD} &=& N_p \int\limits_{E_e^{min}}^{E_e^{max}} dE_e\Phi^\oplus_{\bar{\nu}_e}(E_\nu,t)\sigma_{IBD}(E_\nu)J(E_\nu).
	\ee
	Here $\Phi^\oplus_{\nu_\alpha}$ is the neutrino flux at Earth, we use Eq.~(\ref{eq:EvEeIBD}) to compute $E_\nu(E_e)$ in the integral, and the Jacobian is $J(E_\nu)=(1+E_\nu/m_p)^2/(1+Q/m_p)$.  
	The minimum detectable neutrino energy is $Q+m_e$, for which the positron is emitted at $ E_{e}=m_e$. The details of the choice of $E_e^{min}$ are not very important; in practice, we choose $E_e^{min}=2$~MeV. We take $E_e^{max}=100$~MeV, noting that the actual neutrino spectrum dies off exponentially above a few tens of MeV. The explicit form of $\sigma_{IBD}$ is recalled in App~\ref{app:xs}. For the range of neutrino energies we are interested in, $ \sigma_{IBD} $ is known with less than $ 1\% $ uncertainty. 
	Finally, note that the IBD events can be tagged by, e.g., neutron capture, with tagging efficiency estimated at about 90\%~\cite{Lu:2016ipr}.

\subsection{ pES ($\nu_\alpha+p\to \nu_\alpha+p $)}\label{ssec:pes}
	In a neutrino-proton elastic scattering (pES)~\cite{Beacom:2002hs}, the proton receives some recoil energy $T$. However, due to the quenching process of the scintillator, the observed energy of the proton, $T'$, is smaller. The relation $T'(T)$ is dictated for each detector by the solvent, using the semi-empirical Birk's law~\cite{Birks:1964zz}. In what follows, we use $T'(T)$ taken from~\cite{An:2015jdp}, noting that accurate measurements of this relation~\cite{vonKrosigk:2013sa} can be conducted in-situ once JUNO begins operation.
	
	The rate of detected events per unit reconstructed recoil energy is~\cite{Dasgupta:2011wg} 
	\be\label{eq:pesdiff}
	\frac{d\dot N_{pES}}{dT'}  &=& \frac{N_p}{(dT'/dT)}\int\limits_{E_\nu^{min}(T)}^{\infty}dE_\nu\sum_{\nu_\alpha,\bar\nu_\alpha}\Phi^\oplus_{\nu_\alpha}(E_\nu)\frac{d\sigma_{pES}(E_\nu,T)}{dT}.	\ee
	The sum includes all flavours of neutrinos and anti-neutrinos. 
	The differential cross section $\frac{d\sigma_{pES}(E_\nu,T)}{dT}$, recalled in App~\ref{app:xs}, is independent of incoming neutrino flavour. Therefore, for unitary three-flavour neutrino propagation, the pES channel depends only on the total all-flavour neutrino flux at the source and is independent of the details of neutrino oscillations~\cite{Beacom:2002hs}. 
	
	The minimum detectable neutrino energy is\footnote{Ref.~\cite{Dasgupta:2011wg} used the approximation $ E_\nu^{min}(T)\approx \sqrt{m_p T/2} $. We find that this approximation leads to a $ \sim 15\% $ error in the total event rate, because most of the contribution to the integral is typically close to the lower integration limit in a CCSN scenario.} 
	\be\label{eq:EminpES} E_\nu^{min}(T)&=&\sqrt{\left(1+\frac{T}{2m_p}\right)\frac{m_p\,T}{2}}+\frac{T}{2}.\ee
	The event rate is then
	\be\label{pesevents}
	\dot N_{pES} &=& N_p \int\limits_{T_{min}}^{T_{max}}dT \int\limits_{E_\nu^{min}(T)}^{\infty}dE_\nu\sum_{\nu_\alpha,\bar\nu_\alpha}\Phi^\oplus_{\nu_\alpha}(E_\nu)\frac{d\sigma_{pES}(E_\nu,T)}{dT}.
	\ee
	The integration over proton recoil energy runs between $ T_{min} $ to $ T_{max} $. These recoil energy limits are derived from quenched (observed) reconstructed energy limits, $ T'_{min} $ and $ T'_{max} $, using the relation $ T^{\prime}(T) $. In the current work we set the upper limit $ T'_{max} =2$~MeV, for which the corresponding minimal incoming neutrino energy is $E_\nu^{min}(T(T'_{max}))\approx50$~MeV. We set this  $T'_{max}$ cut in order to decrease the contamination of eES events in the pES sample. As long as $T'_{max}\sim2$~MeV, the details of the choice of $T'_{max} $ are not very important for our analysis, although changing the $T'_{max}$ cut around $1-3$~MeV would make a small (controllable) numerical change to the expected value of our flavour observable. 
	
\begin{figure}[htb!]
\centering
\includegraphics[width=0.485\textwidth]{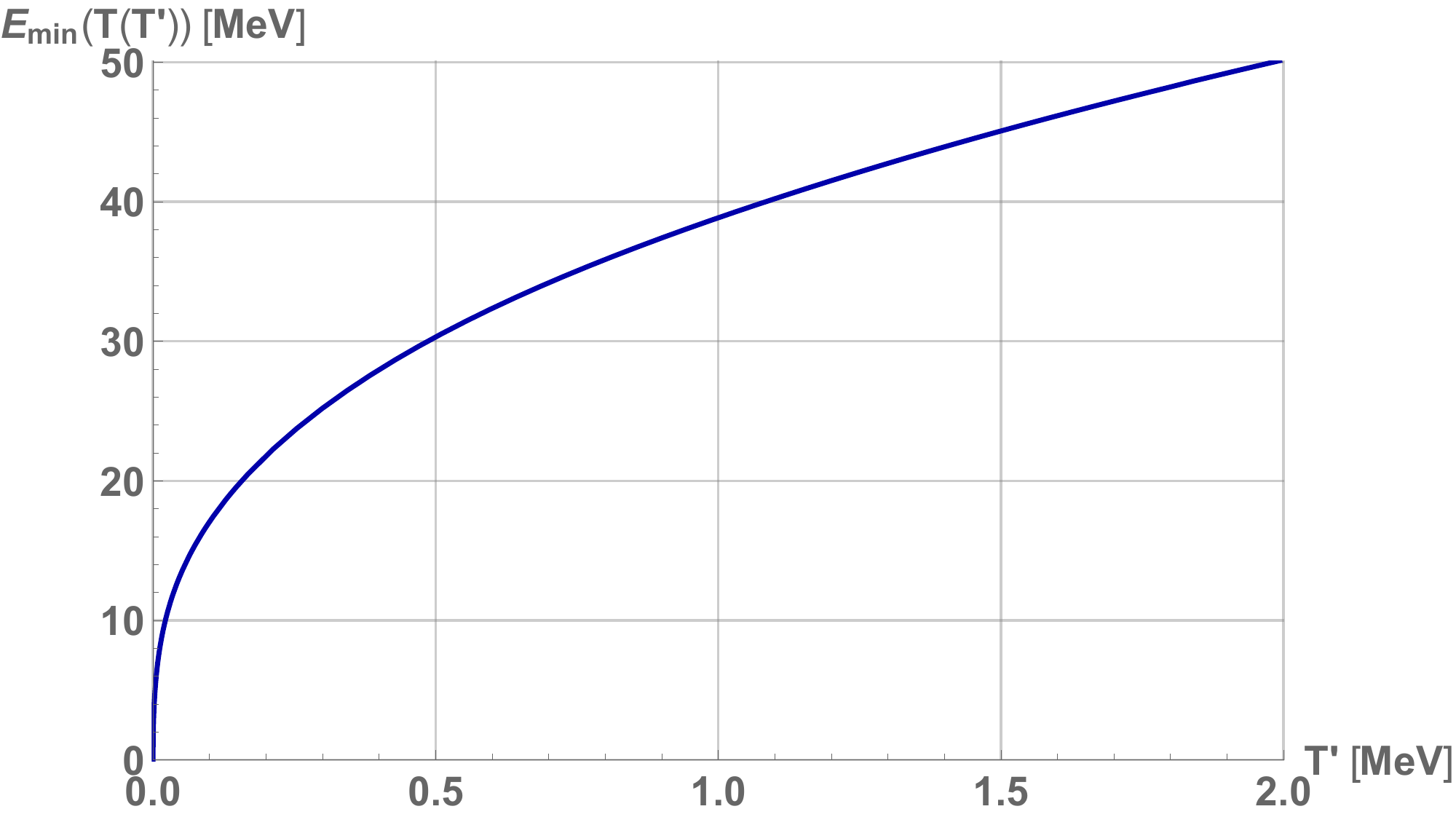}\;\;
\includegraphics[width=0.47\textwidth]{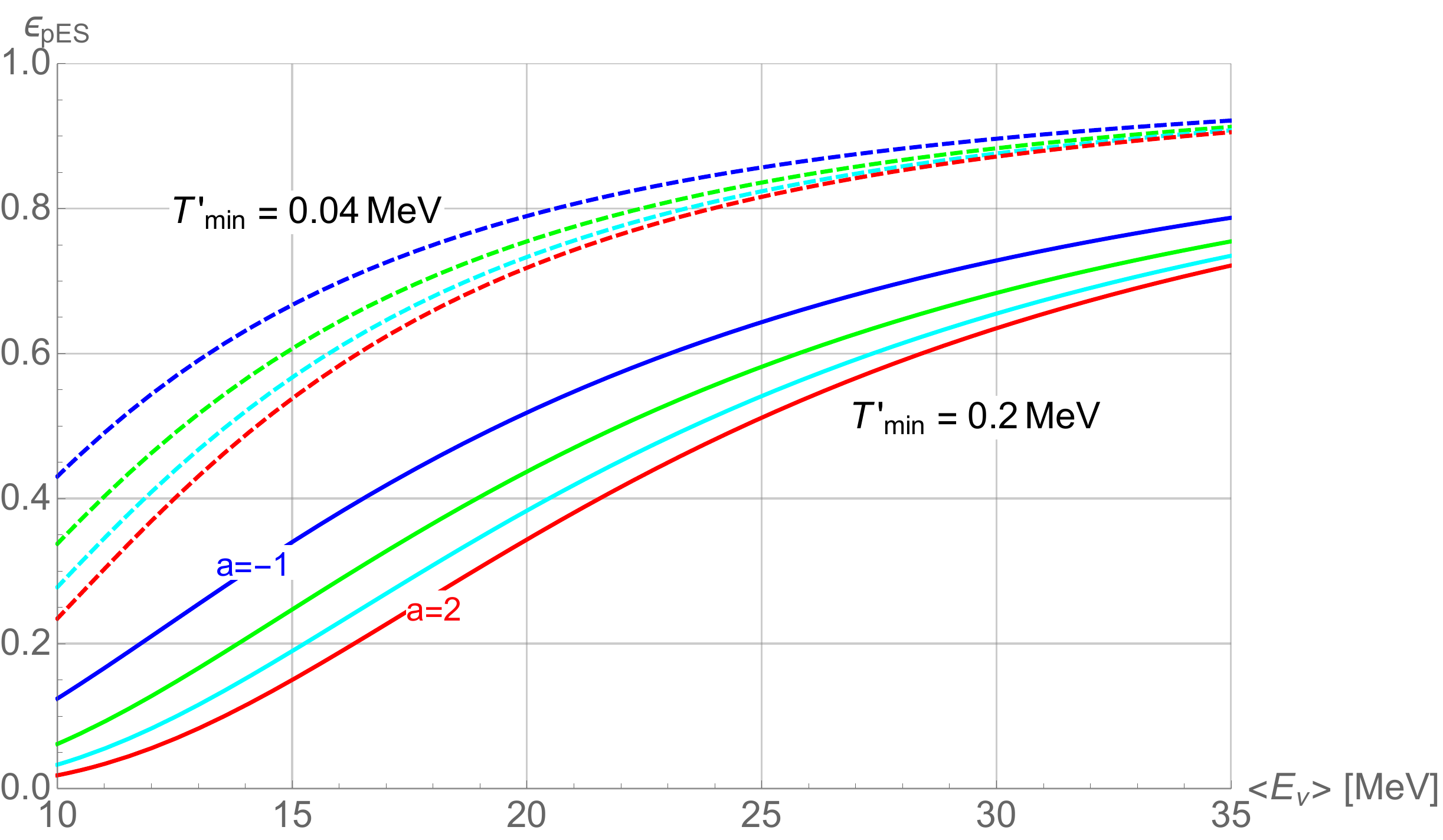}
\caption{{\bf Left:} The minimal neutrino energy required to induce a reaction with observable (quenched) energy $T'$. {\bf Right:} pES efficiency $\epsilon_{\rm pES}$, calculated for the pinched Fermi-Dirac spectrum as function of the mean neutrino energy $\langle E_\nu\rangle$ for recoil energy threshold $T'_{\rm min}=0.2$ (solid) and $T'_{\rm min}=0.04$ (dashed). The four lines (for each value of $T'_{\rm min}$) correspond to $a=-1,0,1,2$, from top (blue) to bottom (red). 
}\label{fig:EminTq}
\end{figure}

	The lower threshold $ T'_{min} $ is more important and requires discussion. The left panel of Fig.~\ref{fig:EminTq} shows the minimal neutrino energy required to induce a reaction with observable quenched energy $T'$. CCSN neutrinos carry characteristic energies of between $\sim10$ to a few 10's of MeV, so the value of the quenched energy threshold $T'_{min}$ can have a significant effect on the detection efficiency.  
	%
	To estimate the effect, consider a pinched Fermi-Dirac representation of the neutrino flux at Earth for a CCSN at distance $D_{SN}$,
	\be\label{eq:simplemodel}\Phi_{\nu,\rm FD}^\oplus&=&\frac{L}{4\pi D_{SN}^2\,c_L(a)\,T^2}\,\frac{\left(E_\nu/T\right)^{2+a}}{\exp\left(E_\nu/T\right)+1}.\ee
%
This spectrum is parametrised by three numbers: the pinch index $a$, the temperature $T$, and the source luminosity $L$. We define $c_L(a)=\left(1-2^{-3-a}\right)\Gamma(4+a)\zeta(4+a)$, and note that the mean neutrino energy $\langle E_{\nu}\rangle$ for this spectrum is given by $\langle E_{\nu}\rangle=c_T(a)\,T_{\nu}$, where $c_T( a )=\frac{\left(2^{3+ a }-1\right)\Gamma(4+ a )\zeta(4+ a )}{2\left(2^{2+ a }-1\right)\Gamma(3+ a )\zeta(3+ a )}$. 
Using Eq.~(\ref{eq:simplemodel}), and neglecting the upper threshold $T'_{\rm max}$, we can calculate the pES efficiency as the ratio between the number of pES events detected with finite $T'_{min}$ threshold to the number of events that would be detected with $T'_{min}\to0$,
\be\label{eq:pESeff}\epsilon_{\rm pES}\left(T'_{\rm min}\right)&\approx&\frac{\int\limits_{T_{min}}^{\infty}dT \int\limits_{E_\nu^{min}(T)}^{\infty}dE_\nu\Phi^\oplus_{\nu,\rm FD}(E_\nu)\frac{d\sigma_{pES}(E_\nu,T)}{dT}}{\int\limits_{0}^{\infty}dT \int\limits_{E_\nu^{min}(T)}^{\infty}dE_\nu\Phi^\oplus_{\nu,\rm FD}(E_\nu)\frac{d\sigma_{pES}(E_\nu,T)}{dT}}.\ee
%
%
In modelling JUNO we will show results using two values of the lower quenched energy threshold, $ T'_{min}=0.2$~MeV (the main value used in~\cite{An:2015jdp}), along with an optimistic $ T'_{min}=0.04$~MeV. The resulting $\epsilon_{\rm pES}$ is shown in the right panel of Fig.~\ref{fig:EminTq}. 

We emphasise that Eq.~(\ref{eq:pESeff}) and the spectral model of Eq.~(\ref{eq:simplemodel}) are not used in our numerical calculations,  in which we take neutrino spectra directly from the simulations and use the full cross section expressions to calculate event rates. Nevertheless, in a realistic CCSN, the flux of neutrinos of flavour $\nu_\alpha$ is  reasonably well described by Eq.~(\ref{eq:simplemodel}) with flavour-dependent parameters, $(a,T,L)\to(a_{\nu_\alpha},T_{\nu_\alpha},L_{\nu_\alpha})$, and with a weight corresponding to the oscillation probability: $\Phi_{\nu_\alpha}^\oplus\approx\sum_{\nu_\beta}P_{\alpha\beta}\Phi_{\nu_\beta,\rm FD}^\oplus$. Thus, the right panel of Fig.~\ref{fig:EminTq} gives a useful illustration of the impact of $T'_{min}$. 

Finally, it is important to note that current uncertainty on the strangeness content of the proton~\cite{Ahrens1987,Aidala2013,Sufian:2018qtw} induces a $ \sim 20\% $ uncertainty on $\sigma_{pES}$. However, there is a proposal to reduce this uncertainty by an order of magnitude~\cite{Pagliaroli2013}. We will assume here that the uncertainty on $\sigma_{pES}$ can indeed be reduced to the few percent level.

\subsection{Additional channels: eES and neutrino-nucleus scattering}

\subsubsection{ eES ($\nu_\alpha+e\to \nu_\alpha+e $)}
	The cross section for neutrino-electron elastic scattering (eES) is summarised in App~\ref{app:xs}. The cross section depends on the incoming neutrino flavour, with $\sigma_{\nu_ee}/\sigma_{\bar\nu_ee}\approx2.38$, $\sigma_{\nu_ee}/\sigma_{\nu_xe}\approx5.81$, $\sigma_{\nu_ee}/\sigma_{\bar\nu_xe}\approx7.15$. 	
%
	The eES cross section is smaller than the pES cross section by a factor of $\approx22,\,43,\,65$ for $E_\nu=10,\,20,\,30$~MeV, respectively, where for this comparison we considered incoming $\nu_e$. Therefore, despite the target enhancement factor of $N_e/N_p\approx4.6$, the total eES event rate is smaller than the total pES event rate by a factor of $\sim10$, provided that the analysis can achieve $T'_{min}\lesssim0.2$~MeV in pES detection.
	
The kinematics of eES is the same as that in pES, up to replacing $m_p\to m_e$. The electron recoil energy is $ T_e\equiv E_e-m_e=\frac{1-\cos\theta}{1+(1-\cos\theta)\frac{E_\nu}{m_e}}\frac{E_\nu^2}{m_e} $, where $ \theta $ is the scattering angle of the neutrino. 
	For a typical CCSN neutrino, $E_\nu\gg m_e$ and thus $T_e \sim E_\nu$. As a result, the deposited energy spectra of eES and pES are separated, the former peaking at higher deposited energy than the latter. With our upper cut of $T'_{max}=2$~MeV, the eES contribution in the elastic scattering sample will be at the percent level of that from pES.

\subsubsection{Neutrino-nucleus scattering}
	The neutrino-nucleus charged current (CC) $\nu_e\,^{12}$C $\to e^-\,^{12}$N, $\bar\nu_e\,^{12}$C $\to e^+\,^{12}$B, $\nu_e\,^{13}$C $\to e^-\,^{13}$N and neutral current (NC) $\nu\,^{12}$C $\to\nu\,^{12}$C$^*$, $\nu\,^{13}$C $\to\nu\,^{13}$C$^*$ interactions would also contribute some events in JUNO~\cite{Laha:2014yua,Lu:2016ipr,An:2015jdp,Nikrant:2017nya,Seadrow:2018ftp} (see also~\cite{GalloRosso:2017hbp,Nakazato:2018xkv} for $\nu$O scattering). The recoil energy of the nucleus is too small to be measured directly. The expected number of NC events is below about 10\% of the IBD sample~\cite{Lu:2016ipr}, and the channel can be effectively identified via the monochromatic deexitation gamma rays. The expected number of CC events is comparable to NC and eES. The deposited energy spectra due to the CC final state $e^-$ or $e^+$ are peaked in approximately the same range as the IBD events, and die off faster than eES towards low deposited energy~\cite{Lu:2016ipr}. With a cut of $T'_{max}=2$~MeV, the contamination of CC events to the pES channel is negligible.
	   
In summary, neutrino-nucleus events are statistically sub-dominant, and with simple event tagging and analysis cuts, would not contaminate the IBD and pES samples. In our analysis we therefore do not consider these channels further, noting only that their detection would add more information to a CCSN signal analysis beyond what we offer in this work~\cite{Laha:2014yua,Lu:2016ipr,An:2015jdp,Nikrant:2017nya}.  

%

\section{Neutrino oscillations}\label{sec:osc}

Neutrino flavour conversion in the supernova is computed using the MSW formalism~\cite{Wolfenstein:1977ue,Mikheev:1986gs} (if self-induced oscillations are not important; however, see discussion below). 
Focusing on the oscillations of anti-neutrinos, important for the IBD channel, the survival probability of  $ \bar{\nu}_e $ is given by~\cite{Dighe:1999bi}
\eqspla{eq:PeeAn2}{
	P_{\bar{e}\bar{e}} & = \begin{cases}
		|U_{e1}|^2(1-P_{\bar L})+|U_{e2}|^2P_{\bar L} ,&{\rm NH},  \\
		 |U_{e1}|^2P_{\bar H}(1-P_{\bar L})+|U_{e2}|^2P_{\bar H}P_{\bar L}+|U_{e3}|^2(1-P_{\bar H}) , & {\rm IH},
	\end{cases}  \\ 
	&  \approx \begin{cases}
		0.68 -0.38P_{\bar L} ,& {\rm NH},  \\  0.02 +(0.66-0.38P_{\bar L})P_{\bar H} , & {\rm IH}.
	\end{cases} 
}
$ P_{\bar H} $ and $P_{\bar L} $ are the level crossing probabilities between propagation eigenstates for the so-called $ H $- and $ L $- (antineutrino) subsystems, respectively. The adiabatic limit amounts to setting $P_{\bar L}=P_{\bar H}=0$. NH and IH refer to normal mass hierarchy and inverted mass hierarchy, respectively.

Even for prolonged accretion, relevant for the CITE scenario, we demonstrate in App.~\ref{app:osc} that the adiabatic limit is justified at least in the first few seconds ($t_{\rm PB}\lesssim5$~sec) of the neutrino burst. For the D$\nu$M, once explosion happens the situation may be more complicated because the matter overburden is swept away such that level-crossing resonance regions may be traversed in the non-adiabatic regime~\cite{Mirizzi:2015eza}. 
Nevertheless, the adiabatic limit is still a useful benchmark, used extensively in analyses of SN1987A~\cite{Loredo2002,Pagliaroli2009,Vissani:2014doa,Blum:2016afe} and in the discussion of a future galactic CCSN~\cite{Horiuchi2018,Horiuchi2017,Tamborra:2014hga,GalloRosso:2017hbp,GalloRosso:2017mdz,Dighe:1999bi,Chakraborty:2012gb,Seadrow2018}. Formulating clear predictions for the neutrino signal in this limit would allow to identify and interpret possible deviations in real data. We therefore adopt this limit in most of our current work.  
Assuming that the x-flavour spectra are equal at the source, the $\bar\nu_e$ flux at earth is related to the fluxes at the source via (note that in our notation, $\Phi_{\bar\nu_x}$ is the average flux per species of $\bar\nu_\mu$, $\bar\nu_\tau$)
\be\Phi^\oplus_{\bar\nu_e}&=&P_{\bar{e}\bar{e}}\,\Phi_{\bar\nu_e}+(1-P_{\bar e\bar e})\,\Phi_{\bar\nu_x}.\ee
In the numerical expressions above and in our main analysis, we use best-fit values for the oscillation parameters, taken from~\cite{Esteban:2016qun,nufit}. 

Self-induced neutrino oscillations may significantly affect the adiabatic MSW prediction~\cite{Duan:2010bg,Duan:2006an,Fogli:2007bk,Raffelt:2007cb,Fogli:2008pt,Dasgupta:2009mg,Fogli:2008fj,Dasgupta:2010cd,EstebanPretel:2008ni}. The problem, however, is not settled yet. In particular, during the accretion phase the large matter density may inhibit the self-induced oscillations effect~\cite{Chakraborty:2011nf,Sarikas:2011am,Saviano:2012yh,Cherry:2012zw,Sarikas:2012vb,Sarikas:2012ad,Raffelt:2013isa,Hansen:2014paa,Mirizzi:2013rla,Mirizzi:2013wda,Chakraborty:2014nma,Mangano:2014zda}. 
Since the jury is still out on the outcome and importance of self-induced oscillations, we study the impact of different oscillation probabilities in Sec.~\ref{ssec:R}, spanning the range $P_{\bar e\bar e}=[0,1]$. This also serves to exhibit the sensitivity of our main results, using Eq.~(\ref{eq:PeeAn2}), to measurement uncertainties in the oscillation parameters.

Finally, note that neutrino propagation is also slightly affected by matter in the Earth~\cite{Dighe:1999bi,Lunardini:2001pb}. The effect depends on directionality, which would be known accurately for a future galactic CCSN event allowing corrections to our formalism to be implemented if needed. We therefore omit Earth matter effects in this work.

%

\section{Numerical simulations of CCSN detection in JUNO}\label{sec:juno}

With the calculations of Secs.~\ref{sec:detect} and~\ref{sec:osc}, we can use numerical simulations of core-collapse to predict pES and IBD event rates at JUNO. We will find the simple counting observable $R$, introduced in Eq.~(\ref{eq:R}), particularly useful. 
In what follows, after some discussion of the spectral shape of the signal, we analyse $R$ for simulated bursts. We then carry out a likelihood analysis of mock  data. 

\begin{figure}[htb!]
\centering
\includegraphics[width=0.85\textwidth]{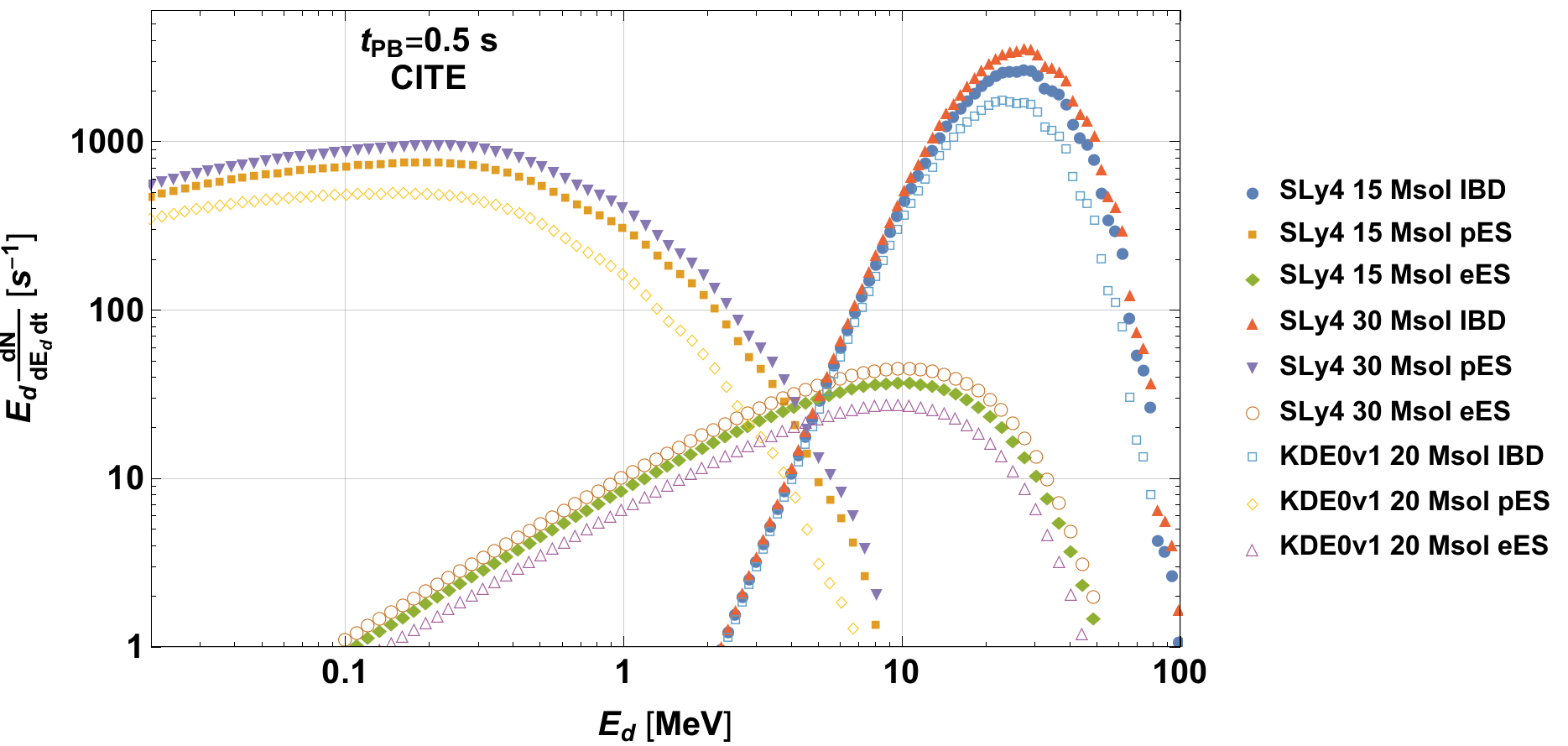}
\includegraphics[width=0.85\textwidth]{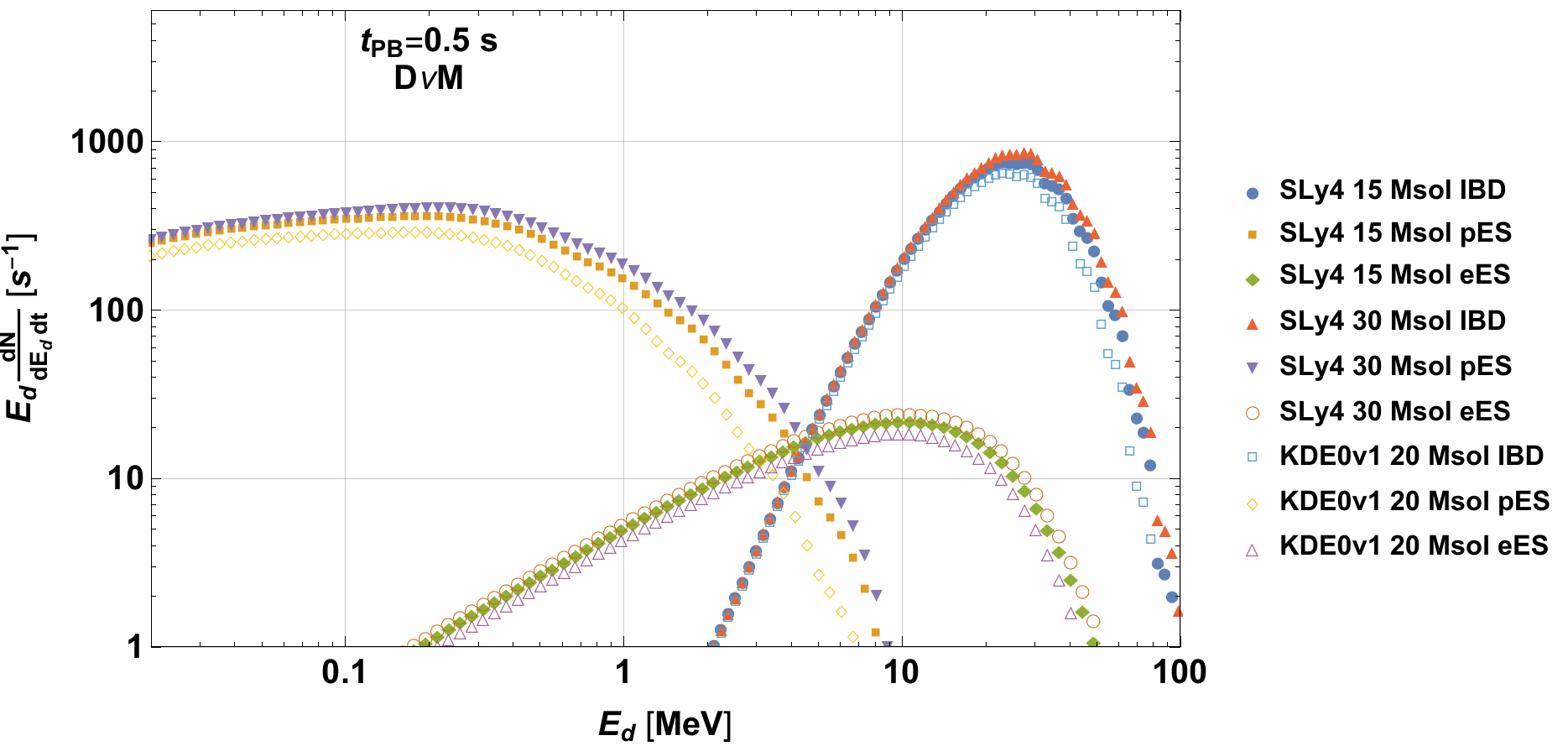}
\caption{
Event rate vs. deposited energy at JUNO. The broad bump peaking around $E_d\approx 0.3$~MeV is due to pES. The dominant, narrower bump peaking around $E_d\approx30$~MeV is due to IBD. The sub-dominant bump (lower than the peak pES rate by a factor of $\sim100$) is due to eES. We plot a snapshot of the predicted detection spectrum at $t_{\rm PB}=0.5$~sec, for adiabatic matter flavour conversion assuming normal hierarchy. The CCSN simulations corresponding to the different markers are explained in the text. {\bf Top:} CITE. {\bf Bottom:} D$\nu$M. 
}\label{fig:diff05cite}
\end{figure}

We first study spectral information. 
Fig.~\ref{fig:diff05cite} shows a temporal snapshot, taken at $t_{\rm PB}=0.5$~sec, of the deposited energy spectra in JUNO. To produce this plot, we assume a CCSN at 10~kpc~\cite{Adams:2013ana} with adiabatic matter-induced flavour conversion and normal hierarchy. We show the results of 3 different simulation runs, computed with GR1D. The first two runs use the SLy4 EoS~\cite{Chabanat:1997un} for 15$~M_{\odot}$ and 30$~M_{\odot}$ stars. The third run uses the KDE0v1 EoS~\cite{Agrawal:2005ix} and a 20$~M_{\odot}$ star. In the top panel we show the deposited energy spectra obtained directly from the simulations. These results correspond to the CITE scenario. In the bottom panel we show an indirect estimate of what the deposited energy spectra are expected to look like in the D$\nu$M. To obtain this estimate, we set the $\nu_e$ and $\bar\nu_e$ luminosities at the source equal to the $\nu_x$ luminosity. 

The event spectra in the top panel of Fig.~\ref{fig:diff05cite} (CITE) are higher by a factor of 2-3 compared to the rates in the bottom panel (D$\nu$M). This overall enhancement is interesting for estimates of the diffuse supernova neutrino background (DSNB) \citep{2010ARNPS..60..439B}. However, it would only give a partial hint towards diagnosing the explosion mechanism underlying a single galactic explosion, because it is degenerate with details of the stellar profile and the nuclear EoS. Rather than the total luminosity, a key difference between the top and bottom panels is the {\it relative} height of the pES and IBD peaks. This information is captured by $R$.

\subsection{Simple counting observable: $R=$pES/IBD.}\label{ssec:R}
In this section we study $R$ as a diagnostic of the neutrino emission process. We start in Sec.~\ref{ssec:Rnum} by analysing numerical simulations, and discuss the sensitivity of our results to modelling uncertainties in Sec.~\ref{ssec:Rwhat}. 

\subsubsection{Analysis of core-collapse simulations}\label{ssec:Rnum}
\begin{figure}[htb!]
\centering
\includegraphics[width=0.75\textwidth]{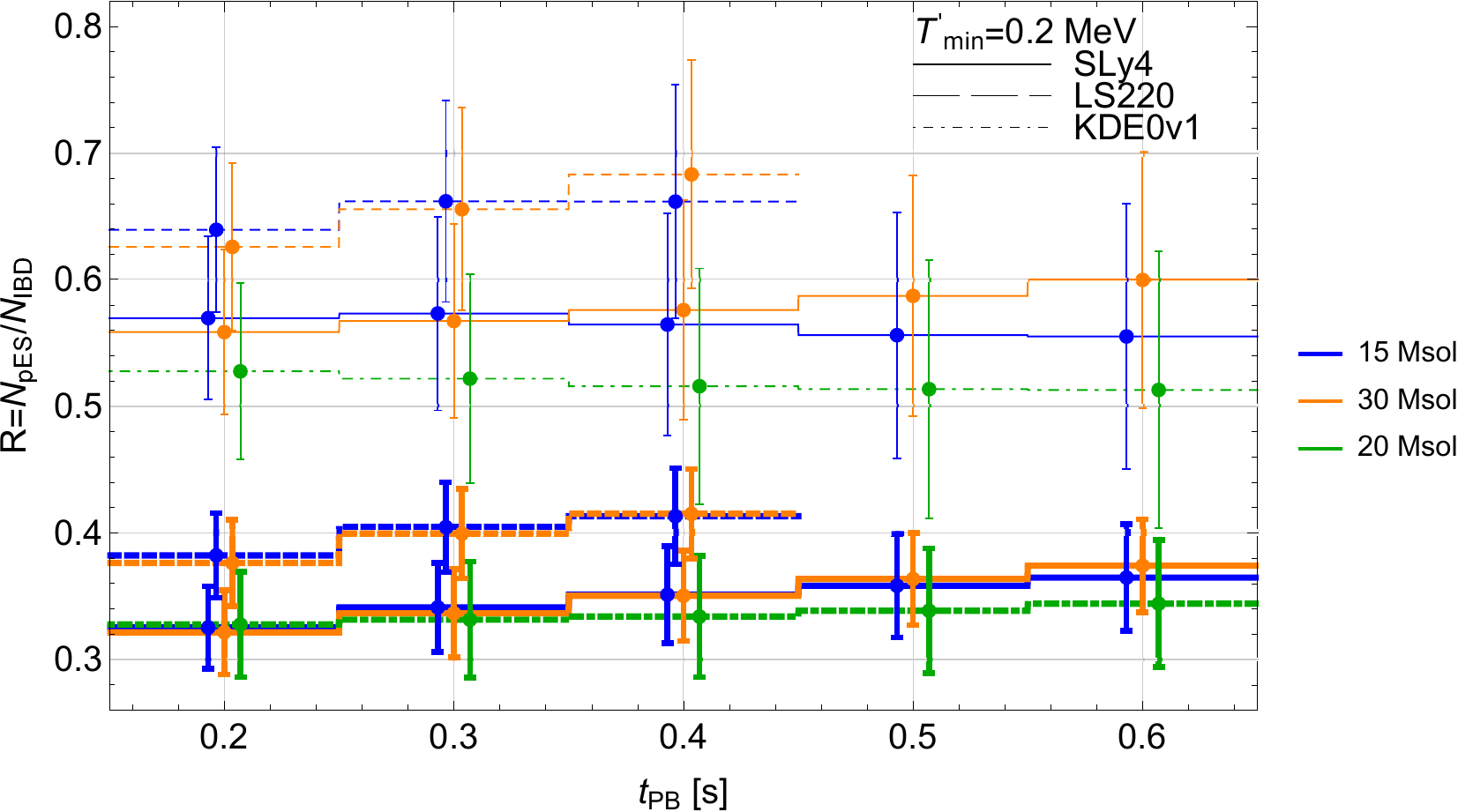}
\includegraphics[width=0.75\textwidth]{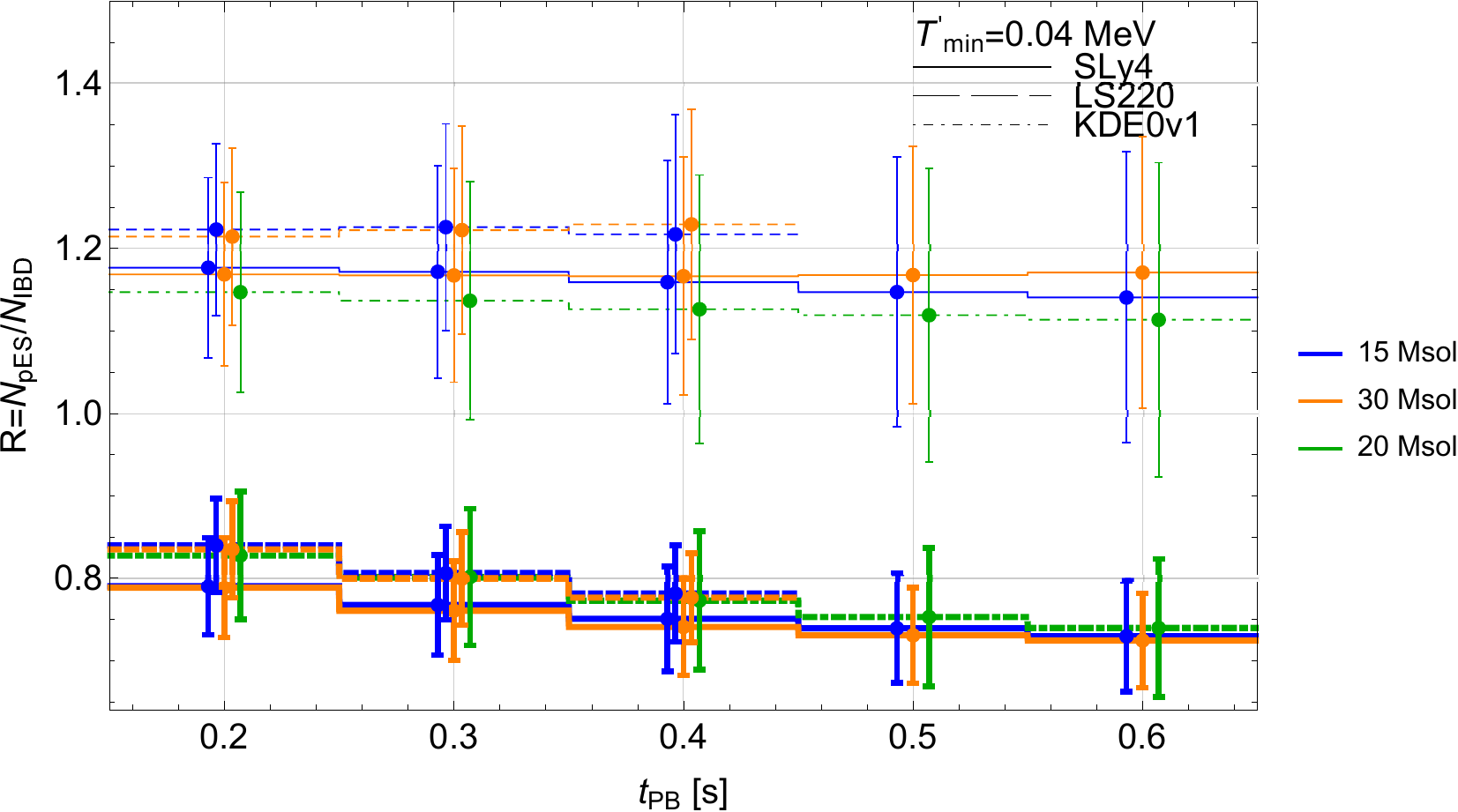}
\caption{
{\bf Top:} The observable $R$ vs. post-bounce time $t_{\rm PB}$, calculated for a galactic CCSN occurring 10kpc away and measured by JUNO with a reconstructed proton recoil energy threshold $T'_{min}=0.2$~MeV. 
Thick lines (all occurring below $R=0.5$) correspond to a direct simulation of CITE. Thin lines (all occurring above $R=0.5$) correspond to our indirect estimate of the D$\nu$M, explained in the text. 
Based on the $15~M_\odot$, $20~M_\odot$, and $30~M_\odot$ progenitor star GR1D simulations, with 3 different assumed nuclear EoS. Error bars denote ${\bf 1\sigma}$ statistical uncertainties. Calculated assuming adiabatic matter-induced flavour conversions with normal neutrino mass hierarchy.
{\bf Bottom:} same as the top panel, but with reconstructed proton recoil energy threshold $T'_{min}=0.04$~MeV.
}\label{fig:rgridEB}
\end{figure}

The top panel of Fig.~\ref{fig:rgridEB} shows the observable $R$ vs. post-bounce time $t_{\rm PB}$, binned in 0.1~sec time segments, calculated for a galactic CCSN at 10~kpc and measured by JUNO with a deposited energy threshold $T'_{min}=0.2$~MeV. 
Thick lines (all below $R=0.5$) correspond to CITE. Thin lines (all above $R=0.5$) correspond to the D$\nu$M. 
We show results from GR1D simulations of $15~M_\odot$, $20~M_\odot$, and $30~M_\odot$ progenitor stars with 3 different assumed nuclear EoS~\cite{daSilvaSchneider:2017jpg}: SLy4~\cite{Chabanat:1997un}, LS220~\cite{Lattimer:1991nc}, and KDE0v1~\cite{Agrawal:2005ix}. Error bars denote ${\bf 1\sigma}$ statistical uncertainties. We assume adiabatic matter-induced flavour conversions with normal neutrino mass hierarchy.

The bottom panel of Fig.~\ref{fig:rgridEB} repeats the calculation of the top panel, but this time assuming $T'_{min}=0.04$~MeV. The low threshold allows significantly better separation of CITE and the D$\nu$M, the separation occurring at $R\approx0.9$,  instead of $R\approx0.5$ found in the upper panel. Importantly, besides from the separation between CITE and the D$\nu$M, low $T'_{min}$ causes the prediction of CITE to become less sensitive to progenitor mass and EoS, an important result that is decoupled from uncertainties related to our indirect estimate of the D$\nu$M. 

\begin{figure}[htb!]
\centering
\includegraphics[width=0.75\textwidth]{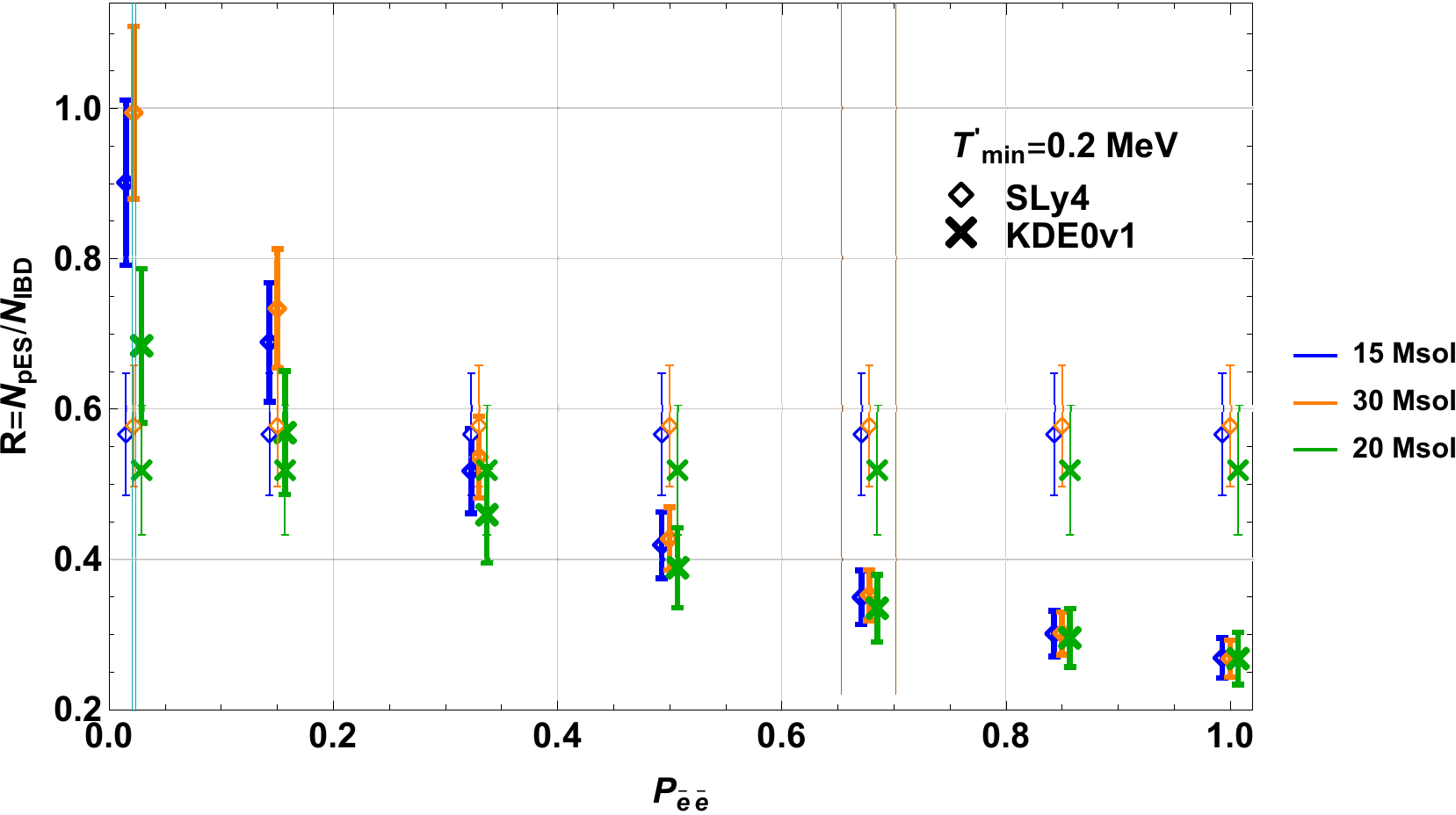}
\includegraphics[width=0.75\textwidth]{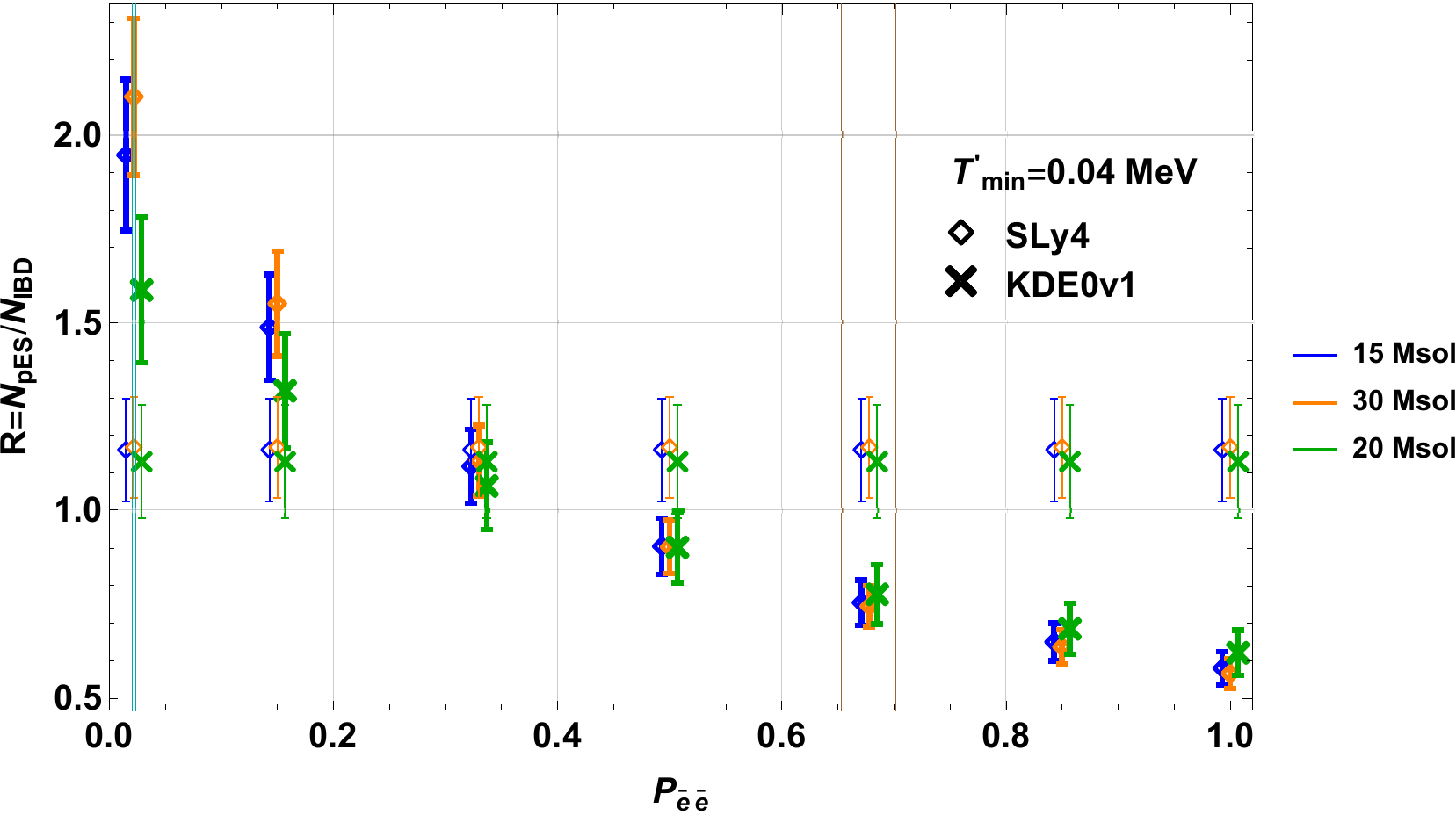}
\caption{{\bf Top:} 
The observable $R$ vs. the conversion probability $P_{\bar e\bar e}$, calculated for a galactic CCSN occurring 10kpc away and measured by JUNO with a reconstructed proton recoil energy threshold $T'_{min}=0.2$~MeV. The counts are obtained by integrating the expected detection rates in the time interval (0.2, 0.65)~sec post-bounce. Based on the $15~M_\odot$, $20~M_\odot$, and $30~M_\odot$ progenitor star GR1D simulations described in the text. Error bars denote ${\bf 2\sigma}$ statistical uncertainties. Thin vertical lines around $P_{\bar e\bar e}=0.02$ and $P_{\bar e\bar e}=0.068$ highlight the allowed region of $P_{\bar e\bar e}$, obtained by varying the neutrino oscillation parameters within their current ${\bf 2\sigma}$ range, for adiabatic matter-induced oscillations with IH and NH, respectively. Results for CITE (D$\nu$M) are shown by thick (thin) markers. {\bf Bottom:} same as in the top panel, but with reconstructed proton recoil energy threshold $T'_{min}=0.04$~MeV. 
}\label{fig:rvspgrid02}
\end{figure}

In Fig.~\ref{fig:rvspgrid02} we integrate the rates over the time interval $t_{\rm PB}=(0.2,0.65)$~sec, and explore the dependence on neutrino oscillation parameters by plotting the result vs. the $\bar\nu_e$ survival probability $P_{\bar e\bar e}$. The top panel shows the results assuming $T'_{min}=0.2$~MeV and the bottom panel shows the results for $T'_{min}=0.04$~MeV. Vertical error bars show ${\bf 2\sigma}$ statistical uncertainties. 
Thin vertical lines around $P_{\bar e\bar e}\approx0.68$ and $P_{\bar e\bar e}\approx0.02$ show the allowed region of $P_{\bar e\bar e}$ assuming NH and IH, respectively, w.r.t. the current ${\bf 2\sigma}$ allowed range of the oscillation parameters. 

We can now appreciate the sensitivity of $R$ to the details of neutrino flavour conversion: to determine the prediction in CITE, the neutrino mass hierarchy would be a crucial ingredient. Fortunately, there are good prospects for determining the hierarchy in the relatively near future (see, e.g.~\cite{Patterson2015}). In fact, JUNO itself can deliver this information within the next decade~\cite{An:2015jdp}. Fig.~\ref{fig:rvspgrid02} shows that (i) given the binary information of NH vs. IH, further uncertainty due to the neutrino oscillation parameters -- even with the current state of the art, not accounting for future improvement -- would not limit the CCSN analysis; and (ii) the NH scenario would be particularly convenient.

Finally, we comment about the time integration in the computation of $R$, which we limited to the interval $t_{\rm PB}=(0.2,0.65)$~sec. The lower end of this interval was motivated by D$\nu$M simulations, which tend to show explosions that start by approximately this time. D$\nu$M explosion is needed to expel the accretion flow and terminate the accompanied accretion luminosity. Only after this happens, the D$\nu$M and the CITE scenarios begin to differ. The upper end was chosen in the interest of simulation run time. We note that CITE allows for a longer period of accretion luminosity, lasting up to a few seconds, with a possible termination due to BH formation and plausible subsequent continuation due to an accretion disk~\cite{Blum:2016afe}. During this entire accretion period, CITE predicts $f_{\bar e}\gtrsim2-3$ and a corresponding value of $R$ similar to the late time bins in Fig.~\ref{fig:rgridEB}. The actual statistical uncertainty on $R$, integrated on a longer period than that shown in Fig.~\ref{fig:rvspgrid02}, would therefore be better than in the plot.

\subsubsection{Sensitivity to modelling uncertainties.}\label{ssec:Rwhat}

Assuming that the oscillation probability is known, our results from Sec.~\ref{ssec:Rnum}, based on GR1D simulations, suggest that measurements of $R$ can discriminate between CITE and the D$\nu$M, despite variances in progenitor and EoS details. It is important to analyse where this discrimination power comes from, and what are its limitations in terms of the sensitivity to the CCSN modelling. The neutrino transport implementation in GR1D (and other codes in the literature) is simplified, and does not include several complications that may modify the neutrino spectra in a realistic CCSN. In what follows, we study the sensitivity of our results to reasonable neutrino spectra that go beyond the range found in the numerical simulation. For concreteness, we focus on the NH adiabatic oscillations scenario, $P_{\bar e\bar e}\approx0.68$.

\begin{figure}[htb!]
\centering
\includegraphics[width=0.6\textwidth]{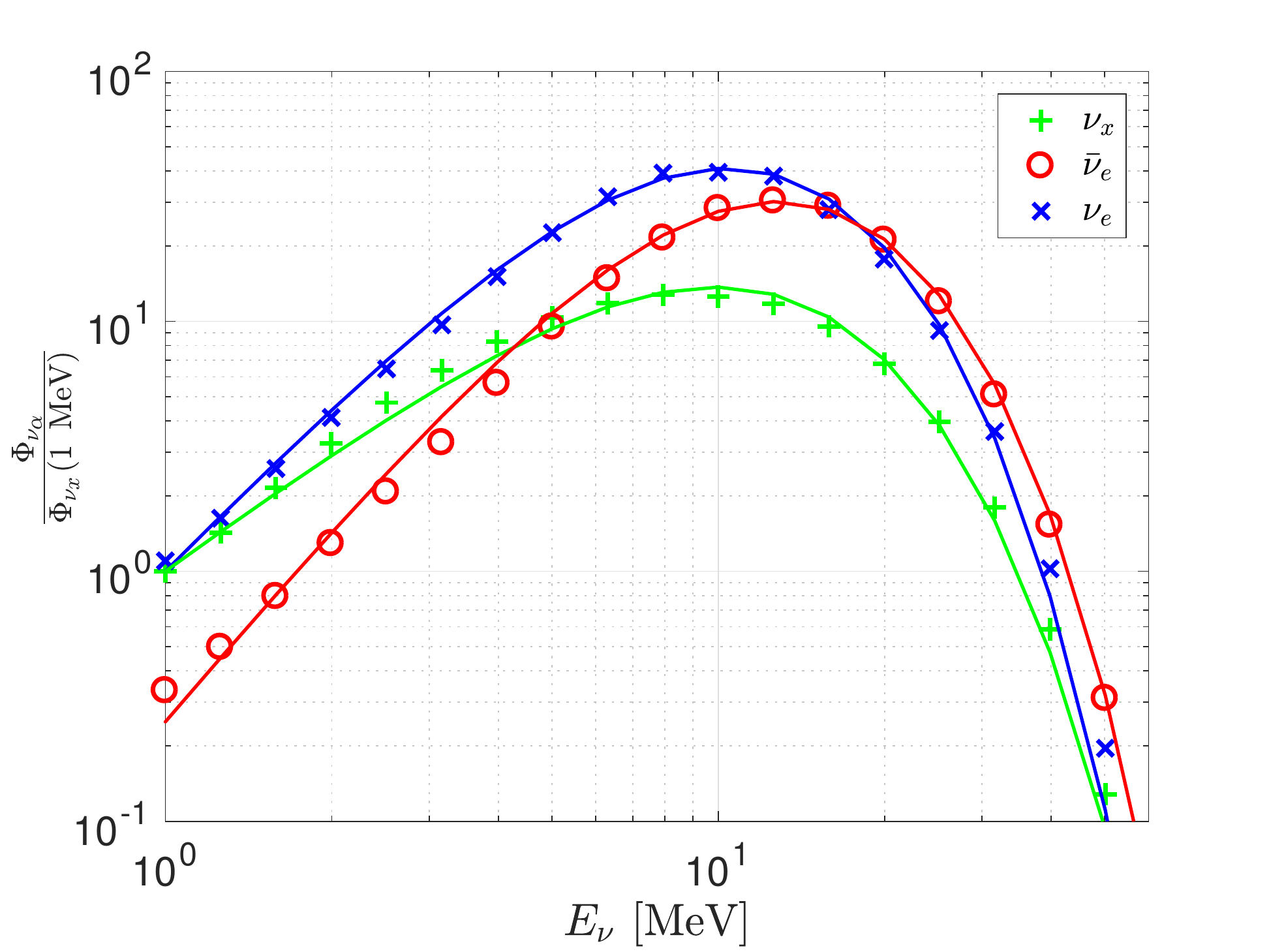}
\caption{
Time-integrated neutrino spectra computed with GR1D, using the KDE0v1~\cite{Agrawal:2005ix} EoS for a 20~M$_\odot$ progenitor. Markers show the numerical result from the simulation. Solid lines show a fit to the pinched Fermi-Dirac form of Eq.~(\ref{eq:simplemodel}). 
}\label{fig:KDE0v1_s20_spectra}
\end{figure}

For time-integrated spectra, the accretion-phase GR1D simulations are roughly consistent with the pinched Fermi-Dirac spectrum of Eq.~(\ref{eq:simplemodel}). This is demonstrated in Fig.~\ref{fig:KDE0v1_s20_spectra}, using the KDE0v1 EoS~\cite{Agrawal:2005ix} for a 20~M$_\odot$ progenitor. The spectra are integrated over time in the range $t_{\rm PB}=(0.2,0.72)$~sec, and are normalised to the $\nu_x$ fluence at $E_\nu=1$~MeV. Markers show the numerical result from the simulation. Solid lines show a fit to the pinched Fermi-Dirac form of Eq.~(\ref{eq:simplemodel}), with $a_{\nu_x}\approx-0.3$, $\langle E_{\nu_x}\rangle\approx14.6$~MeV; $a_{\bar\nu_e}\approx0.7$, $\langle E_{\bar\nu_e}\rangle\approx17.2$~MeV; $a_{\nu_e}\approx0.35$, $\langle E_{\nu_e}\rangle\approx14.3$~MeV. The ratios of (time-averaged) luminosities are $L_{\bar\nu_e}/L_{\nu_x}\approx2.74$, $L_{\bar\nu_e}/L_{\nu_e}\approx1.03$. 

To study the sensitivity of $R$ to modelling uncertainties, we therefore use Eq.~(\ref{eq:simplemodel}) and vary the parameters $a_{\nu_\alpha}$, $\langle E_{\nu_\alpha}\rangle$, and $L_{\nu_\alpha}$ within a reasonable range. For simplicity, we fix the $\bar\nu_e$ and $\nu_e$ spectral parameters to be equal, setting $a_{\bar\nu_e}=a_{\nu_e}$, $\langle E_{\bar\nu_e}\rangle=\langle E_{\nu_e}\rangle$, and $L_{\bar\nu_e}= L_{\nu_e}$. We allow the $\nu_x$ parameters to vary independently from those of $\nu_e$, apart from the constraint $0.5<\langle E_{\nu_x}\rangle/\langle E_{\nu_e}\rangle<2$. Comparing to the numerical simulations, we note that the approximations $a_{\nu_e}\approx a_{\bar\nu_e}$ and $\langle E_{\nu_e}\rangle\approx\langle E_{\bar\nu_e}\rangle$ are, in fact, not very good: the GR1D $\nu_e$ spectra are softer than those of $\bar\nu_e$ (the approximation $L_{\bar\nu_e}\approx L_{\nu_e}$ appears, however, accurate\footnote{A similar trend is seen, e.g., in the simulations of Ref.~\cite{Perego:2015zca}. We suspect that the $\nu_e$ spectrum may be softer than the $\bar\nu_e$ spectrum, despite having almost equal luminosities, because of $\nu_ee^-\to\nu_ee^-$ elastic scattering distorting the shape of the $\nu_e$ spectrum on its way out of the star, without appreciably affecting the total energy carried by it. However, we did not investigate this issue further.}). Nevertheless, relaxing the $\bar\nu_e-\nu_e$ spectral identification does not affect our results significantly.

\begin{figure}[htb!]
	\centering
	\includegraphics[width=0.495\textwidth]{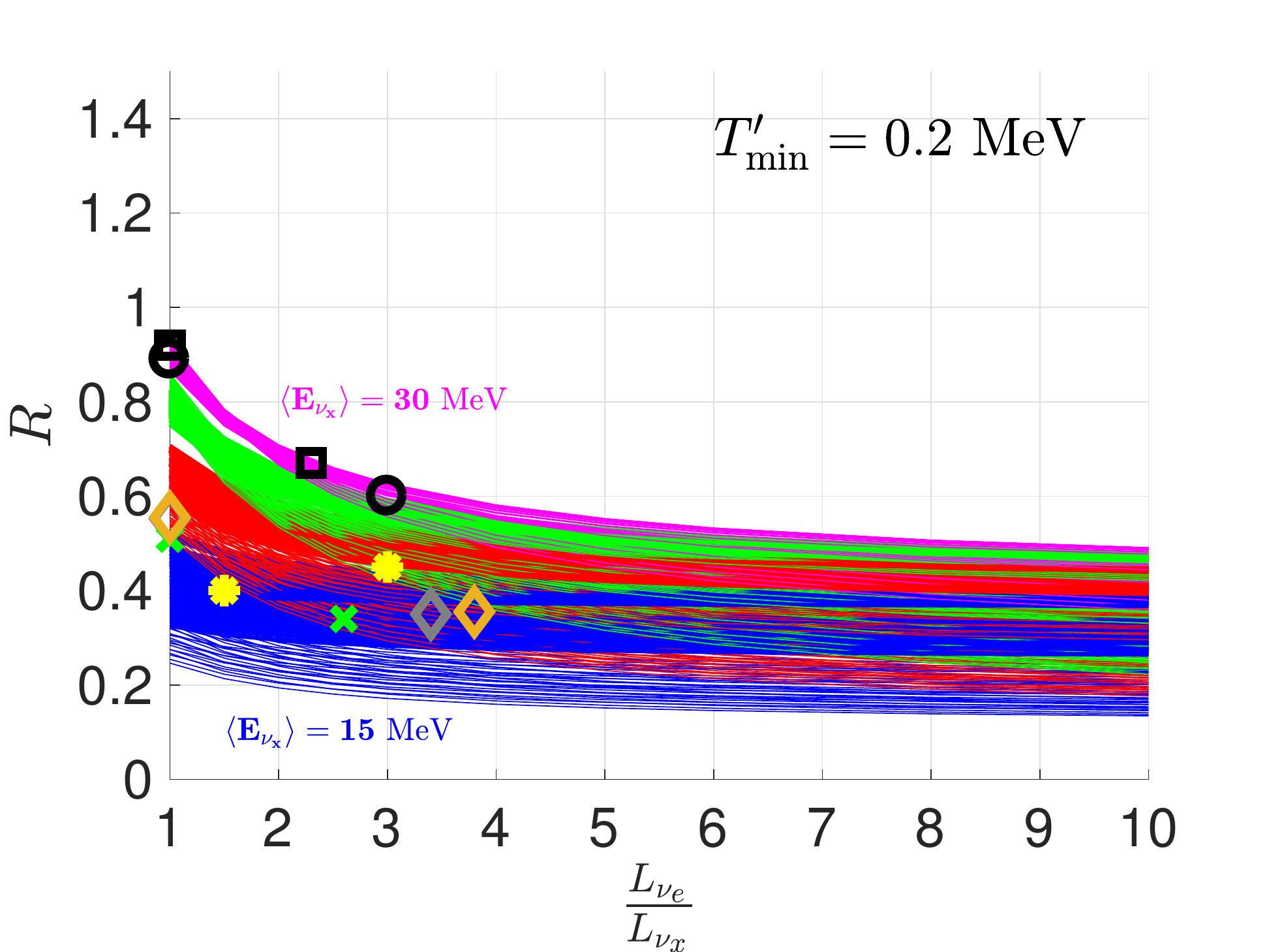}
		\includegraphics[width=0.495\textwidth]{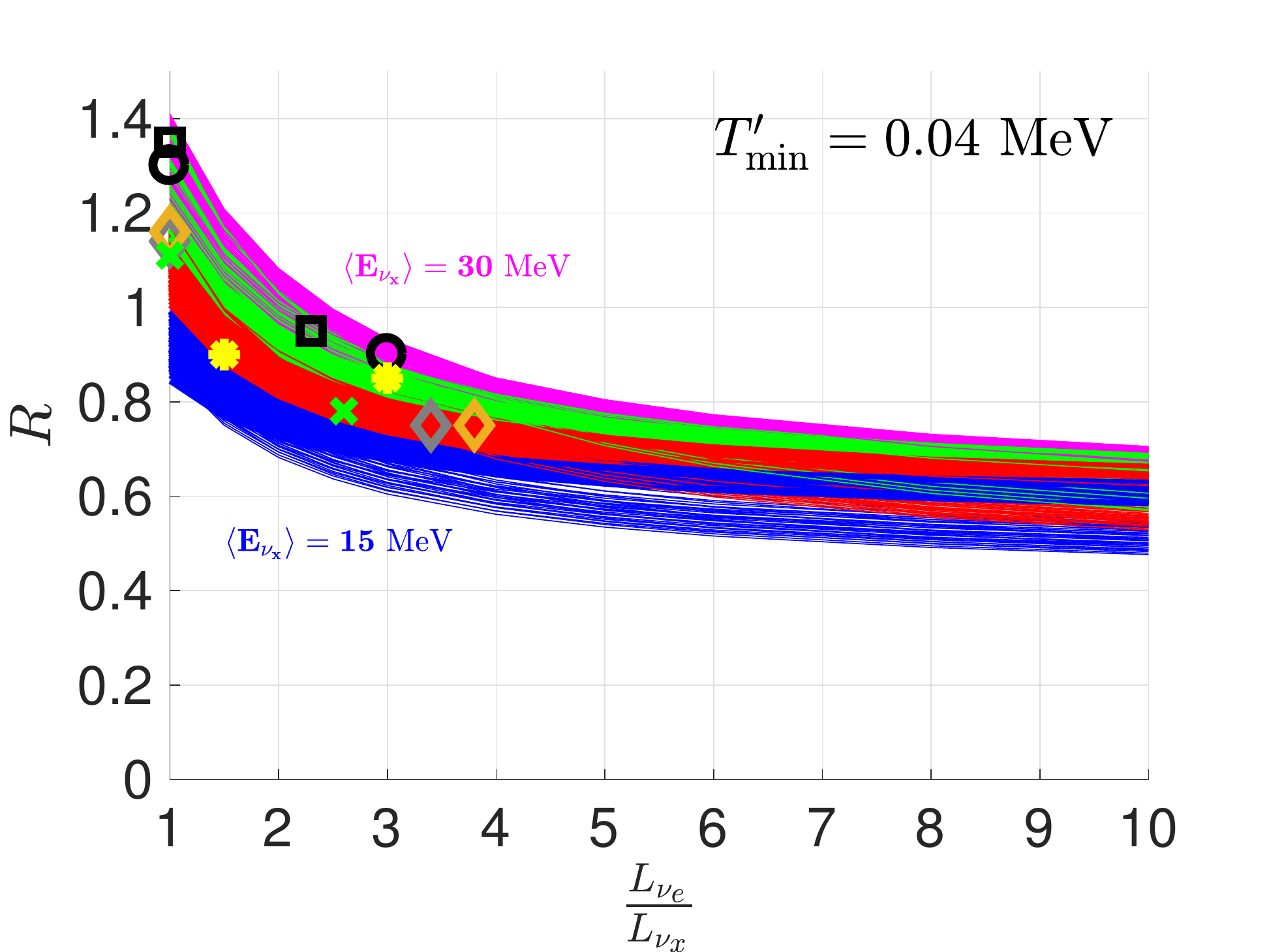}
	\caption{$R$ vs. the luminosity ratio $L_{\nu_e}/L_{\nu_x}$, calculated for the pinched Fermi-Dirac spectral parametrisation of  Eq.~(\ref{eq:simplemodel}) with varying model parameters (see text for details). {\bf Left:} lower (quenched) proton recoil energy threshold $T'_{min}=0.2$~MeV. {\bf Right:} $T'_{min}=0.04$~MeV. The results of three GR1D calculations are shown, denoted by the same markers as in Fig.~\ref{fig:rvspgrid02} with the some marker colors besides from for the 15~M$_\odot$ case, where the blue is replaced by grey for clarity in the plot. We also show estimated results from external simulations, discussed in App.~\ref{app:comp}. Two results from Ref.~\cite{Perego:2015zca} are shown by black circle (square) markers, denoting the HC (LC) progenitors there. Results from Ref.~\cite{Seadrow:2018ftp} are estimated by yellow star.}\label{fig:Rscan}
\end{figure}

In Fig.~\ref{fig:Rscan} we calculate $R$ as a function of the luminosity ratio $L_{\nu_e}/L_{\nu_x}$, using Eq.~(\ref{eq:simplemodel}) as prescribed above and scanning over the range:
\be a_{\nu_x},\,a_{\nu_e}=(-0.5,2),\;\;\;\langle E_{\nu_x}\rangle,\,\langle E_{\nu_e}\rangle=(15,30)~{\rm MeV}.\ee
Each individual line in Fig.~\ref{fig:Rscan} corresponds to one set of the combination $\left\{a_{\nu_x},\,a_{\nu_e},\,\langle E_{\nu_x}\rangle,\,\langle E_{\nu_e}\rangle\right\}$. Blue lines (generally corresponding to smaller numerical values for $R$) show the result for $\langle E_{\nu_x}\rangle=15$~MeV. As $\langle E_{\nu_x}\rangle$ is increased, the numerical value of $R$ also increases; the magenta lines correspond to $\langle E_{\nu_x}\rangle=30$~MeV. On the {\bf left}, we show the result for $T'_{min}=0.2$~MeV. On the {\bf right}, we use $T'_{min}=0.04$~MeV. 

The results of three GR1D calculations are superimposed in Fig.~\ref{fig:Rscan}, denoted by the same markers as in Fig.~\ref{fig:rvspgrid02} with the same marker colors besides from for the 15~M$_\odot$ case, where the blue is replaced by grey for clarity in the plot. We also show estimated results from external simulations~\cite{Perego:2015zca,Seadrow:2018ftp}, discussed in App.~\ref{app:comp}. We do not have detailed neutrino spectra from these simulations. We therefore estimate $R$ for these codes by using the mean neutrino energies and luminosity ratios reported in~\cite{Perego:2015zca,Seadrow:2018ftp}. Two results from Ref.~\cite{Perego:2015zca} are shown by black circle (square) markers, denoting the HC (LC) progenitors there\footnote{The PUSH (``D$\nu$M") simulations in~\cite{Perego:2015zca} actually get $L_{\bar\nu_e}/L_{\nu_x}$ very near, but slightly smaller than unity.}. Results from Ref.~\cite{Seadrow:2018ftp}, approximately applicable to the 11, 17, and 19~M$_\odot$ progenitors considered in App.~\ref{app:comp}, are estimated collectively by yellow star. 
%

We make a few comments regarding Fig.~\ref{fig:Rscan}: 
\begin{itemize}
\item At a given value of $L_{\nu_e}/L_{\nu_x}$, Fig.~\ref{fig:Rscan} shows a significant spread in $R$. The dominant variable that correlates with this spread is the neutrino mean energy $\langle E_\nu\rangle$, and the cause for the spread is  the pES detection efficiency $\epsilon_{\rm pES}$ (see Eq.~(\ref{eq:pESeff})): the right panel of Fig.~\ref{fig:EminTq} shows that for $\langle E_\nu\rangle$ varying in the range $15-30$~MeV, $\epsilon_{\rm pES}$ changes by a factor of $\sim2$ for $T'_{min}=0.2$~MeV, or a factor of $\sim1.5$ for $T'_{min}=0.04$~MeV. 
The same effect is seen in the semi-analytical calculation (solid lines) and the numerical simulations (markers). For example, the $\nu_x$ spectra from Ref.~\cite{Perego:2015zca} are generically characterised by higher mean energy than predicted by GR1D, which brings these simulations to predict high values of $R$. 

It is important to note, that the mean energy of the $\bar\nu_e$ component of the flux at Earth would be well determined by the IBD event spectra. This means that some of the $\langle E_\nu\rangle$-related spread in the $R$ vs. $L_{\nu_e}/L_{\nu_x}$ relation can be mitigated by a spectral analysis, going beyond $R$ alone. We return to this point extensively below, in this and the next sections.
\item The 2D ``exploding" simulations of Ref.~\cite{Seadrow:2018ftp} demonstrate a deviation from our simplified treatment of the D$\nu$M prediction: the luminosity ratio found in these 2D simulations is $L_{\nu_e}/L_{\nu_x}\sim1.5$, rather than $L_{\nu_e}/L_{\nu_x}\sim1$ as we have used here to define the D$\nu$M. The reason for this behaviour, as noted in~\cite{Seadrow:2018ftp}, is that accretion onto the PNS is observed to continue, albeit at a reduced level, even after the D$\nu$M-driven explosion is ongoing. 
\end{itemize}
We conclude that in order to mitigate modelling uncertainties, the consideration of $R$ needs to be combined with a spectral analysis of the IBD data, supplemented by the less precise information from the quenched pES recoil energy spectrum. 

\begin{figure}[htb!]
\centering
\includegraphics[width=0.495\textwidth]{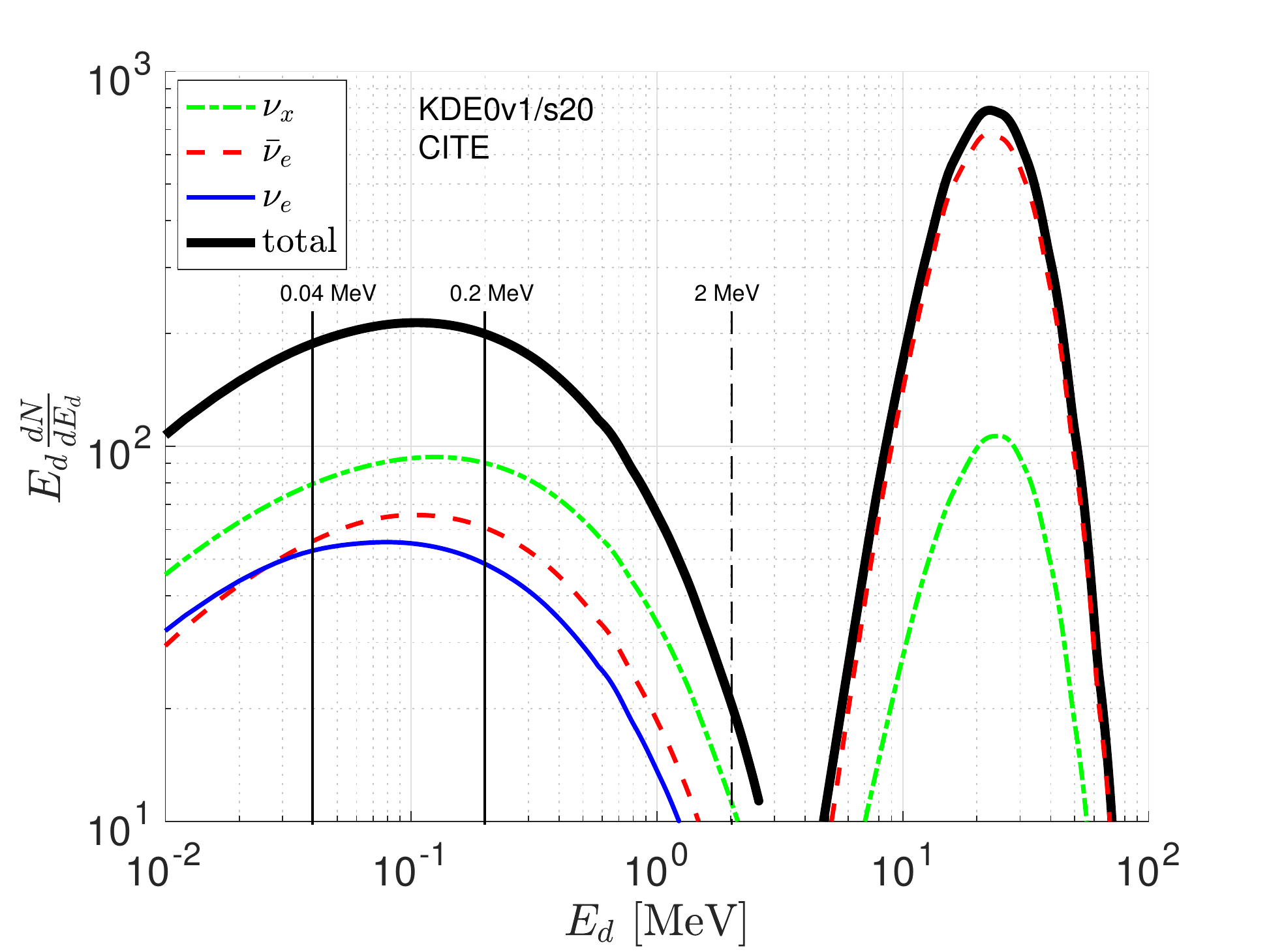}
\includegraphics[width=0.495\textwidth]{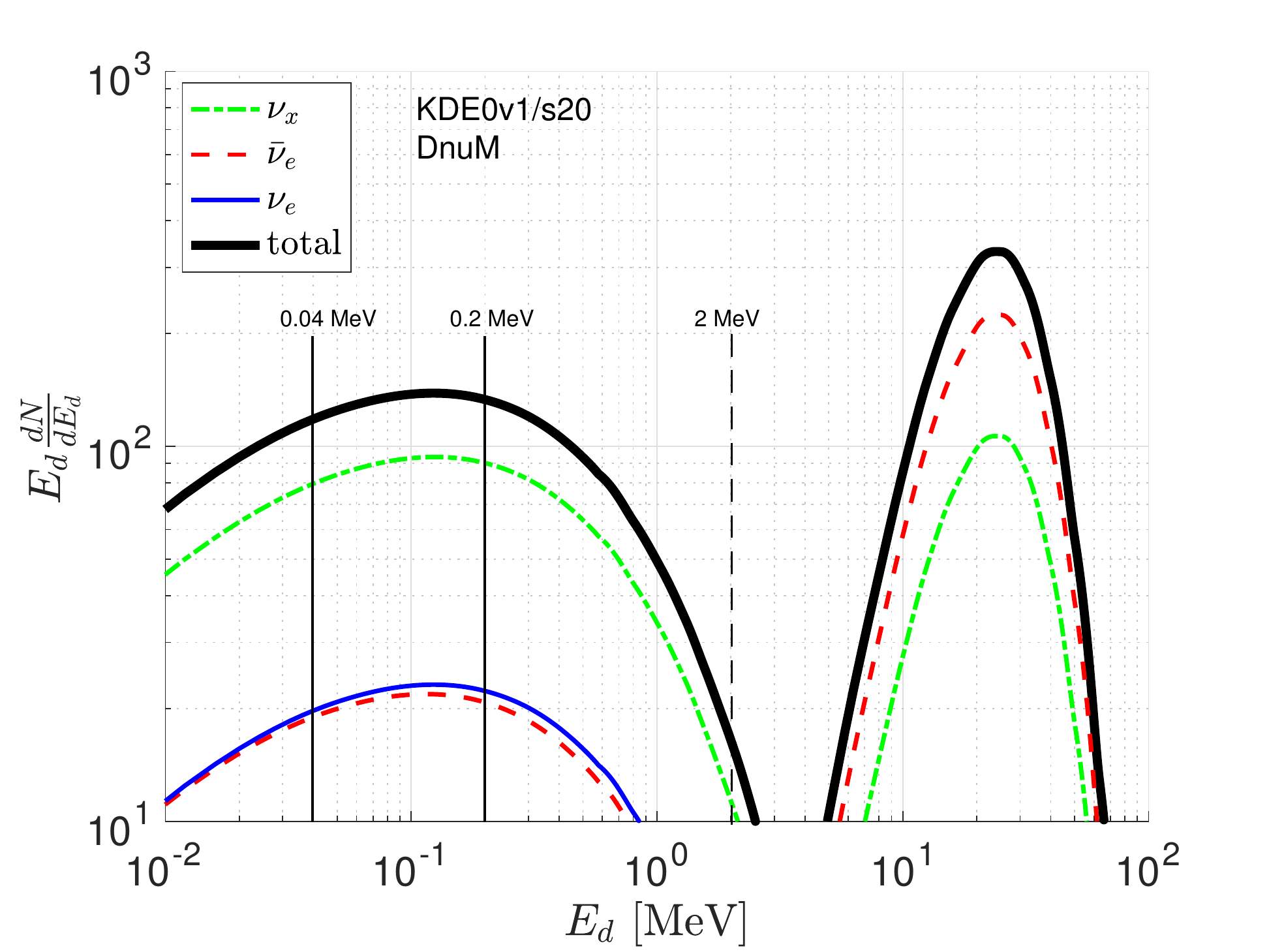}
\caption{
Time-integrated deposited energy spectra, for the simulation shown in Fig.~\ref{fig:KDE0v1_s20_spectra}. IBD peaks at $E_d\sim20$~MeV, while pES peaks at $E_d\sim0.1$~MeV. The pES spectra are quenched. {\bf Left:} CITE scenario. {\bf Right:} D$\nu$M scenario (estimated by setting the flux of $\nu_e$ and $\bar\nu_e$ equal to the simulated flux of $\nu_x$). The pES lower thresholds are marked by solid vertical lines. The pES upper threshold is marked by a dashed vertical line.
}\label{fig:KDE0v1_s20_EddNdEd}
\end{figure}

The parameter reconstruction exercise is illustrated in Fig.~\ref{fig:KDE0v1_s20_EddNdEd}. Here, we show detected energy spectra of IBD and pES for the 20~M$_\odot$ GR1D simulation (the same simulation that was used in Fig.~\ref{fig:KDE0v1_s20_spectra}). On the {\bf left} panel we show the CITE scenario. On the {\bf right} we show the D$\nu$M scenario (estimated by setting the flux of $\nu_e$ and $\bar\nu_e$ equal to the simulated flux of $\nu_x$). The IBD peak is seen at $E_d\sim20$~MeV, while the pES quenched energy spectrum peaks at $E_d\sim0.1$~MeV. The two pES lower threshold values discussed in this work, $T'_{min}=0.04$ and $0.2$~MeV, are marked by solid vertical lines. The pES upper threshold $T'_{max}=2$~MeV is marked by a dashed vertical line. The contributions due to individual neutrino flavours are shown in green, red, and blue for $\nu_x$, $\bar\nu_e$, and $\nu_e$, respectively. These colors correspond to neutrino flavour at the source, before oscillations are taken into account (with $P_{\bar e\bar e}=0.68$).

Considering Fig.~\ref{fig:KDE0v1_s20_EddNdEd}, the problem of diagnosing the emission process boils down to telling apart between the left and the right panels. The name of the game is to find the correct balance between the $\nu_x$ (green) and $\bar\nu_e$ (red) contributions. Because of modelling uncertainties, this should better be done without imposing strong theoretical bias on the individual parametrisation of each of the spectra. Note that accurate reconstruction of the $\nu_e$ contribution, which only affects pES, is not crucial; setting the $\nu_e$ spectrum to match that of $\bar\nu_e$ is a reasonable approximation that does not affect the results. Further, our analysis in this and the previous sections suggests that if the mean energies of $\nu_x$ and $\bar\nu_e$ can be deduced from the data with precision of a few tens of percent or better, than the pES/IBD event count ratio (with appropriate, not too high, threshold $T'_{min}$) provides the remaining information needed to determine the $L_{\bar\nu_e}/L_{\nu_x}$ luminosity ratio, and thus identify accretion-dominated emission. We explore this reconstruction problem in the next section.

\subsection{Likelihood analysis.}\label{ssec:likelihood}
%

In this section we go beyond simple counting, and attempt to reconstruct directly $L_{\bar\nu_e}/L_{\nu_x}$ from the experimental information. We base our analysis on GR1D as a concrete example.  
Given a numerical simulation, we construct multiple mock realisations of deposited energy spectra collected over a time interval $\Delta t$ at JUNO, including the effects of detector energy resolution and analysis cuts. For each of the mock detections, using the un-binned Poisson  likelihood method of~\cite{Ianni:2009bd,Pagliaroli2009,Vissani:2014doa}, we fit a pinched Fermi-Dirac model as in Eq.~\ref{eq:simplemodel}.  
We keep the fit parameters constant during the time interval $\Delta t$; time-dependent information can be obtained by analysing different time segments along the burst.

For simplicity, in fitting the mock data, we set equal the $\bar\nu_e$ and the $\nu_e$ spectra: $L_{\nu_e}=L_{\bar\nu_e},\;a_{\nu_e}=a_{\bar\nu_e},\;T_{\nu_e}=T_{\bar\nu_e}$. We stress that this simplification, just like the simplified spectral form given in Eq.~(\ref{eq:simplemodel}), is taken only in the fitting procedure and {\it not} in the calculation of the mock detection data being fitted. The mock data uses the full flavour- and energy-dependent information from the simulations. We thus have in total 6 free parameters, $\left\{L_{\nu_x},a_{\nu_x},T_{\nu_x}\right\}$ and $\left\{L_{\bar\nu_e},a_{\bar\nu_e},T_{\bar\nu_e}\right\}$. 
We present results in terms of the distribution of reconstructed best-fit parameters. Specifically, we focus on the ratio of best-fit $\bar\nu_e$ and $\nu_x$ luminosity parameters, $\left(L_{\bar\nu_e}/L_{\nu_x}\right)_{\rm BF}$. This distribution captures the key information we are after, considering $\left(L_{\bar\nu_e}/L_{\nu_x}\right)_{\rm BF}$ as one possible test-statistic with direct physical interpretation.

\begin{figure}[htb!]
\centering
\includegraphics[width=0.495\textwidth]{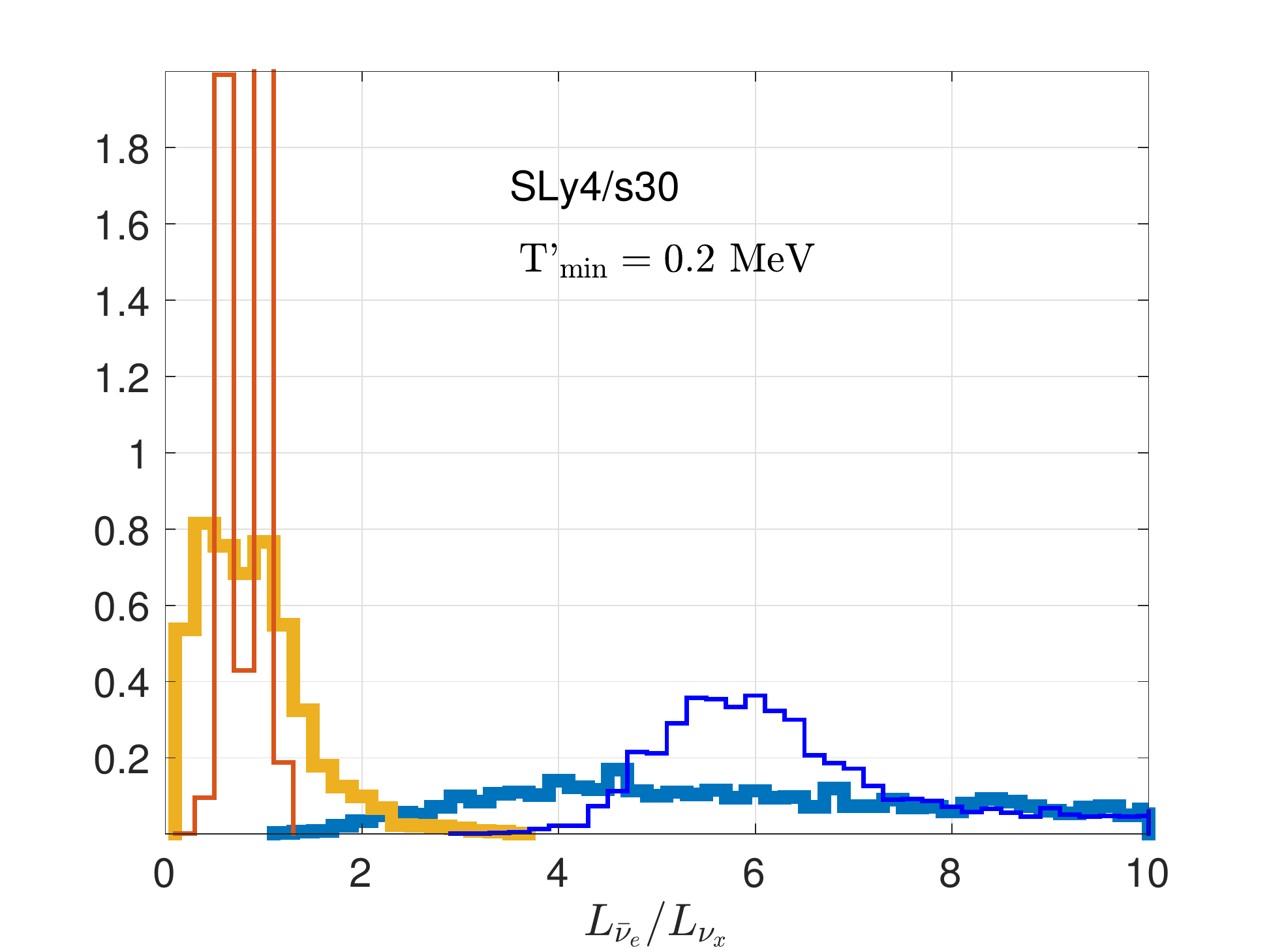}
\includegraphics[width=0.495\textwidth]{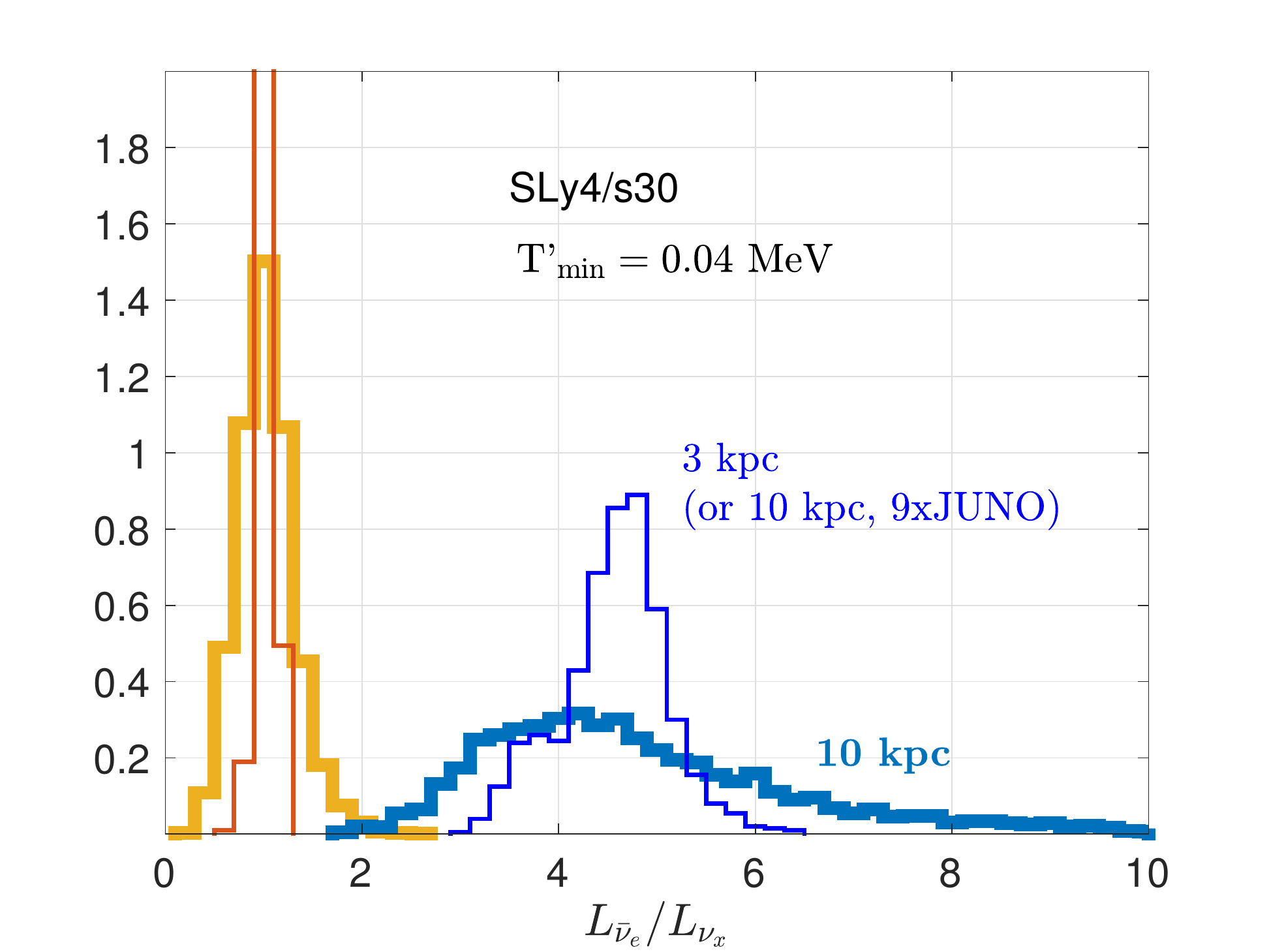}
\includegraphics[width=0.495\textwidth]{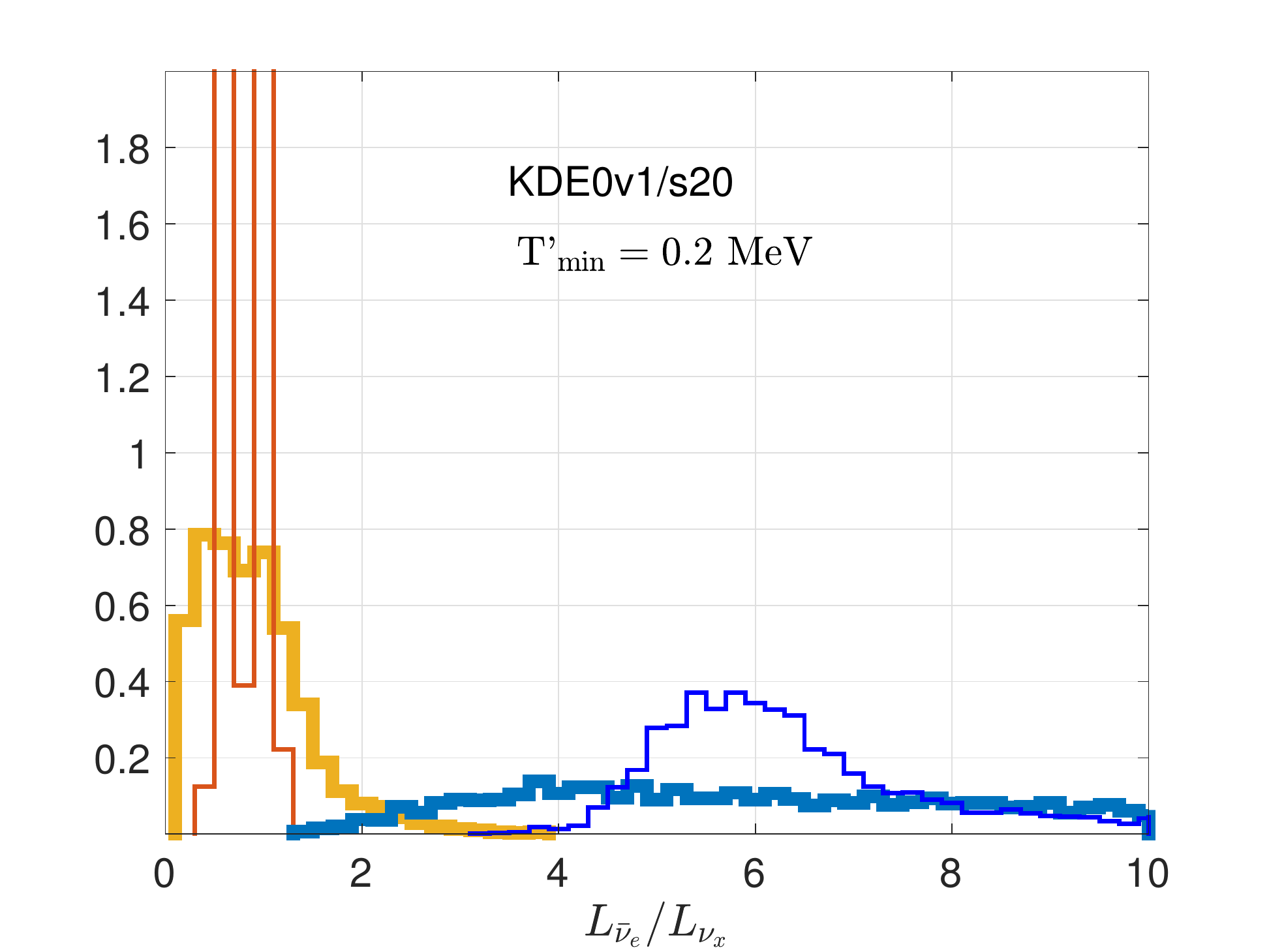}
\includegraphics[width=0.495\textwidth]{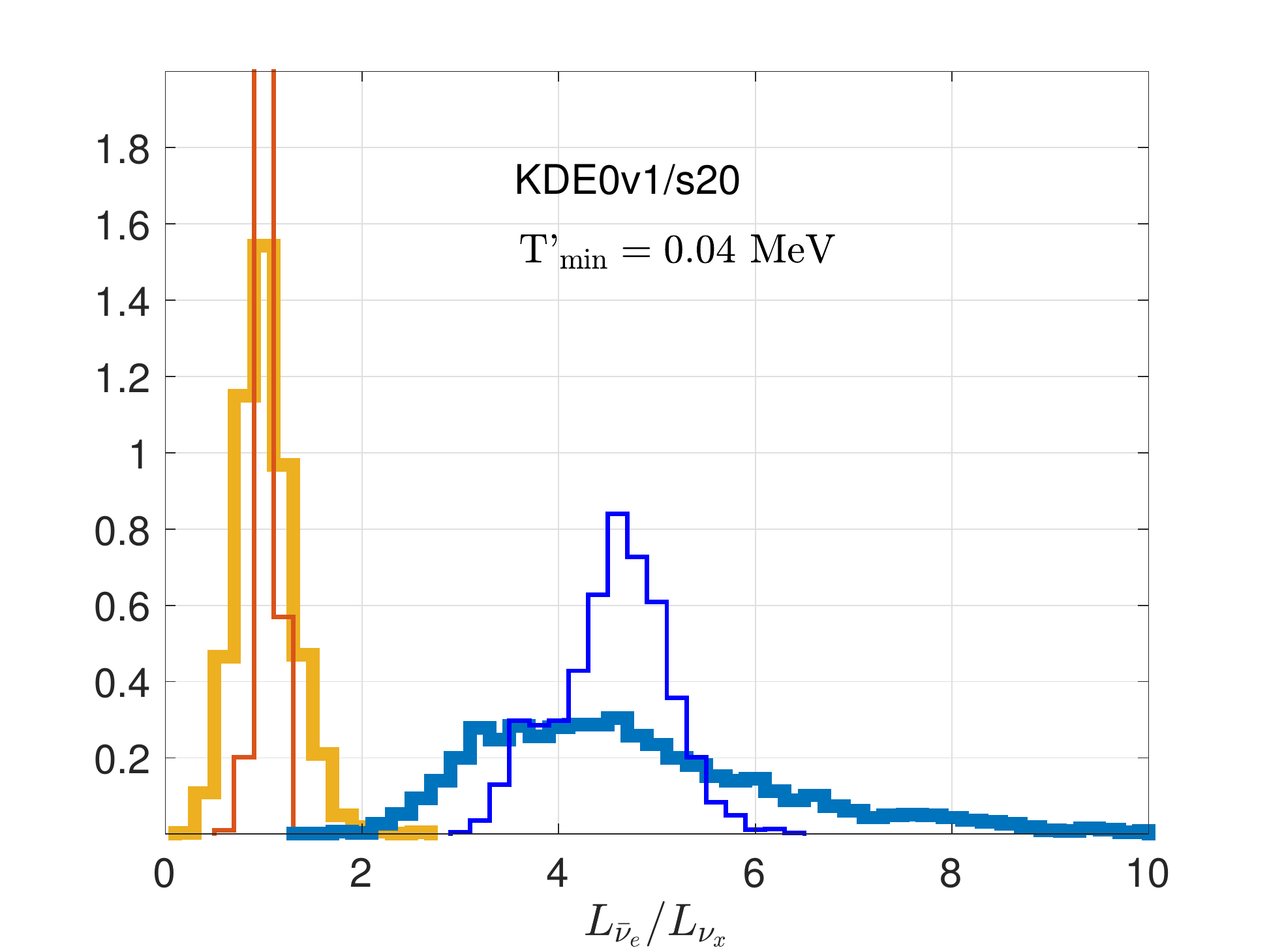}
\caption{
Distribution of reconstructed best-fit luminosity ratios for 150 mock data realisations, computed from GR1D simulations of CITE (blue) and the D$\nu$M (orange). {\bf Top, Bottom:}  30 and 20~M$_\odot$ progenitors, respectively. The analysed time interval is [0.2-0.7]~sec post-bounce. {\bf Left:} Proton recoil deposited energy threshold $T'_{min}=0.2$~MeV. {\bf Right:} $T'_{min}=0.04$~MeV.
}\label{fig:lik}
\end{figure}

Fig.~\ref{fig:lik} shows the probability distribution (PDF) of $\left(L_{\bar\nu_e}/L_{\nu_x}\right)_{\rm BF}$ obtained for 150 mock data realisations with CITE (blue) and with the D$\nu$M (orange). The neutrino source is a GR1D simulation for a 30 and 20~M$_\odot$ star (top and bottom panels, respectively). The analysed time segment was [0.2,0.7]~sec post-bounce. In the left (right) panel, the proton deposited recoil energy threshold was $T'_{min}=0.2$~MeV (0.04~MeV). Thick lines show the result when setting the distance to the CCSN to $D_{\rm SN}=10$~kpc. Thin lines show the result when the distance is $D_{\rm SN}=3$~kpc; equivalently, these results correspond to a detector with $9$ times the effective volume of JUNO. We assume $P_{\bar e\bar e}=0.68$. 

%

The likelihood modelling of Figs.~\ref{fig:lik} generalises the simple analysis of the counting observable $R$. Without imposing strong modelling bias, and despite degeneracies in the neutrino spectral parameters, a global analysis of the deposited energy spectra at JUNO within the first $\sim1$~sec of the CCSN burst would allow to constrain the luminosity ratio $L_{\bar\nu_e}/L_{\nu_x}$. Accretion-dominated emission with $L_{\bar\nu_e}/L_{\nu_x}\gtrsim3$ can be discriminated from $L_{\bar\nu_e}/L_{\nu_x}=1$ with high confidence. The analysis in Sec.~\ref{ssec:Rwhat} makes clear how this discrimination works: (i) the IBD spectrum effectively constrains the mean neutrino energies; (ii) the pES/IBD event ratio then allows to reconstruct $L_{\bar\nu_e}/L_{\nu_x}$, with relative insensitivity to details of the spectral pinch parameter ($a_\nu$ in Eq.~(\ref{eq:simplemodel})). Degeneracy towards high values of $L_{\bar\nu_e}/L_{\nu_x}\gg1$ is seen, and is understood from, e.g., Fig.\ref{fig:Rscan}: once $L_{\bar\nu_e}/L_{\nu_x}\gtrsim3$ is achieved, the pES/IBD ratio approaches its asymptotic value obtained at $L_{\bar\nu_e}/L_{\nu_x}\to\infty$. This asymptotic value, $R\to2\times\frac{1.7}{6.1}\frac{\epsilon_{\rm PES}}{P_{\bar e\bar e}}$ for $L_{\bar\nu_e}/L_{\nu_x}\to\infty$, can be understood from the approximate cross section formula in Eqs.~(\ref{eq:xsecapp1}-\ref{eq:xsecapp2}), combined with the pES efficiency $\epsilon_{\rm pES}$ shown in the right panel of Fig.~\ref{fig:EminTq}.

\section{Summary}\label{sec:sum}

Most of the theoretical efforts in solving the explosion mechanism of core-collapse supernovae (CCSNe) revolves around the hypothesis, that a small fraction (of order percent) of the large gravitational binding energy of the core (few~$10^{53}$~erg) couples somehow to the stellar envelope and powers the explosion (with kinetic energy $E_K\sim10^{51}-10^{52}$~erg). 
The delayed neutrino mechanism (D$\nu$M) is the most well studied example of this kind~\cite{Bethe:1984ux,Kotake:2005zn,Burrows:2012ew,Janka:2016fox}. 
%
A competing hypothesis is that the supernova is powered by a collapse-induced thermonuclear explosion of the envelope~\cite{Burbidge:1957vc,Hoyle:1960zz,Fowler:1964zz,Kushnir:2014oca,Kushnir:2015mca} (CITE). In this case, the explosion energy does not come from a low-efficiency coupling of the envelope to the core. Instead, the explosion energies of CCSNe are reproduced by the $\sim$MeV nuclear binding energy per nucleon, released in burning a few M$_\odot$ of the progenitor star~\cite{Kushnir:2015vka}, with $\mathcal{O}(1)$ efficiency.

The explosion time scale in the D$\nu$M is of the order of the dynamical time scale of the $\sim1-2$~M$_\odot$ inner core, less than 1~sec. In contrast, in CITE the explosion time scale is of the order of the dynamical time scale of the He-O layer, about 10~sec. As a result, the D$\nu$M predicts a short period of accretion-dominated neutrino luminosity, transiting at $t<1$~sec into proto-neutron star (PNS) cooling, while CITE predicts a prolonged accretion phase that can last for a few sec.  

In both models, the total neutrino luminosity could vary significantly with time on hundreds of ms timescale. Such variations follow from the accretion of composition transition layers through the PNS accretion shock. In the D$\nu$M, a luminosity drop occurs upon explosion when the accretion component is blown away. In CITE, a feature could follow from BH and/or accretion disk formation.
Both models also exhibit nontrivial time evolution of the mean neutrino energy. In the D$\nu$M, neutrino energies rise during the first $\sim1-2$~sec but then decrease smoothly as the PNS cools. In CITE, neutrino energies are expected to rise towards BH formation, but the formation of a dominant accretion component from an accretion disc could lead the observed mean energy to decrease. 
In case of a galactic CCSN, these different features would be manifest with high statistics and could help to diagnose the explosion mechanism. In particular, a sharp drop in total luminosity accompanied by a few sec of quiescence, as a result of BH formation, would allow clean identification of a CITE-like scenario. Such abrupt luminosity gap is consistent with data for SN1987A~\cite{Blum:2016afe}, but statistics for that event were too sparse to allow robust conclusions. 

It is possible, however, that the next galactic CCSN would not provide us with a clear smoking gun like prompt BH formation followed by a luminosity gap; and the various luminosity and energy features discussed above could be nontrivial to interpret. 
In this work, we therefore focused on another characteristic -- the relative neutrino flavour composition at the source -- as a diagnostic of the explosion. 
Accretion-dominated neutrino emission at the source is characterised by enhanced $\nu_e,\,\bar\nu_e$ compared to $\nu_x$ (where $x$ stands collectively for $\mu$ and $\tau$ neutrinos and anti-neutrinos). Focusing on the scintillation detector JUNO, currently under construction, we showed that the ratio $R={\rm pES}/{\rm IBD}$ between inverse-beta decay (IBD) and proton-neutrino elastic scattering (pES) event rates -- the two dominant detection channels, assuming realistic detection energy thresholds -- is a robust probe of such an accretion-dominated source. Using 1D numerical simulations, we studied the impact of progenitor star profile and of nuclear equation of state (EoS) on the source prediction. Given that the mean neutrino energy can be reconstructed with $\mathcal{O}(10\%)$ accuracy, from the spectrum of detected IBD events, we find that the accretion-dominated scenario makes a distinctive prediction for $R$. Theoretical model uncertainties shrink if we allow optimistic lower quenched proton energy threshold, $T'_{min}<0.1$~MeV, as compared to the nominal $T'_{min}=0.2$~MeV currently suggested for JUNO. 


The main caveats in the analysis are the unknown impact of neutrino self-induced flavour oscillations at the source, and the current uncertainty on the pES cross section. With a clear prediction for the  accretion phase value of $R$ at JUNO, assuming standard adiabatic MSW propagation, the possible effect of self-induced oscillations can be read from the early CCSN data itself.

\section*{Acknowledgements}
We are grateful to Evan O'Connor and Christian Ott for making their numerical simulation code GR1D publicly available, and to Evan O'Connor for help in technical issues related to running the code. We thank Avital Dery, Yossi Nir, and Eli Waxman for useful discussions, John Beacom for comments on the manuscript, and Doron Kushnir for many insightful discussions and for providing MESA profiles and CITE numerical simulation data.  
The work of NB and KB was supported by grant 1937/12 from 
the I-CORE program of the Planning and Budgeting Committee and the Israel
Science Foundation and by grant 1507/16
from the Israel Science Foundation. KB is incumbent
of the Dewey David Stone and Harry Levine career
development chair. GDA is supported by the Simons Foundation Origins of the Universe program (Modern Inflationary Cosmology collaboration).

\begin{appendix}

\section{Comparing accretion-dominated and PNS cooling-dominated neutrino emission in different simulations}\label{app:comp}

In the body of this work, we used GR1D to simulate the accretion phase of a CCSN neutrino burst. If CITE causes the explosion, then this ``accretion phase" luminosity with its characteristic neutrino flavour pattern is expected to describe the entire neutrino burst, with a possible temporary pause on BH formation~\cite{Blum:2016afe}. 

On the other hand, because the 1D simulation does not explode, we cannot directly simulate the neutrino flavour content of the D$\nu$M. This caveat could only be truly resolved with numerically converged, demonstrated D$\nu$M explosions. Until such tools become available, below we show two D$\nu$M examples from the literature where explosions are either started artificially by hand, or where explosions are reported albeit with no demonstration of numerical convergence. 

\subsection{``PUSH" simulations of Ref.~\cite{Perego:2015zca}}
\begin{figure}[htb!]
\centering
\includegraphics[width=0.495\textwidth]{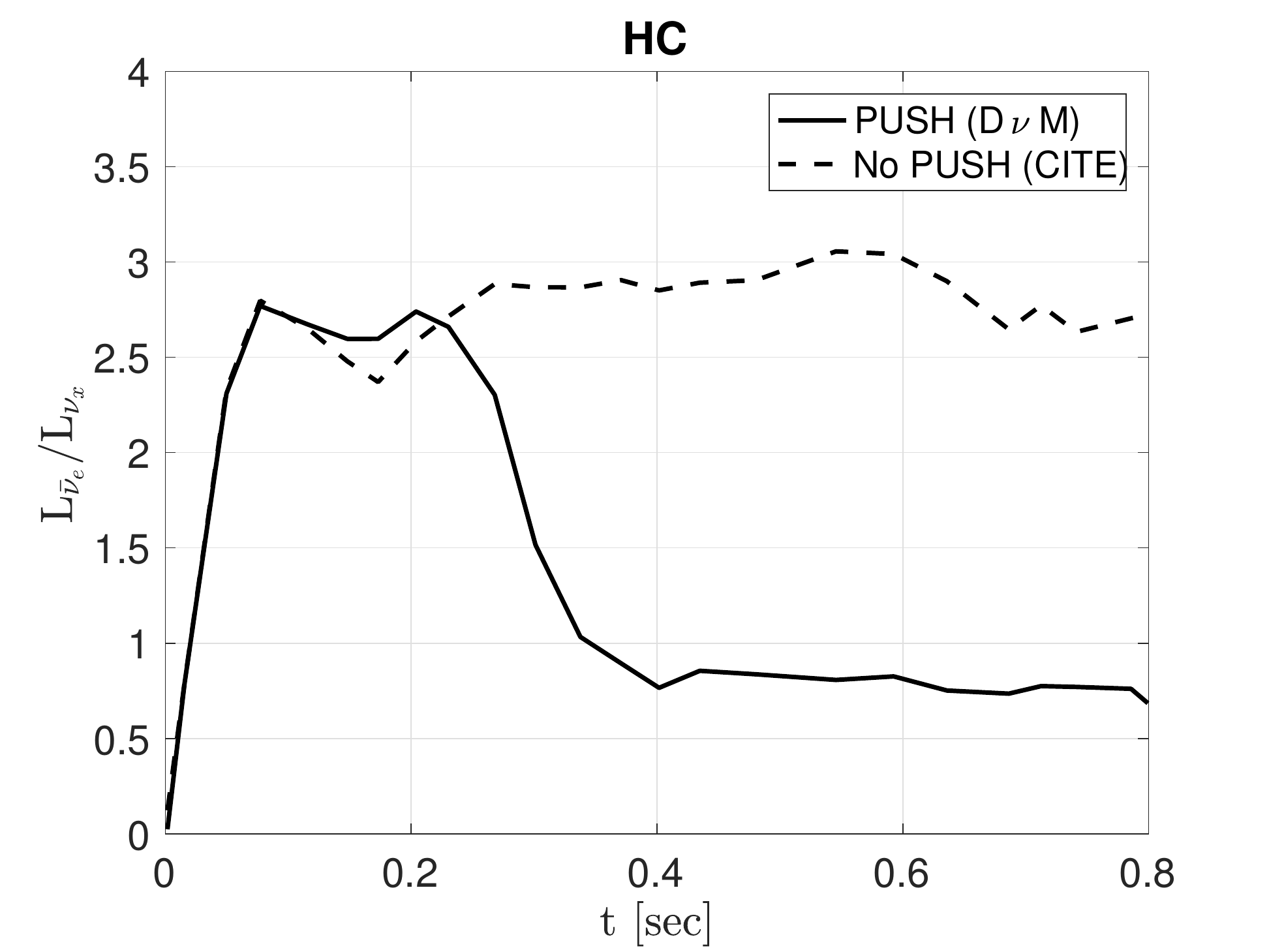}
\includegraphics[width=0.495\textwidth]{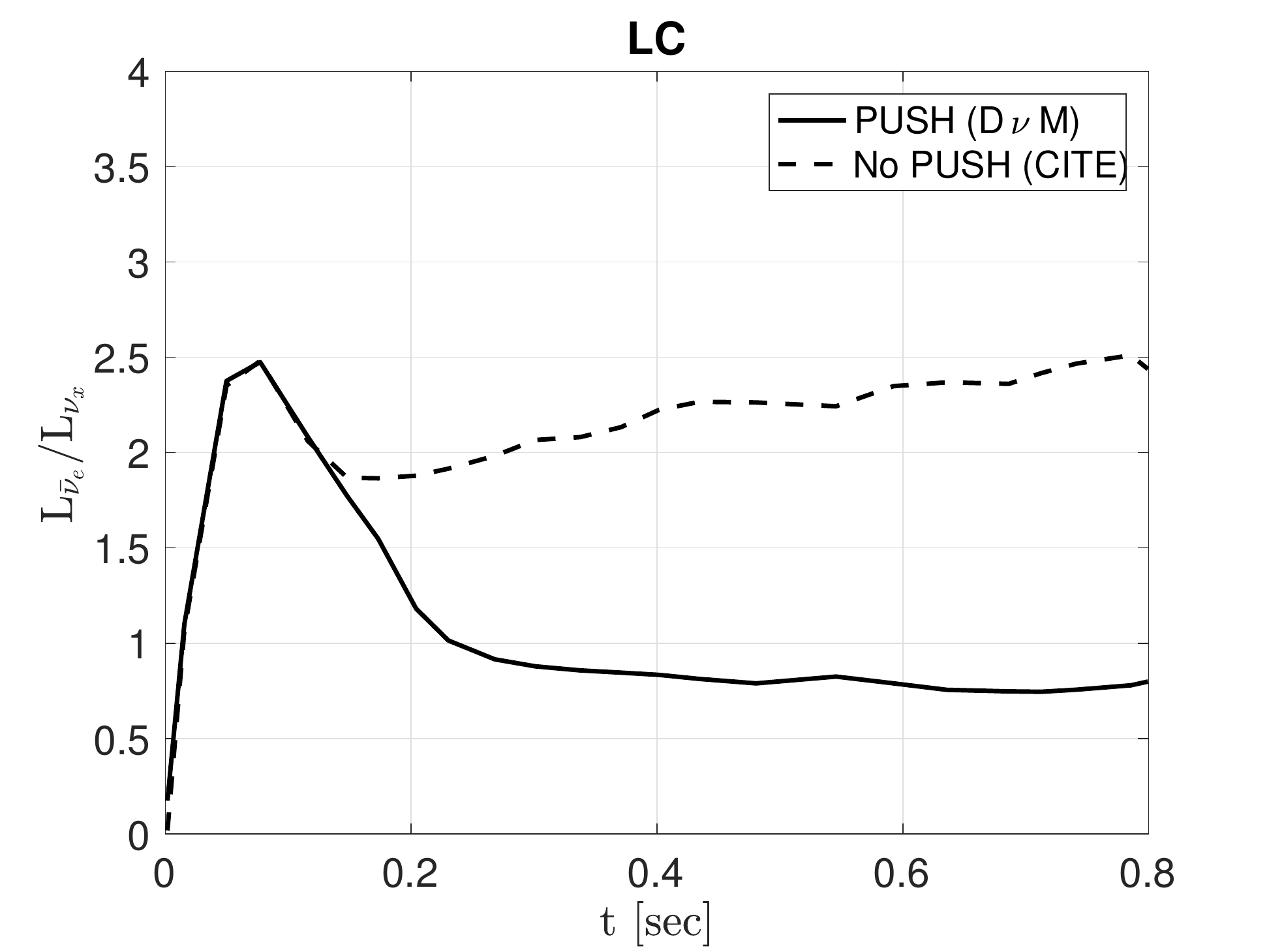}
\caption{
Temporal dependence of the neutrino luminosity, taken from the simulations of Ref.~\cite{Perego:2015zca}. {\bf Left:} The HC progenitor of~\cite{Perego:2015zca}. {\bf Right:} The LC progenitor of~\cite{Perego:2015zca}. In both panels, solid curves (denoted PUSH) show the case where an artificial explosion is triggered by hand at $t\sim0.2$~sec, while dashed curves (denoted No PUSH) show the result where the simulation is allowed to run without modification, so that no explosion occurs within the simulated time. The solid curve demonstrates the prediction of the D$\nu$M, while the dashed curve demonstrates the prediction of CITE.
}\label{fig:Lpush}
\end{figure}

Ref.~\cite{Perego:2015zca} reported the results of state of the art simulations in which artificial ``D$\nu$M explosions" were triggered by hand. Conveniently for our purpose, Ref.~\cite{Perego:2015zca} also reported, for the same initial conditions, the results of simulation runs in which the explosion was not triggered, allowing to demonstrate the difference in flavour content between the two cases. The results are shown in Fig.~\ref{fig:Lpush}, where the left and right panels depict two different progenitor star profiles, denoted high compactness (HC) and low compactness (LC) in~\cite{Perego:2015zca}. These progenitor profiles correspond to zero-age main sequence (ZAMS) mass of $20$ and $19.2$~M$_\odot$, respectively.

\subsection{``1D non-explosion" vs. ``2D exploding" simulations of Ref.~\cite{Seadrow:2018ftp}}

\begin{figure}[htb!]
\centering
\includegraphics[width=0.6\textwidth]{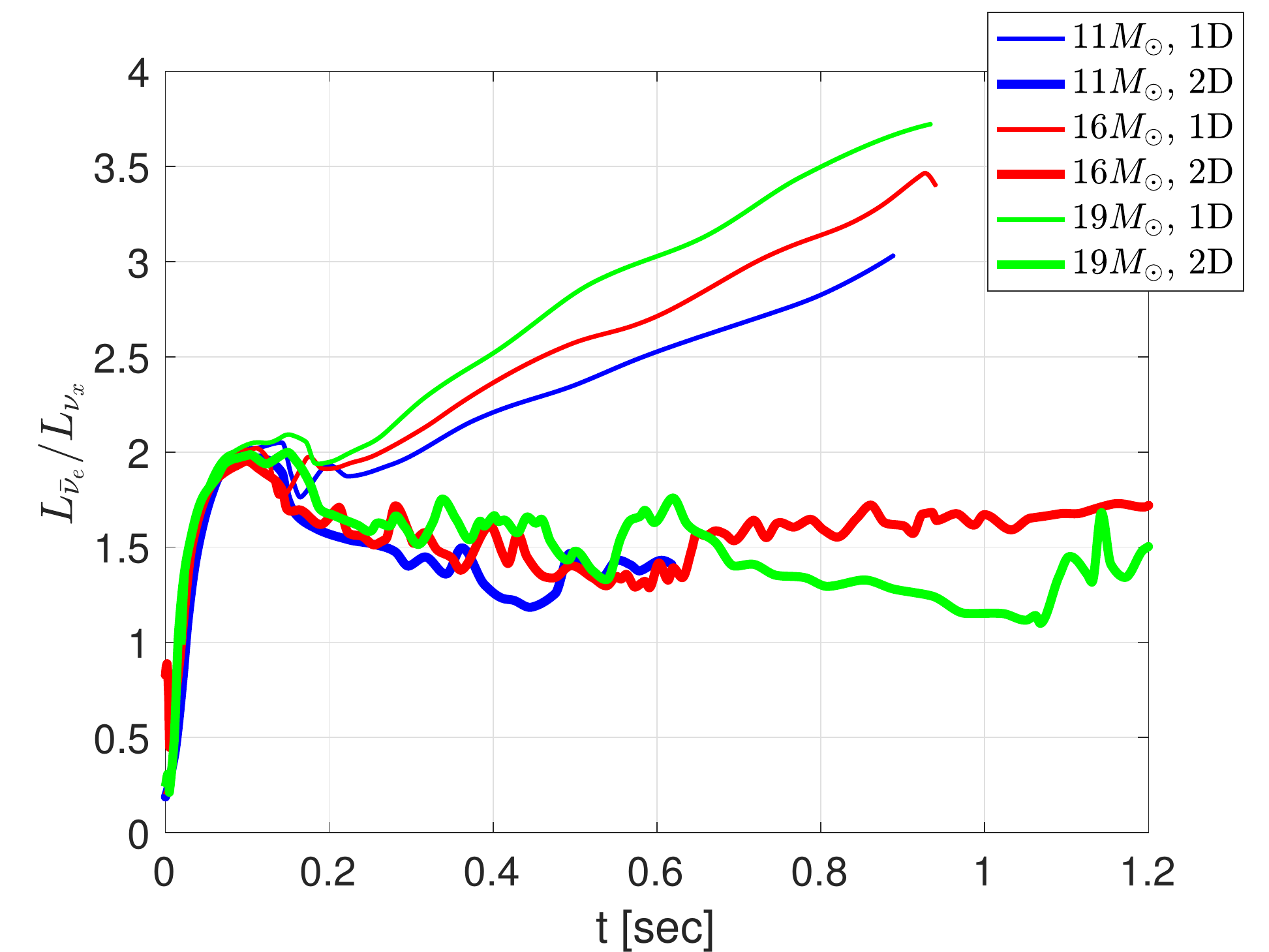}\caption{
Temporal dependence of $f_{\bar e}=L_{\bar\nu_e}/L_{\nu_x}$, calculated from the published plots in~\cite{Seadrow:2018ftp}.
}\label{fig:bur}
\end{figure}

Ref.~\cite{Seadrow:2018ftp} studied the neutrino signal of a galactic CCSN, highlighting -- as we did here -- the signature of the accretion phase. In~\cite{Seadrow:2018ftp}, 1D and 2D simulations of the same progenitor star were performed. The 2D simulations where reported as exploding, while the 1D simulations did not explode. 
These two sets of simulations give a useful demonstration of the flavour information we study in this work. In Fig.~\ref{fig:bur} we show the $L_{\bar\nu_e}/L_{\nu_x}$ ratio calculated from the published plots in~\cite{Seadrow:2018ftp} for three progenitor masses. 

\section{Neutrino scattering cross sections}\label{app:xs}
The differential cross section for neutrino-proton elastic scattering (pES) with proton recoil energy $T$ and initial neutrino energy $E_\nu$, is given by~\cite{Bilenky:1982ms}
\be\!\!\label{eq:dxspES}\frac{d\sigma_{\nu(\bar\nu) p}}{dT}&=&\frac{G_F^2m_p}{2\pi}\left(\left(c_{Vp}\pm c_{Ap}\right)^2+\left(c_{Vp}\mp c_{Ap}\right)^2\left(1-\frac{T}{E_\nu}\right)^2+\left(c_{Ap}^2-c_{Vp}^2\right)\frac{m_pT}{E_\nu^2}\pm4c_{Mp}c_{Ap}\left(1\pm\xi_{Mp}\right)\frac{T}{E_\nu}\right),\\
c_{Vp}&=&\frac{1-4s^2_W}{2},\;\;\;\;c_{Ap}=\frac{1.27-\Delta s}{2},\no\\
c_{Mp}&=&\frac{\mu_p}{2}\approx1.4,\;\;\;\;\xi_{Mp}=\frac{T}{E_\nu}\frac{c_{Vp}+\frac{1}{4}c_{Mp}\mp c_{Ap}}{c_{Ap}}+\frac{c_{Mp}}{2c_{Ap}}\left(1-\frac{T}{E_\nu}\right)\frac{E_\nu}{m_p}.\no
\ee
Here, in $\pm$, the upper sign refers to $p+\nu\to p+\nu$ and the lower sign refers to $p+\bar\nu\to p+\bar\nu$. For the Weinberg angle we use $s^2_W=0.23$~\cite{Tanabashi:2018oca}. 
For the $\Delta s$ correction in the expression for $c_{Ap}$, due to the strange quark contribution to the proton spin, a recent result from~\cite{Sufian:2018qtw} is $\Delta s=-0.196\pm0.127$. The uncertainty on $\Delta s$ leads to an uncertainty of $\approx\pm9\%$ on $c_{Ap}$ which, in turn, implies a systematic cross section uncertainty of $\approx\pm20\%$. This would be a large systematic effect, directly entering the pES/IBD ratio. There is a proposal to reduce this uncertainty by an order of magnitude~\cite{Pagliaroli2013}, and as explained in the main text we assume that the uncertainty on $\sigma_{pES}$ can indeed be reduced to the few percent level. Since the uncertainty is still very large, we use for concreteness $ \Delta s=0 $ in all calculations.

The maximal proton recoil energy is
\be T_{max}&=&\frac{2E_\nu^2}{m_p+2E_\nu},\ee
implying $T/E_\nu<2E_\nu/m_p\ll 1$ for CCSN neutrinos. In all of the core-collapse simulations we study, neutrinos with $E_\nu>100$~MeV contribute no more than a few percent of the detected events. For $E_\nu=100$~MeV, $T/E_\nu<0.2$, implying sub-percent contribution to the detected events coming from terms in Eq.~(\ref{eq:dxspES}) that are linear in $T/E_\nu$ (not including the third term in Eq.~(\ref{eq:dxspES}), where the factor $T/E_\nu$ is compensated by a factor $m_p/E_\nu$).
The weak magnetism contribution, proportional to $c_{Mp}\,c_{Ap}$, is suppressed both by an uncompensated factor of $T/E_\nu$ and in addition by the fact that it reverses sign (up to the numerically insignificant $\xi_{Mp}$ correction) when changing between $\nu$ and $\bar\nu$. Core-collapse simulations and analytic estimates predict nearly equal spectra for $\nu_{\mu,\tau}$ and $\bar\nu_{\mu,\tau}$ (which, therefore, are usually collectively denoted $\nu_x$), such that the weak magnetism contributions cancel out for the $\nu_x$ part of the flux. The $\nu_e$ and $\bar\nu_e$ spectra are not equal, but are still observed to follow each other to about 20\% level in simulations (this discussion pertains to the post-bounce phase of the collapse).

The cross section for inverse beta decay (IBD), $p+\bar\nu_e\to n+e^+$, is given to $\sim1\%$ accuracy by~\cite{Strumia:2003zx}
\be\sigma_{IBD}&=&\kappa_{IBD}\,p_e\,E_e,\\
\kappa_{IBD}&=&10^{-43}\,\exp\left(-0.07056\,x+0.02018\,x^2-0.001953\,x^4\right)~{\rm cm^2},\ee
where $p_e=\sqrt{E_e^2-m_e^2}$ and $x=\log\left(E_\nu/{\rm MeV}\right)$.

The cross section for neutrino-electron elastic scattering, $e^-+\nu\to e^-+\nu$, depends on the neutrino flavour. For $1<E_\nu<100$~MeV, to $\sim10\%$ accuracy, it is given by (see, e.g.,~\cite{Marciano:2003eq})
\be\sigma_{\nu_ee}=1.86\times10^{-43}\left(\frac{E_\nu}{20~{\rm MeV}}\right)~{\rm cm^2},\;\;\;\;
\sigma_{\bar\nu_ee}=0.78\times10^{-43}\left(\frac{E_\nu}{20~{\rm MeV}}\right)~{\rm cm^2},\\
\sigma_{\nu_{\mu,\tau}e}=0.32\times10^{-43}\left(\frac{E_\nu}{20~{\rm MeV}}\right)~{\rm cm^2},\;\;\;\;
\sigma_{\bar\nu_{\mu,\tau}e}=0.26\times10^{-43}\left(\frac{E_\nu}{20~{\rm MeV}}\right)~{\rm cm^2}.\nonumber
\ee

\begin{figure}[htb!]
\centering
\includegraphics[width=0.6\textwidth]{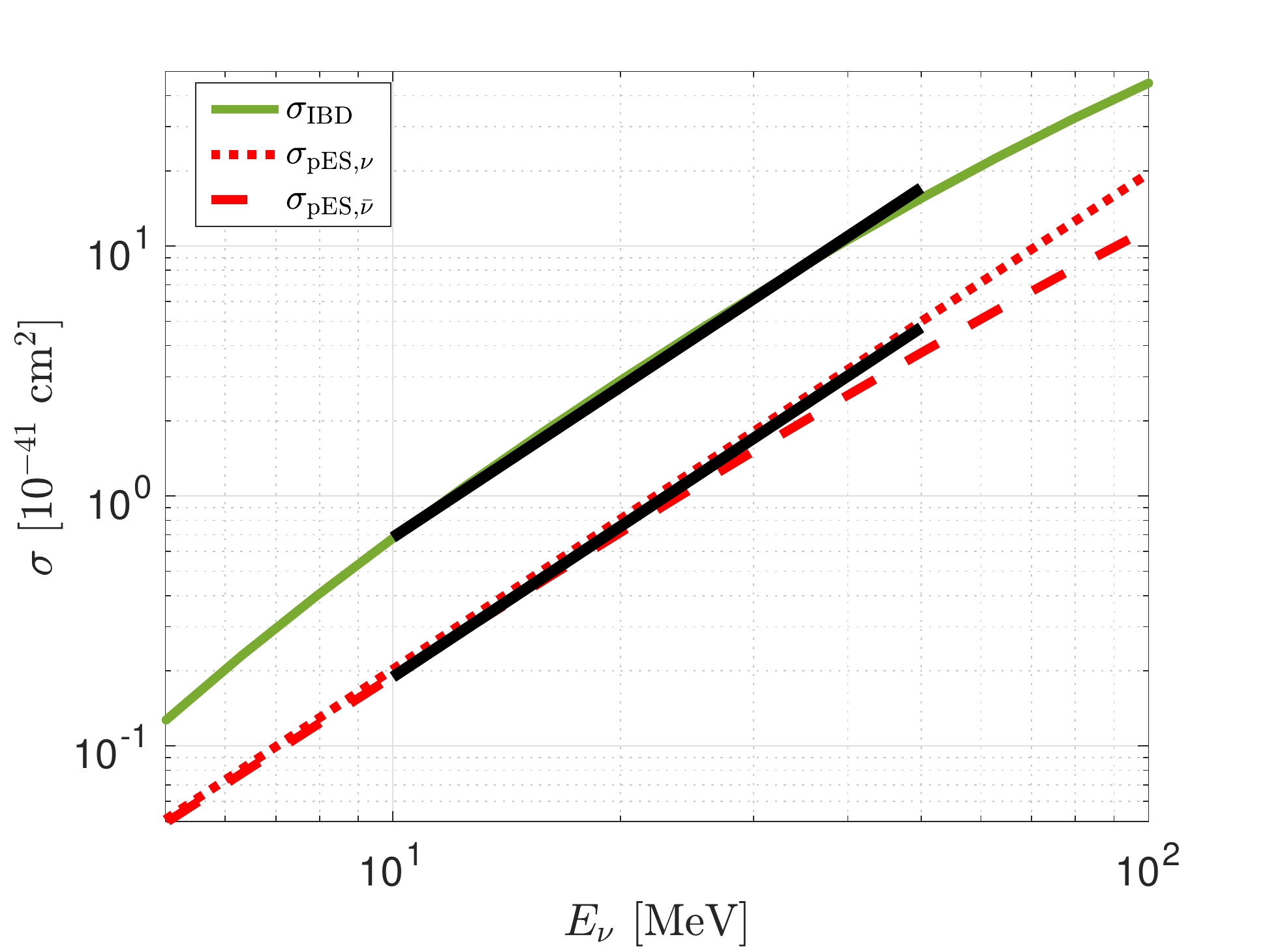}
\caption{
Total pES and IBD cross sections. The approximate formulae of Eqs.~(\ref{eq:xsecapp1}-\ref{eq:xsecapp2}) are shown by black lines. This plot is done using $\Delta s=0$.
}\label{fig:xsecs}
\end{figure}

For $E_\nu$ between 10 to 50~MeV, the total pES and IBD cross sections are well approximated by 
\be
\label{eq:xsecapp1}\sigma_{\rm pES}&\approx&1.7\times10^{-41}\left(\frac{E_\nu}{\rm30~MeV}\right)^2\left(1-1.5\Delta s\right),\\
\label{eq:xsecapp2}\sigma_{\rm IBD}&\approx&6.1\times10^{-41}\left(\frac{E_\nu}{\rm30~MeV}\right)^2.
\ee
The IBD and pES total cross sections are shown in Fig.~\ref{fig:xsecs}. Here, for concreteness, we use $\Delta s=0$.

\section{On the adiabaticity condition for neutrino oscillations in an accretion-dominated CCSN neutrino burst}\label{app:osc}

In the absence of self-induced oscillations, the propagation of neutrinos in matter reduces to a Schr\"{o}dinger equation describing the flavour transformation of the wave function of a neutrino with energy $E$ along the radial direction (for a review, see, e.g.~\cite{GonzalezGarcia:2002dz}),
\be\label{eom2}
i\dfrac{d}{dr}\begin{pmatrix}
\nu_e \\ \nu_\mu \\ \nu_\tau 
\end{pmatrix} & =  &  \left[\dfrac{1}{2E}U\begin{pmatrix}
-\delta m^2/2 & 0 & 0 \\ 0 & \delta m^2/2 & 0 \\ 0 & 0 & \pm\Delta m^2
\end{pmatrix}U^\dagger +\begin{pmatrix} \lambda_m & 0 & 0 \\ 0 & 0 & 0\\0 & 0 & 0\end{pmatrix}\right]\begin{pmatrix}
\nu_e \\ \nu_\mu \\ \nu_\tau 
\end{pmatrix}.
\ee
Here, $U$ is the PNMS matrix~\cite{Maki1962a,Kobayashi1973}, $\delta m^2\equiv m_2^2-m_1^2$, positive by convention, is the solar mass-squared difference and $\Delta m^2\equiv  \left|m_3^2-(m_1^2+m_2^2)/2\right|$ is the atmospheric mass-squared difference. The sign $ \pm $ on $\Delta m^2$ corresponds to the mass hierarchy, with $(+)$ for NH and $(-)$ for IH. In our main analysis we use the best-fit values from~\cite{Esteban:2016qun,nufit} for the oscillation parameters. The matter-induced potential~\cite{Wolfenstein1978,Mikheev1986} is given by $ \lambda_m=\sqrt{2}G_F(n_{e^-}-n_{e^+}) $, where $n_{e^\pm}$ is the $e^\pm$ density in the medium. For anti-neutrinos one replaces $ \lambda_m\to -\lambda_m $. 
 In the limit $ |\lambda_m|\gg \Delta m^2/2E $ the electron and anti-electron neutrino states are propagation eigenstates. This limit holds very well in the emission region of CCSN neutrinos ($ r\lesssim 100$~km), where $ \lambda_m \gtrsim 10^5\,\text{km}^{-1} $ and $ \Delta m^2/2E \lesssim 1\,\text{km}^{-1} $.

In a region with smoothly-varying matter density, and away from resonances of Eq.~(\ref{eom2}), an adiabatic limit applies in which a neutrino prepared in a propagation eigenstate will remain in it. When the adiabatic limit does not apply, probability leaks into other states.  Since $ \Delta m^2\gg \delta m^2 $, one can factor the analysis into two regimes, where in each regime there is a pair of states which may approach a resonance region and a third decoupled state which remains far from resonance. Defining 
\be\omega_H&=& \frac{\Delta m^2}{2E},\;\;\;\;\omega_L= \frac{\delta m^2}{2E} , \ee
the region where $\omega_H\sim\lambda_m$ is known as the $ H $ resonance, and the region where $\omega_L\sim\lambda_m$ the $ L $ resonance.
In an effective two-level system with mixing angle $ \theta_t $ and mass difference $ \Delta m_t^2 $,  the degree of adiabaticity can be analysed using the adiabaticity parameter $\gamma$, defined by
\be\label{adiabf}
\gamma &=&  \frac{\Delta m_t^2 \sin^22\theta_t}{2E\cos 2\theta_t \left|\frac{1}{\lambda_m}\frac{d\lambda_m}{dr}\right|}.
\ee
The crossing probability for a neutrino traversing the resonance region is given by
\be\label{Padiab}
P&\approx& \exp\left(-\frac{\pi}{2}\gamma F\right),
\ee
where $ F $ is an $\mathcal{O}(1)$ parameter that depends on the matter distribution~\cite{Kuo1989}.

Shock fronts present a singular density profile, requiring special treatment using an abrupt approximation. The crossing probability when passing through a shock front is given by~\cite{Kuo1989}
\be\label{Psform}
P=\sin^2(\theta_m^+-\theta_m^-),
\ee
where $\theta_m$ denotes the effective mixing angle in matter and the superscript $ + $ ($ - $) denotes the mixing angle after (before) crossing the shock. In a 2-flavour case, the effective mixing angle is given by
\eqspla{aneffangle}{
\tan 2\theta_m = \dfrac{\sin 2\theta }{\cos 2\theta -2E \lambda_m/\Delta m^2} \; .
}
In the general $ 3 $-flavour case, the amplitude for crossing between different propagation eigenstates is given by
\be
{\rm Amp}(\nu_i^{m(-)}\to \nu_j^{m(+)}) & =& \left[U^{m(-)\dagger}U^{m(+)}\right]_{ij}\ee
such that the crossing probability is
\be\label{eq:Pproj3}
P(\nu_i^{m(-)}\to \nu_j^{m(+)}) & =& \left|\left[U^{m(-)\dagger}U^{m(+)}\right]_{ij}\right|^2.\ee
In practice we find $ U^m(r) $ by diagonalizing the right hand side of Eq.~(\ref{eom2}).

\subsection{Resonance region analysis with MESA}\label{as:mesa}
\begin{figure}[h!]
	\begin{center}
		\includegraphics[width=0.7\textwidth]{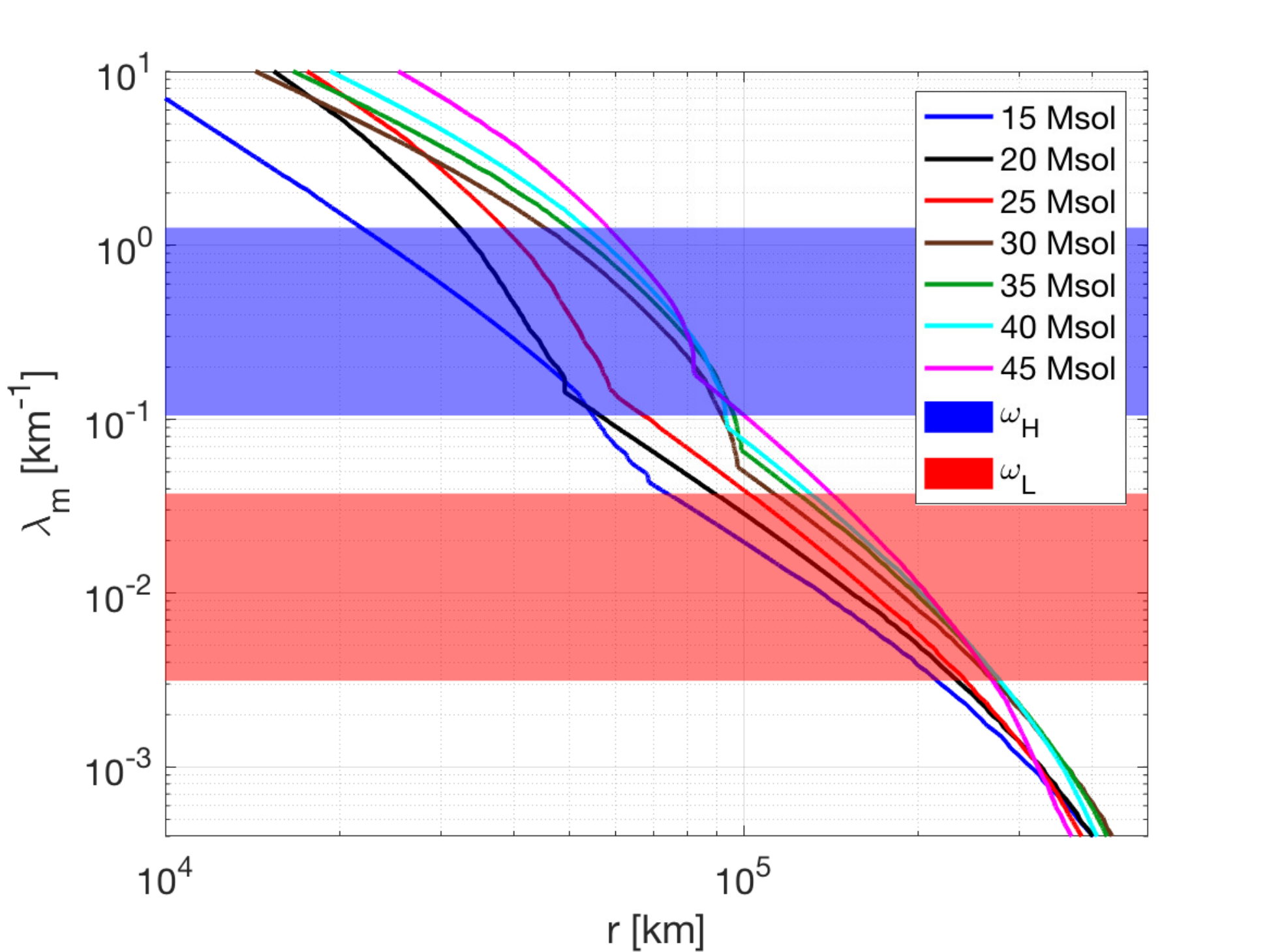}	
	\end{center}
	\caption{Matter-induced potential $\lambda_m$ for stellar profiles calculated with MESA. Each curve corresponds to a progenitor, distinguished by its ZAMS mass. The shaded areas are $ \omega_H $ and $ \omega_L $, as defined in the text. For the width of the shaded area we set $ E\in [5,60]$~MeV.}\label{mesapot}
\end{figure}

We now investigate the adiabaticity parameter $\gamma$ during the first few seconds of the CCSN.  The first tool we employ is a set of non-rotating pre-collapse progenitor stellar profiles, calculated by Roni Waldman using the MESA stellar evolution code \cite{Paxton2011}\footnote{We thank Doron Kushnir for providing us with the MESA profiles.}. The zero-age main-sequence (ZAMS) stellar masses vary from $ 15 $ to $ 45\,M_{\odot} $. The masses at the time of collapse vary non-monotonically between $ \approx 13 $ to $ 24.4\,M_{\odot} $ due to mass loss by winds. 
In Fig.~\ref{mesapot} we plot the matter-induced potential $ \lambda_m $ for the MESA profiles, and compare it with the vacuum inverse-oscillation wavelengths $ \omega_L $ and $ \omega_H $.

Fig.~\ref{mesapot} gives a stationary picture of the resonance regions before core-collapse. However, in the first few seconds of the collapse, this picture is sufficient for the analysis of flavour conversion in the CITE scenario. The reason for this is that for all of the profiles in Fig.~\ref{mesapot}, resonance regions ($\omega\sim\lambda_m$) begin only at $ r \gtrsim 10^4$~km. When collapse starts, a shell located initially at radius $r_0$ in the envelope of the star starts collapsing when the rarefaction wave reaches it, in a time~\cite{Kushnir:2014oca}\footnote{This timing analysis can be refined by integrating the speed of sound $ c_s $ from MESA simulations, $ t_{fall}(r)=\int_0^r c_s^{-1}(r^{\prime})dr^{\prime} $. This agrees with Eq.~(\ref{eq:tfall}) to 10\% accuracy.}\footnote{Note that $t_{fall}$ is measured from the onset of core-collapse, when the rarefaction wave heads out towards the envelope, and not from core-bounce (when the $\bar\nu_e$ burst begins). Expressed as post-bounce time, we should subtract the homologous core-collapse time so that $t_{fall,PB}\approx t_{fall}-(0.3~{\rm sec})$. This small correction does not affect the results in this section.}
\be\label{eq:tfall}
t_{fall}(r_0)&\approx&\frac{\pi\,r_0^{\frac{3}{2}}}{\sqrt{8GM(r_0)}}.
\ee
Using the MESA output and plugging for $r$ the first occurrence of the ($H$) resonance, we find a minimal time for the $ 15\,M_{\odot} $ progenitor of $t_{fall} \approx 6 $~sec.  Thus, Fig.~\ref{mesapot} shows the correct structure of the resonance regions during the first few seconds after core-collapse.

Evaluating Eq.~(\ref{adiabf}) for the MESA profiles, we find $ \gamma>100 $ in both of the $\omega_{L,H}$ resonance bands, except at composition jumps, which are visible in Fig.~\ref{mesapot} mostly in the $ \omega_H $ band. Treating these jumps in the abrupt approximation (as outlined in Eqs.~\ref{Psform}-\ref{eq:Pproj3}) may result in crossing probability $ P_{\bar{H}} \sim \mathcal{O}(0.1)$ (therefore somewhat affecting the IH prediction for the IBD channel, cf. Eq.~\ref{eq:PeeAn2}). Direct numerical integration of Eq.~\ref{eom2} with the MESA profiles does not show any significant crossing probability. Composition jumps deserve a more thorough analysis, taking into account the detailed physical conditions in these regions. Such analysis, beyond the scope of the current work, is needed in order to verify that indeed composition jumps do not alter the adiabatic MSW prediction. We conclude that matter-induced flavour conversion in the CITE scenario is most-likely adiabatic, supporting our choice of $P_L=P_H=0$ in Sec.~\ref{sec:osc}.

\subsection{PNS accretion shock}\label{as:pns}

\begin{figure}[h!]
\begin{center}
	\includegraphics[width=0.7\textwidth]{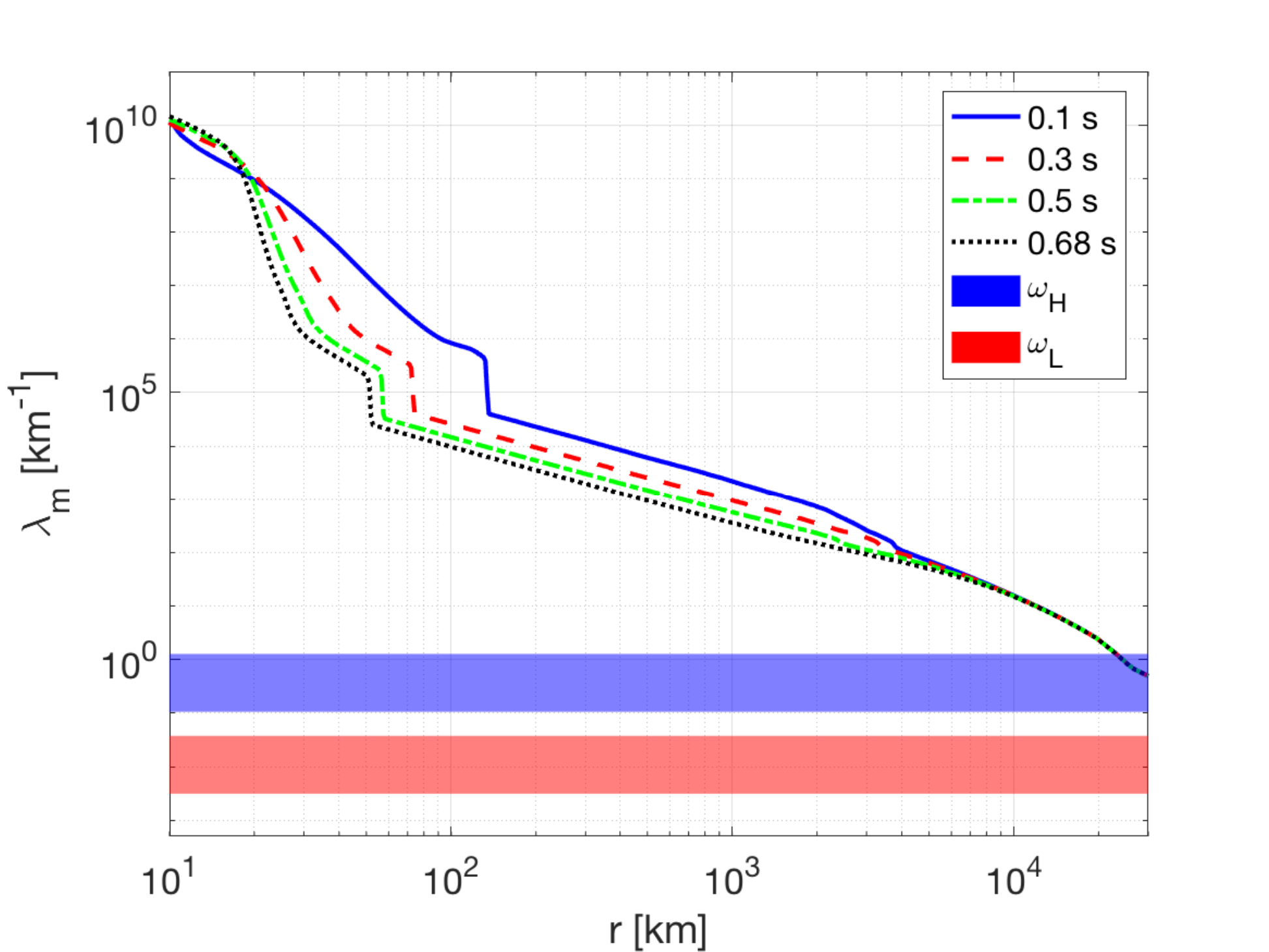}
\end{center}
\caption{Matter-induced potentials at different times based on the output of a single GR1D simulation of a $ 15\,M_{\odot} $ progenitor with \texttt{SLy4} nuclear equation of state. Listed times are post-bounce. The shaded areas are $ \omega_H $ and $ \omega_L $, as defined in the text. We use $ E\in [5,60]\, $MeV. The step-like feature at $ r\sim100\, $km is the stalled bounce-shock. Since it is high above the resonance bands, any crossing probability is extremely negligible.}\label{grid1}
\end{figure}

We complement the stationary MESA analysis with a dynamical study of the PNS accretion shock during the first 1~sec from core-collapse, using GR1D. In Fig.~\ref{grid1} we plot the matter-induced potential and the resonance bands for different post-bounce times in the core collapse simulation of a 15~$M_\odot$ star. The PNS accretion shock is seen as a sharp density drop, receding inwards slightly from $r\gtrsim100$~km at $t_{\rm PB}=0.1$~sec to $r\sim50$~km at  $t_{\rm PB}=0.68$~sec. The shock occurs far from resonance, at $\lambda_m/\omega_H\gtrsim10^4$. Eq.~(\ref{Psform}) then leads to negligible crossing probability, $P_L<P_H=\mathcal{O}\left(\omega_H^2/\lambda_m^2\right)$.

\subsection{Rotation-induced accretion shock}\label{as:rias}

Refs.~\cite{Kushnir:2014oca,Kushnir:2015mca,Kushnir:2015vka} found that successful CITE requires rotating stars. Rotation was needed to produce a rotation-induced accretion shock (RIAS), which propagates outward through the infalling material and ignites the thermonuclear detonation once it encounters a mixed O-He layer. The RIAS is formed during the collapse of the envelope, when a rotating mass shell hits a centrifugal barrier. At the base of the RIAS, an accretion disk forms where the matter density is compressed and heated. Accretion disk neutrino emission in this scenario was studied in Ref.~\cite{Blum:2016afe} using 2D numerical simulations.

CITE appears not to be very sensitive to the precise time, $t_{disk}$, at which the RIAS is launched. Ref.~\cite{Kushnir:2015mca} found strong explosions with $t_{disk}>10$~sec, but in~\cite{Blum:2016afe} CITE was demonstrated to operate also with $t_{disk}\sim5$~sec. We do not know if, e.g., $t_{disk}\sim1$~sec would be consistent with a successful SN in CITE. 

If $t_{disk}$ is larger than a few seconds, then the continued accretion onto the PNS leads to BH formation on a time $t_{BH}<t_{disk}$. BH formation abruptly halts (within a dynamical time $\sim$~ms) the neutrino emission at $t_{BH}$. The neutrino burst can be rekindled later on, with reduced but potentially observable luminosity, at $t=t_{disk}$. Analysing the neutrino burst of SN1987A, Ref.~\cite{Blum:2016afe} showed that an abrupt break in the neutrino light curve at $t_{BH}\approx2.7$~sec, followed by accretion disk luminosity starting at $t_{disk}\approx5$~sec, is consistent with SN1987A data.

However, as discussed above, low $t_{disk}\sim1$~sec cannot be excluded on theoretical grounds at this point. In this case, disk formation may precede BH formation, $t_{disk}<t_{BH}$. If that happens, then detectable neutrino emission from the disk may continue across $t>t_{BH}$ without being completely cut-off by the BH formation, smoothing out the spectacular signal of BH formation promised for $t_{disk}>t_{BH}$. 
We now estimate the impact of disk formation for neutrino flavour conversion, relevant for $t_{disk}<t_{BH}$. Again, the important question to address is the level of adiabaticity for neutrino propagation. 

Consider a rotating star, with the rotation on the $z=0$ plane parametrised by a function $f(r)$, specifying the ratio between the centrifugal force and the gravitational force. We assume $f\ll1$. Once collapse begins, a shell starting at radius $r_0$ begins to fall at time $t_{fall}$ given by Eq.~(\ref{eq:tfall}). If no shock meets it, the shell hits a centrifugal barrier when it attains a radius $r_{disk}=(f/2)r_0$. To lowest order in $f$, this occurs at a time
\be t_{disk}&\approx&2\,t_{fall},\ee
independent of $f$. Just upstream of $r_{disk}$ the matter density is compressed by the factor
\be\frac{\rho(r_{disk})}{\rho_0(r_0)}&\approx&\frac{8\sqrt{2}}{\pi\gamma_\rho(r_0)}f^{-\frac{3}{2}},\label{eq:comp}\ee
where $\rho_0(r)$ is the pre-collapse density at $r_0$ and $\gamma_\rho=3-\frac{d\log M}{d\log r}$. (For $\rho_0(r)\propto r^{-\Gamma}$, for example, we have $\gamma_\rho=\Gamma$.) 
For reasonable values of $f$ between 0.01 to 0.1 and $\gamma_\rho$ between 1 to 3, Eq.~(\ref{eq:comp}) gives a compression factor $\rho(r_{disk})/\rho_0(r_0)$ ranging between 40 to $3000$. 

 Referring back to Fig.~\ref{mesapot} and the discussion around Eq.~(\ref{eq:tfall}), and noting that the pre-collapse matter potential $\lambda_m$ in Fig.~\ref{mesapot}, when evaluated in the disk region at $t_{disk}$ during the collapse, is enhanced by the compression factor Eq.~(\ref{eq:comp}), we conclude that if an accretion disk forms at $t_{disk}<5$~sec, then it is located in a deep adiabatic region for neutrino propagation, $\lambda_m\gg\omega_H,\omega_L$.

\end{appendix}

\bibliography{ref}

\end{document}